\documentclass{article}
\usepackage{amsmath}
\usepackage{amsfonts}
\usepackage{makeidx}
\usepackage{cite}
\usepackage{hyperref}
\usepackage[toc,page]{appendix}
\usepackage{graphicx}

\hypersetup{linktoc=page}
\newtheorem{theorem}{Theorem}
\newtheorem{acknowledgement}[theorem]{Acknowledgement}

\newenvironment{proof}[1][Proof]{\noindent\textbf{#1.} }{\ \rule{0.5em}{0.5em}}
\input{tcilatex}
\begin{document}

\title{Unification of Mixed Hilbert-Space Representations in Condensed
Matter Physics and Quantum Field Theory}
\author{Felix A. Buot,$^{1,2,3}$Gibson T. Maglasang,$^{1,3}$and Allan Roy B.
Elnar$^{1,3}$ \\
$^{1}$CTCMP, Cebu Normal University, Cebu City 6000, Philippines\\
$^{2}$C\&LB Research Institute, Carmen, Cebu 6005, Philippines\\
$^{3}$LCFMNN, University of San Carlos, Talamban,\\
Cebu City 6000, Philippines}
\maketitle

\begin{abstract}
We present a unification of mixed-space quantum representations in Condensed
Matter Physics (CMP) and Quantum Field Theory (QFT). The unifying formalism
is based on being able to expand any quantum operator, for bosons, fermions,
and spin systems, using a universal basis operator $\hat{Y}\left( \mathfrak{%
u,v}\right) $ involving mixed Hilbert spaces of $\hat{P}$ and $\hat{Q}$,
respectively, where $\hat{P}$ and $\hat{Q}$ are momentum and position
operators in CMP (which can be considered as a bozonization of free Bloch
electrons which incorporates the Pauli exclusion principle and Fermi-Dirac
distribution), whereas these are related to the creation and annihilation
operators in QFT, where $\hat{\psi}^{\dagger }=-i\hat{P}$ and $\hat{\psi}=%
\hat{Q}$. The expansion coefficient is the Fourier transform of the Wigner
quantum distribution function (lattice Weyl transform) otherwise known as
the characteristic distribution function.

Thus, in principle, fermionization via Jordan-Wigner for spin systems, as
well as the Holstein--Primakoff transformation from boson to the spin
operators can be performed depending on the ease of the calculations.
Unitary transformation on the creation and annihilation operators themselves
is also employed, as exemplified by the Bogoliubov transformation. Moreover,
whenever $\hat{Y}\left( \mathfrak{u,v}\right) $ is already expressed in
matrix form, $M_{ij}$, e.g. the Pauli spin matrices, the Jordan--Schwinger
transformation is a map to bilinear expressions of creation and annihilation
operators which expedites computation of representations.

We show that the well-known coherent states formulation of quantum physics
is a special case of the present unification. \ A new formulation of QFT
based on Q-distribution of functional-field variables is suggested. The case
of nonequilibrium quantum transport physics, which not only involves
non-Hermitian operators but also time-reversal symmetry breaking, is
discussed in the Appendix.

\pagebreak
\end{abstract}

\tableofcontents

\section{Introduction}

The canonical conjugate variables formulation of quantum physics has been
thoroughly established and seems to characterizes classical limits and
quantum dynamics. Indeed, the discrete quantum mechanics and mixed
representations, namely, the lattice Weyl-Wigner (W-W) quantum physics has
been developed in condensed matter physics with its various successful
applications \cite{buot1, buot2, buot3, buot3-1,buot4, buot5, buot6, buot7,
buot8, buot9, direct,jb,comments}. Here we extend the formalism to
non-Hermitian operators, where translation operators or quantum state
generators are still well-defined, as well as transition function between
the dual Hilbert spaces. The resulting general formalism unifies all
mixed-space representations on quantum mechanics, Hermitian and
non-Hermitian operators and includes coherent state representation \cite%
{coher} as a special case. It has been employed by one of the authors in
calculating the magnetic susceptibility of interacting many-body Bloch
electrons \cite{many}, and in direct construction of fermionic path
integrals \cite{direct}. First, we give an introduction on the physical
basis of the discrete W-W nonequilibrium quantum transport theory \cite%
{transport}. One of the important aspects of the formalism is the bijective
pairing of the elements of the dual spaces, known in mathematics as
Pontryagin duality, which allows generalization to other dual spaces. The
present unification naturally leads us to propose a new formulation of QFT
based on the Q-distribution of functional-field conjugate variables, which
is expected to avoid infinities and the need of artificially imposed cut-off
ubiquitous in QFT. Consequently, time-dependent problems can then be dealt
as a nonequilibrium QFT transport physics in place of S-matrix theory.

The case of nonequilibrium quantum transport physics of many-body in CMP,
which not only involves non-Hermitian operators but also time-reversal
symmetry breaking, deserves a separate treatment. This is given in the
Appendices. By virtue of the doubling of degrees of freedom, there it is
expected that the dual spaces consist of chronological and
anti-chronological solutions of the quantum Liouville equation, instead of
dual spaces made through ordinary operator conjugation in many-body physics
or quantum field theory.

\subsection{Physical models for discrete quantum mechanics}

Historically, there are already l\textit{ong-standing} \textit{existing
quantum models in physics} that have physically guided the discrete
phase-space physics of the lattice W-W formulation, namely, (a) \textit{%
Localized Wannier function} and \textit{extended Bloch function} for
discrete lattices in solid state physics, obeying the Born-von Karman
boundary condition (strictly speaking, obeying \textit{modular arithmetic
based on finite fields, }akin to a group theory of integers), (b) Dirac
delta function and plane waves in the continuum limit, i.e., \textit{only
for continuous coordinate space can one have continuous momentum space}
(this bijective quantum-mechanical canonical variables is disregarded in
some of the W-W formalism of lattice models, where \textit{discrete lattice}
coordinates is unphysically paired with \textit{compact continuous} momentum
space \cite{fial, liga, kasper}).

In both (a) and (b), we have the eigenvector for positions (or discrete
lattice position), $\left\vert q\right\rangle $, and eigenvector for
momentum (or discrete crystal momentum), $\left\vert p\right\rangle $. Of
course, all respective eigenspaces go to continuum spaces in the limit of
lattice constant goes to zero as in (b).

These respective eigenvectors, $\left\vert q\right\rangle $and $\left\vert
p\right\rangle $, in (a) and (b) are bijectively related by Fourier
transformation, $\left\vert q\right\rangle $to $\left\vert p\right\rangle $
via the transition function $\left\langle p\right. \left\vert q\right\rangle
,$and often produces results \textit{akin} to quantum uncertainty principle
in their probabilistic continuous coordinate components. Thus, to construct
a physically-based discrete phase space or discrete W-W quantum physics one
must be guided by the following observations.

\subsection{Discrete phase space on finite fields}

The original formulation of lattice-space discrete Weyl transform and
discrete WDF\footnote{%
The original formulation is entitled, "Method for Calculating $Tr$$\mathcal{H%
}^{n}$in Solid State Theory", Phys. Rev., \textbf{B10}, 3700-3705 (1974).}
is based on crystalline solid with inversion symmetry, and hence based on an
odd number of discrete lattice points, $(q,p)$, obeying the Born-von Karman
boundary condition. Thus, this formulation is generally based on a finite
field represented by a finite prime number, $N$, of lattice points obeying
modular arithmetic, closed under addition and multiplication\footnote{%
There are\textit{\ infinitely} many prime numbers. Examples of prime numbers
are: $2$, $3$, $5$, $7$, $11$, $13$, $17$, $19$, $23$, $29$, $73$, $79$, $83$%
, $89$, $97$, $101$, $103$, $107$, $179$, $181$, $191$, $193$, $197$, $199$, 
$211$, $223$, $227$, and $229$, etc. When written in base $10$, all prime
numbers except $2$and $5$end in $1$, $3$, $7$or $9$.}. The presence of
multiplicative inverses in the formulation assumes that the finite number of
lattice points is a prime number, since primality is required for the
nonzero elements to have multiplicative inverses. Indeed, in the limit that
the number of lattice points is very large, the lattice point coordinates
obeying the Born-von Karman boundary condition, assumes the field of prime
integers\footnote{%
A review article by Kasperkovitz and Peev, Ann. Phys. \textbf{230}, 21
(1994) and more receently by Fialkovsky et al \cite{fial} makes several
misleading statements about the discrete phase-space formulation of the
quantum theory of solids by failing to recognize the finite-field aspects of
the theory.}

\subsection{Transition functions between dual Hilbert spaces}

The existence of transition function between dual spaces can be easily
established for Hermitian operators leading to bijective discrete Fourier
transformation. The physically-based construction of W-W formalism in
condensed-matter physics is guided by well-known aspects of solid-state
physics \cite{buot1}, namely, (a) The invariance of this physical and
canonical scheme of \textit{complete and orthogonal} set of \ $\left\{
\left\vert q\right\rangle \right\} $and \textit{complete and orthogonal set}
of $\left\{ \left\vert p\right\rangle \right\} ,$in going from discrete to
continuum physics, and (b) in the compactification of the lattice, the
number of discrete lattice points (and hence the number of discrete momentum
points) must be an \textit{odd} prime number for obvious inversion symmetry
reason. Moreover, all arithmetic oprerations on this group of numbers must
be closed, i.e., all arithmetic operation must be a modular arithmetic with 
\textit{prime number modulus} (\textit{akin to a group operation on prime
number of integers}). In short, all arithmetic operation is a modular
arithmetic based on finite fields, since only for finite fields with prime
number modulus does every nonzero element have well-defined multiplicative
inverse, and hence modular division operation also provides closure. Thus,
these dual set of spaces, $\left\{ \left\vert q\right\rangle \right\} $and $%
\left\{ \left\vert p\right\rangle \right\} $, is connected by transition
function which defines the lattice bijective Fourier transform.

\subsection{The direct and reciprocal lattice in condensed-matter physics}

We first summarize the Bravais lattice vectors and their corresponding
reciprocal lattice vectors, since this points to symmetry properties of the
discrete W-W formulation in condensed matter physics. In $2$-D lattice, we
have the reciprocal lattice vectors, $\vec{b}_{1}$and $\vec{b}_{2}$given by
the matrix,%
\begin{equation}
\left( 
\begin{array}{cc}
\vec{b}_{1} & \vec{b}_{2}%
\end{array}%
\right) =\frac{1}{\hat{n}\cdot \left( \vec{a}_{1}\times \vec{a}_{2}\right) }%
\left( 
\begin{array}{cc}
\vec{a}_{2}\times \hat{n} & \hat{n}\times \vec{a}_{1}%
\end{array}%
\right)  \label{eq. 2-4}
\end{equation}%
which geometrically means that $\vec{b}_{1}$is perpendicular to $\vec{a}_{2} 
$and $\vec{b}_{2}$perpendicular to $\vec{a}_{1}$. The $\hat{n}$is the unit
vector normal to the $2$-D lattice plane.

In $3$-D lattice, the reciprocal lattice vectors, $\vec{b}_{1}$, $\vec{b}%
_{2} $, and $\vec{b}_{3}$are given by,%
\begin{equation}
\left( 
\begin{array}{ccc}
\vec{b}_{1} & \vec{b}_{2} & \vec{b}_{3}%
\end{array}%
\right) =\left( 
\begin{array}{ccc}
\frac{\vec{a}_{2}\times \vec{a}_{3}}{\vec{a}_{1}\cdot \left( \vec{a}%
_{2}\times \vec{a}_{3}\right) } & \frac{\vec{a}_{3}\times \vec{a}_{1}}{%
\left( \vec{a}_{3}\times \vec{a}_{1}\right) \cdot \vec{a}_{2}} & \frac{\vec{a%
}_{1}\times \vec{a}_{2}}{\left( \vec{a}_{1}\times \vec{a}_{2}\right) \cdot 
\vec{a}_{3}}%
\end{array}%
\right)  \label{eq 2-5}
\end{equation}%
The above formulas are independent of any choosen coordinate system. We have 
\begin{equation}
\left( 
\begin{array}{ccc}
\vec{a}_{1} & \vec{a}_{2} & \vec{a}_{3}%
\end{array}%
\right) \left( 
\begin{array}{ccc}
\vec{b}_{1} & \vec{b}_{2} & \vec{b}_{3}%
\end{array}%
\right) ^{T}=I  \label{eq. 2-7}
\end{equation}

\subsection{Translation symmetry}

The one eclectic concept in solid state physics is the concept of
translation symmetry along any symmetry directions of the lattice. Inversion
symmetry of the lattice structure is also among the common symmetry
properties. Thus, the discrete phase space W-W formalism in condensed matter
physics \cite{trHn} is compatible with any of the lattice structures defined
by Eq. (\ref{eq. 2-4}) -- (\ref{eq 2-5}), i.e., not limited to cubic lattice
structures only.

\subsubsection{Generalization to other discrete quantum and classical systems%
}

The guidance of (a) and (b) allow us to generalize discrete phase space
based on finite fields to be useful when the quantum numbers specifying the
quantum states, are discrete configurations other than the particle position
and momentum \cite{buot8},\ even to a system describable by a smallest prime
number $2$. A simplest example is that of quantum bit or qubit, like a
two-level atomic systems. Another example lies in the two-state entangled
diagrams of mutli-qubit systems \cite{mechanical}, where the entangled basis
states are connected by the Hadamard lattice transform.

\subsubsection{Generalized operator basis}

Another generalization has to do in the construction of\textit{\ }the Wigner
distribution function. This is done through the generalized projection
operators, sometimes referred to as the generalized Pauli spin operators for
both othogonal and biorthogonal spaces. This will be made clear later.
Indeed, the construction can be generalized to the spaces of non-Hermitian
operators or biorthogonal systems. The power of using finite fields is that
one can also generalized the discrete Wigner distribution construction based
on the algebraic concept of finite fields, which are extension of prime
fields, where $q$and $p$are field elements ($mod$\textit{irreducible
polynomial}), useful in quantum computing, visualization, and
communication/information sciences. Here we have $p^{n}$elements for some
prime $p$and some integer $n$$>1$useful for constructing the Wigner
distribution function for spin-$\frac{1}{2}$systems \cite{gibbons, buot8}.

\subsubsection{One-to-one mapping of discrete momentum and lattice position
spaces}

Each energy band corresponds to the splitting of the energy levels of one
atomic site into $N$levels where $N$is the number of lattice sites in a 
\textit{compactified} Bravais lattice obeying the Born-von Karman boundary
condition in a given symmetry direction. This is the basis of energy band
quantum dynamics. Therefore, the number of crystal momentum states in each
band (in Brillouin zone) is exactly equal to the number of lattice sites.
Hence, there has to be a bijective mapping between number of discrete
lattice sites and the number of discrete crystal momentum states. This is
the essence of the powerful theoretical concept of localized function around
each lattice site, the \textit{Wannier function }$\left\vert q\right\rangle $%
, and the extended function over all lattice points, the \textit{Bloch
function }$\left\vert p\right\rangle $, related through the bijective
discrete Fourier transformation. The bijectivity and finite fields aspects
are crucial in the mathematical manipulations to avoid ambiguities.

It is worth cautioning that the use of discrete lattice models coupled with
the \textit{compact} \textit{continuous} momentum space in some W-W
formulations renders an ill-defined transirion function, by incurring a
non-bijective canonical conjugate dynamical variables, e.g., absence of
bijective transformation between momentum $\left\vert p\right\rangle $and
coordinate $\left\vert q\right\rangle $. This, and \textit{together }with
the lack of modular arithmetic on finite fields not invoked at all is adding
to several more ambiguities.

\subsubsection{Bijective Fourier transformation (Pontryagin duality)}

In short, the crystal momentum space is essentially discrete and yield a
bijective mapping to the discrete lattice sites through a discrete Fourier
transformation. Moreover, dual Hilbert spaces related by generalized Fourier
transform also endows quantum uncertainty relations, well known in classical
probability theory \cite{stratonovich}.

The Bloch function space is orthogonal and complete and so is the Wannier
function space. Observe that these eigenfunctions of phase space operators
are well established for gapped energy band structures, or energy band far
removed from the other energy bands. Generalized Wannier function can also
be defined for coupled energy bands, using decoupling scheme like the
Foldy-Woutheysen transformation for relativistic Dirac electrons. Indeed, it
has been shown that counterparts of Wanner function and Bloch function exist
for the decoupled positive energy states of relativistic electrons, with the
'\textit{Dirac}-\textit{Wannier function}' localization about the size of
Compton wavelength \cite{buot3, buot3-1}. Electric Wannier function and
magnetic Wannier function also exist, as well as their respective Bloch
functions, for uniform external electromagnetic fields \cite{wannier}. This
physical idea has been extended to formally construct the discrete phase
space quantum mechanics \textit{based on finite fields} for cases where $q$
and $p$are not position and momentum variables, useful for quantum computing 
\cite{buot8}.

In general, the discrete phase space in condensed matter physics is based on
the mathematics of finite fields \cite{gibbons} and holds for any prime
number\footnote{%
Although every finite field, with $p^{n}$elements for some prime $p$and some
integer $n$$\succeq 1$, often deals with irreducible polynomials over ring $%
\mathbb{Z}
$of integers, or over\ field $%
\mathbb{Q}
$of rational numbers, or over field $%
\mathbb{R}
$of real numbers, or over field $%
\mathbb{C}
$of complex numbers, the role of irreducible polynomials can be played by
prime numbers themselves for $n=1$: prime numbers (together with the
corresponding negative numbers of equal modulus) are the irreducible
integers. They exhibit many of the general properties of the concept
'irreducibility' that equally apply to irreducible polynomials, such as the
essentially unique factorization into prime or irreducible factors: Every
polynomial $p(x)$in ring of polynomials with coefficients in $F$, denoted by 
$F\left[ x\right] $, can be factorized into polynomials that are irreducible
over $F$. This factorization is unique up to permutation of the factors and
the multiplication of constants from $F$to the factors.
\par
The simplest case of interest in Buot discrete W-W formulation is when $n=1$%
. In this case the finite field $GF\left( p\right) $is the ring $\frac{Z}{pZ}
$. This is a finite field with $p$elements, usually labelled $0,1,2,...p-1$,
where arithmetic is performed modulo $p$, where nonzero elements have
multiplicative inverses$.$} of lattice points obeying the Born-von Karman
boundary condition. i.e., with prime modulus. It even holds for the most
elementary prime number $2$of lattice points, to yield the $2\times 2$ Pauli
spin matrices and the well-known \textit{Hadamard transformation} between
\textquotedblleft Wannier function\textquotedblright\ and \textquotedblleft
Bloch function\textquotedblright , i.e., discrete Fourier transformation of
two points in \textquotedblleft phase space\textquotedblright\ leading to 
\textit{transformation of qubits }\cite{buot8}. Here the \textquotedblleft
Wannier function\textquotedblright\ and \textquotedblleft Bloch
function\textquotedblright\ has acquired the status of a simple theoretical
device for discrete quantum physics. Indeed, the Buot discrete phase-space
formalism also gives the generalized Pauli spin operators for any given
prime number of lattice points, which can be generalized to spaces of
non-Hermitian operators in coherent state representation. It has also
yielded all the entangled basis states, e.g., for two, three and \ four
qubits, crucial to the physics of quantum teleportation \cite{buot8}. In
other words, bijective pairing of different Hilbert spaces is crucial to
various generalization discussed in this paper.

\section{Mixed Space Representations in CMP}

The mixed $q$-$p$ representations is very useful in making analogy with
classical dynamics in terms of position and momentum eigenvalues. This has
also ushered a better understanding of the path integral formulations of
quantum field theory \cite{direct} to make analogy with classical partition
function of statistical physics, especially with the compactification of
time by the aid of Matsubara technique. This quantum field theory aspects
will be discussed later in this paper.

\subsection{\label{q-p_represent}The $q$-$p$ representation}

The mixed $q$-$p$ representation basically start by expanding any quantum
operator, $\hat{A}$, in terms of mutually unbiased basis states, namely the
eigenvector of position operator, $\hat{Q}$, and the eigenvector of momentum
operator, $\hat{P}$. We have%
\begin{eqnarray}
\hat{A} &=&\sum\limits_{p,q}\left\vert q\right\rangle \left\langle
q\right\vert \hat{A}\left\vert p\right\rangle \left\langle p\right\vert 
\notag \\
&=&\sum\limits_{p,q}\left\langle q\right\vert \hat{A}\left\vert
p\right\rangle \ \left\vert q\right\rangle \left\langle p\right\vert
\label{any_op}
\end{eqnarray}

\subsubsection{The completeness of mixed-space projector}

Thus, the set $\left\{ \left\vert q\right\rangle \left\langle p\right\vert
\right\} $is the basis operator for the mixed $q$-$p$representation. From
the completeness relations of the unbiased basis states, $\left\{ \left\vert
q\right\rangle \right\} $and $\left\{ \left\vert p\right\rangle \right\} $,
the set $\left\{ \left\vert q\right\rangle \left\langle p\right\vert
\right\} $obeys the completeness relation%
\begin{equation}
\sum\limits_{q,p}\left\langle q\right\vert \left\vert p\right\rangle
\left\vert q\right\rangle \left\langle p\right\vert =1  \label{complete1}
\end{equation}%
Substituting the expression for $\left\langle q\right\vert \left\vert
p\right\rangle $, 
\begin{eqnarray}
\left\langle q\right\vert \left\vert p\right\rangle &=&C_{o}e^{\frac{i}{%
\hbar }\vec{p}\cdot \vec{q}}  \label{qp_eq} \\
\left\langle p\right\vert \left\vert q\right\rangle &=&C_{o}^{\ast }e^{\frac{%
-i}{\hbar }\vec{p}\cdot \vec{q}}  \label{pq_eq}
\end{eqnarray}%
we obtained, for the completeness relation, 
\begin{equation}
\left( N\hbar ^{3}\right) ^{-\frac{1}{2}}\sum\limits_{q,p}e^{\frac{i}{\hbar }%
\vec{p}\cdot \vec{q}}\left\vert q\right\rangle \left\langle p\right\vert =1
\label{completeness}
\end{equation}%
where $C_{o}$is choosen as 
\begin{equation*}
C_{o}=\left( N\hbar ^{3}\right) ^{-\frac{1}{2}}
\end{equation*}%
Equation (\ref{completeness}) can be rewritten as%
\begin{equation}
\left( N\hbar ^{3}\right) ^{-\frac{1}{2}}\sum\limits_{q,p}\frac{\left\vert
q\right\rangle \left\langle p\right\vert }{\left\langle p\right\vert
\left\vert q\right\rangle }=1  \label{completeness2}
\end{equation}

Here we use the transformation identities in the mixed $q$-$p$
representations, 
\begin{eqnarray}
\left\vert p\right\rangle &=&\sum\limits_{q}\left\langle q\right\vert
\left\vert p\right\rangle \ \left\vert q\right\rangle  \label{eq_p} \\
\left\langle p\right\vert &=&\sum\limits_{q}\left\langle q\right\vert \
\left\langle p\right\vert \left\vert q\right\rangle \   \label{eq_p1} \\
\left\vert q\right\rangle &=&\sum\limits_{p}\left\langle p\right\vert
\left\vert q\right\rangle \ \left\vert p\right\rangle  \label{eq_q} \\
\left\langle q\right\vert &=&\sum\limits_{q}\left\langle p\right\vert \
\left\langle q\right\vert \left\vert p\right\rangle \   \label{eq_q1} \\
\left\langle p\right\vert \left\vert q\right\rangle &=&\exp \left( -\frac{i}{%
\hbar }p\cdot q\right)  \label{p2q}
\end{eqnarray}

\subsubsection{Covariant state vector and contravariant components}

The position operator, $\hat{Q},$and momentum operator, $\hat{P}$, obey the
following commutation relations. When operating on components, we have,%
\begin{equation*}
\left[ \hat{Q},\hat{P}\right] \left\langle x\right\vert \left\vert
q\right\rangle =i\hbar \ \left\langle x\right\vert \left\vert q\right\rangle
=i\hbar \ \psi \left( x,q\right)
\end{equation*}%
On the other hand, when operating on the eigenvector, $\left\vert
q\right\rangle $, 
\begin{equation*}
\left[ \hat{Q},\hat{P}\right] \left\vert q\right\rangle =-i\hbar \
\left\vert q\right\rangle
\end{equation*}%
The translation operator, $T\left( q^{\prime }\right) $, is defined as%
\begin{equation}
\exp \left( -\frac{i}{\hbar }q^{\prime }\cdot \hat{P}\right) \left\vert
q\right\rangle =\left\vert q+q^{\prime }\right\rangle  \label{coord_trans}
\end{equation}%
where%
\begin{eqnarray*}
\hat{P}\left\vert p\right\rangle &=&p\left\vert p\right\rangle \\
\hat{P}\left\vert q\right\rangle &=&i\hbar \frac{\partial }{\partial q}%
\left\vert q\right\rangle
\end{eqnarray*}%
whereas, 
\begin{equation}
\exp \left( ip^{\prime }\cdot \hat{Q}\right) \left\vert p\right\rangle
=\left\vert p+p^{\prime }\right\rangle  \label{mom_trans}
\end{equation}%
where%
\begin{eqnarray*}
\hat{Q}\left\vert q\right\rangle &=&q\left\vert q\right\rangle \\
\hat{Q}\left\vert p\right\rangle &=&-i\hbar \frac{\partial }{\partial p}%
\left\vert p\right\rangle
\end{eqnarray*}%
Equations (\ref{coord_trans}) and (\ref{mom_trans}) can be deduced from Eqs.
(\ref{eq_p}) - (\ref{p2q}).

\section{Physics of Commutators and Anti-Commutators}

Here, we proposed a couple of simple physical interpretation of commutators
and anti-commutators in quantum physics. To the author's knowledge, only
Schwinger \cite{schwinger} seems to have highlighted the physical
implications for bosons and fermions.

\subsection{ Conjugate field operators and quantum state generators}

Indeed, conjugate operators in CMP and QFT serve to generate new quantum
states.

\subsubsection{Space translation, time evolution, and state generation}

It is not often emphasized in the literature, as well as in physics
textbooks, that commutation and anti-commutation relation of canonically
conjugate operators do carry very important physical meanings \cite%
{schwinger}, ie., quantum state generation, and imply the concept of either
space translation (Wannier state generation) or time evolution, and in
quantum field theory it implies energy or excitation-number state generation.

\paragraph{Generator of Wannier functions}

We have for localized Wannier states at lattice position $q$, Eq. (\ref%
{coord_trans}),%
\begin{equation}
\left\vert q\right\rangle =\exp \left\{ -\frac{i}{\hbar }q\cdot P\right\}
\left\vert 0\right\rangle ,
\end{equation}%
where the operator $P=i\hbar \nabla _{q}$ is operating on the basis
eigenvector $\left\vert q\right\rangle $, since $P=-i\hbar \nabla _{x}$
acting on the components behaves contravariantly. Here, we basically made an
assumption in the above expression that there exist continuous \ function of 
$q$having an infinite radius of convergence, which are equal to $\left\vert
q\right\rangle $at the lattice points, i.e., we expand the exponential in
Eq. (\ref{translate-op-position}) as a well-defined Taylor series.

\paragraph{Canonical conjugate operators implies state generators}

All these results are reminiscent of well-known relations with other
canonical conjugate variables operating on wavefunctions, for example, we
have%
\begin{equation*}
\Psi \left( t+t_{0}\right) =\exp \left( -i\frac{t\ \mathcal{H}}{\hbar }%
\right) \Psi \left( t_{0}\right)
\end{equation*}%
and%
\begin{equation*}
\Psi \left( q+q_{0}\right) =\exp \left( i\frac{\hat{P}\cdot q}{\hbar }%
\right) \Psi \left( q_{0}\right)
\end{equation*}%
In other words, canonical conjugate variables acts as displacement operators
acting on its dual eigenvector space , respectively. This operation is also
known as generation of eigenvectors. This principle is useful when we deal
with quantum field theory later. In fact we shall see that the commutation
relation of annihilation and creation operator given by%
\begin{equation*}
\left[ \hat{a},\hat{a}^{\dagger }\right] =1
\end{equation*}%
imply the generation of eigenvector of $\hat{a}$as follows,%
\begin{equation}
\exp \left( \alpha \hat{a}^{\dagger }\right) \left\vert 0\right\rangle
=\left\vert \alpha \right\rangle  \label{alpha_gen}
\end{equation}%
where $\alpha $ is an eigenvalue of annihilation operator $\hat{a}$. This
implication towards quantum state generator holds for commutator and
anti-commutator relations, either fermions or bosons \cite{schwinger}.

As a side remark, it is not clear why time is not elevated to an energy
displacement operator in elementary quantum mechanics since it appears in
Poisson bracket operator on the same level as the position and momentum
variables, although energy displacement operator are ubiquitous in quantum
field theory.

\subsection{Anti-commutator and commutator}

\subsubsection{Entanglement-induced localization and delocalization}

In quantum transport theory of interacting \textit{chiral} (axial or
pseudovectors) degrees of freedom \cite{transport}, e.g., Landau orbits,
spin and pseudospins, the anti-commutator of scattering terms are identified
as a localization, characterized by a fixed point in phase space, whereas
commutator terms are associated with nonlocalization or quantum diffusion.
There, the concept of locality and nonlocality have been extended to
entangled chiral degrees of freedom in nonequilibrium quantum transport
physics.

\subsubsection{Particles and forces}

\begin{figure}[h]
\centering
\includegraphics[width=1.791in]{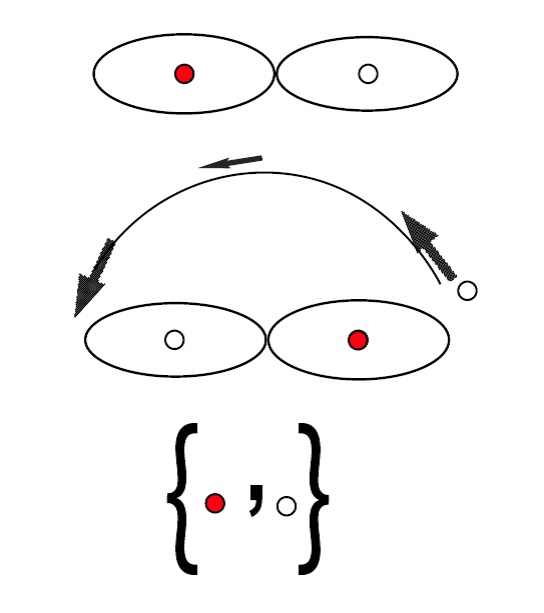}
\caption{A very simple intuitive explanation of spinors. The anti-commutator
represents local or non-diffusive term in quantum transport since repetition
of the anti-commutation operation just result in pure rotation of $\Psi $%
with a phase change of $\protect\pi $. Fermions are spinors and behave like
billiard balls under exchange, as indicated in the figure.}
\label{fig1}
\end{figure}

Indeed, anti-commutation is associated with a wavefunction property of a
rotating spinors where excitations have spatial sizes and behaving like
billiard balls upon exchange, resulting in the wavefunction rotation and
phase change of $\pi $. This is illustrated in Fig. \ref{fig1}.

\begin{figure}[h]
\centering
\includegraphics[width=2.4448in]{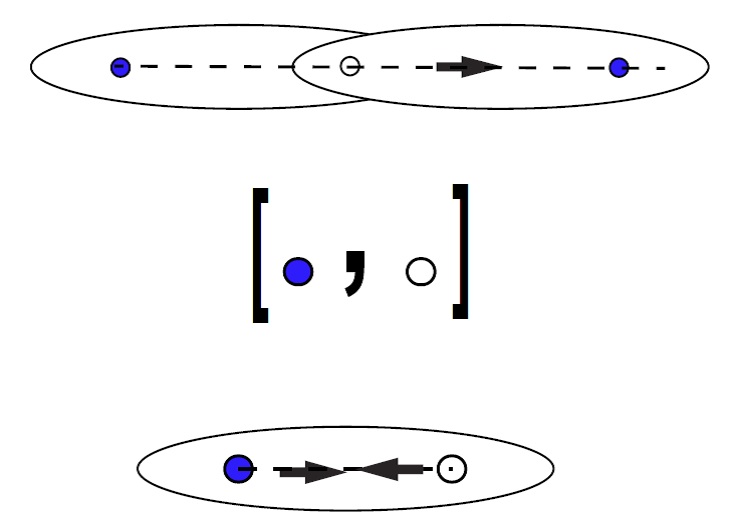}
\caption{A qualitative explanation and schematic representation of bosons as
obeying a nonlocal commutation relation. Bosons can pass through each other
so the wavefunction does not have to rotate to affect an exchange of
particles. Thus, there is no wavefunction phase change. The commutation
terms in quantum transport usually implies unitary evolution as well as a
nonlocal diffusive terms, since repeated commutation will result in
diffusion as schematically indicated in the figure. Note that the boson
particle can pass through the other boson particle in straight line, unlike
the fermions which must go around since the cannot occupy the same state as
indicated in Fig. (\protect\ref{fig1}). This is the reason that boson are
often nonlocal excitations with small masses. There is also an implied
overlap of their wavefunctions which allows for quantum diffusion. The
wavefunctions maintain their phase after exchange with effective diffusion.
The lower graph is a na\"{\i}ve interpretation of zero-point dynamics of
nonlocal bosons. Bosons have quantized energy and momentum and hence
essentially nonlocal or cannot be localized even for Bose condensate which
cannot be a solid.}
\label{fig2}
\end{figure}

On the other hand, a commutator of conjugate operators implies spatial
nonlocality of excitation wavefunctions, where energy is the one quantized
with indefinite non-local spatial dimensions, as depicted in Fig. \ref{fig2}%
. Even the zero-point energy is still non-local in space, which is the main
reason the lowest energy Bose condensate cannot form a solid. Clearly,
commutator cannot be characterized by a point, i.e., it is nonlocal and
excitation can conceptually pass through each other in the process of
exchange \cite{tonks, girardeau, lieb}. Indeed, Fig. \ref{fig2} seems to
suggest that only boson quasiparticles can move independently in one
dimensions.

\subsubsection{Particles versus quantized force fields: Nambu-Goldstone
bosons and the Higgs bosons}

\begin{figure}[h]
\centering
\includegraphics[width=3.2482in]{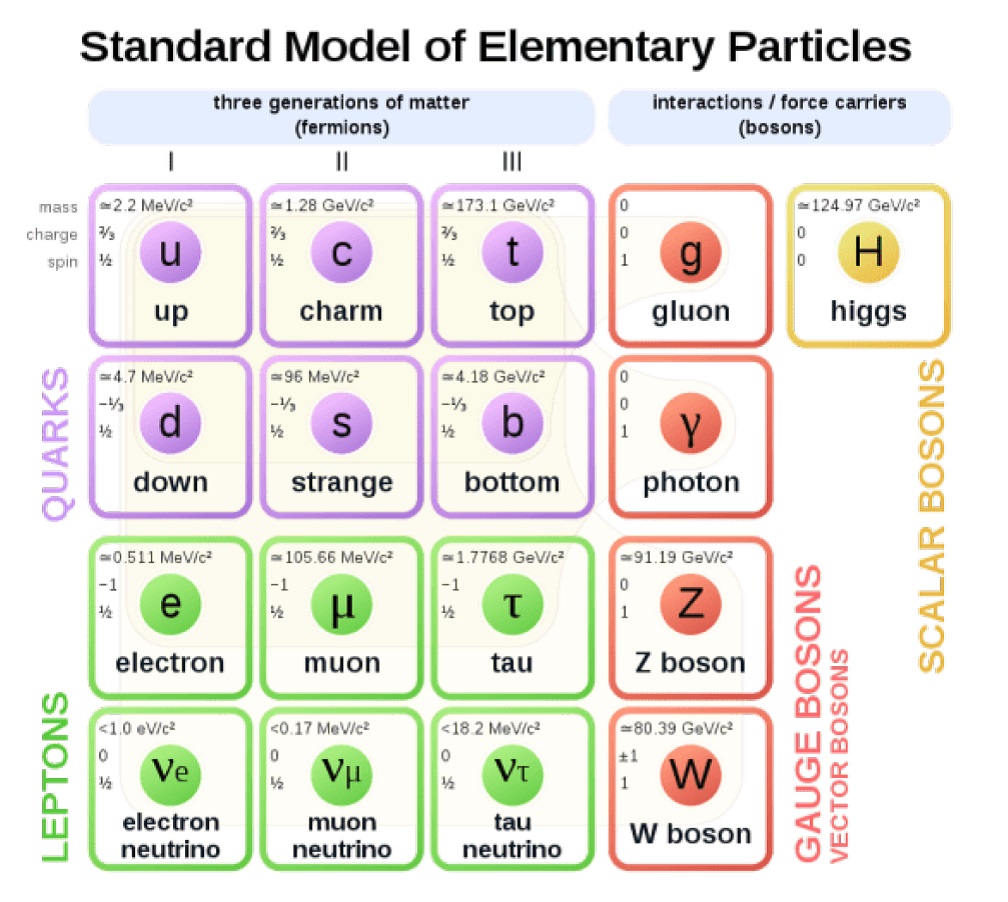}
\caption{In the standard model of particle physics, there are 4 major types
of gauge bosons namely, photons, W bosons, Z bosons, and gluons. Photons are
particles that carry electromagnetic interactions, while W and Z bosons tend
to carry weak interactions, and gluons can carry strong interactions.
Reproduced from:
[https://www.differencebetween.com/difference-between-baryons-and-mesons/%
\#Mesons].}
\label{fig6}
\end{figure}

\begin{figure}[h]
\centering
\includegraphics[width=5.2001in]{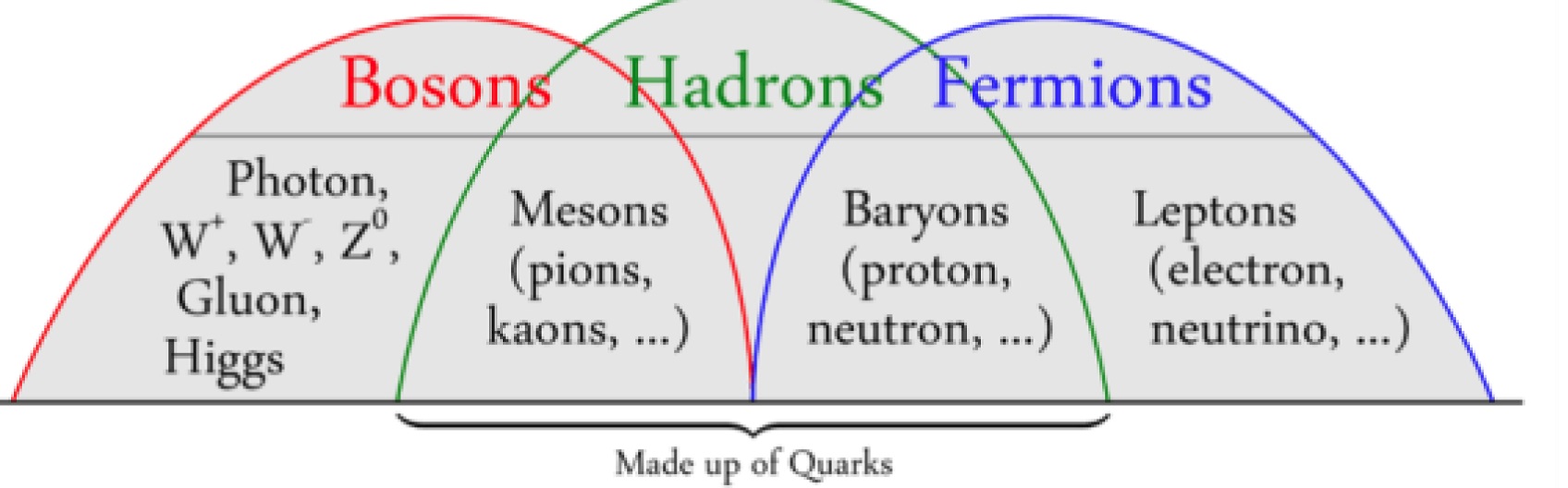}
\caption{All meson particles are unstable. These particles tend to decay,
forming electrons and neutrinos if the meson has a charge. But uncharged
mesons undergo decay forming photons. The mesons have an integer spin
(baryons have half-integer spin) Reproduced from: {\protect\small %
[https://www.differencebetween.com/difference-between-baryons-and-mesons/%
\#Mesons].}}
\label{fig3A}
\end{figure}

The spatial size implications of the anticommutation relations is the reason
why elemetary particles are generally fermions whereas quantized exitations
of force fields are bosons, Figs. \ref{fig6}, \ref{fig3A}. Newly discovered
Higgs fields are thought to provide the short-range force field and very
heavy short-range bosons. Higgs fields are also responsible for giving
masses to particles \cite{standard}.

\subsubsection{Fermionization and bosonization: Symmetry breaking
excitations, spinons, magnons}

Fermionization and bosonization are discused in several works \cite%
{borstnik, bera, miranda, senechal, wilson}. The Jordan-Wigner
transformation is discussed in Refs. \cite{zatloukal, knowen}. Indeed, there
is duality between free bosons and free fermions, or free Bloch fermions in
CMP.

Symmetry breaking excitations of ground states can generally pass through
each other, even in lower dimensions, and are characterized as bosons. In
fact spinons are bosons which carry spin-$\frac{1}{2}$in $1$-D and $2$-D
antiferromagnets \cite{spinon, senthil1, senthil2, discovery, spholorbit}.
Magnons are also bosons \cite{magnon}. Within superconductivity in
condensed-matter physics and electroweak symmetry breaking in particle
physics, some of the intriguing phenomena can be understood within the
regime of spontaneous breaking of symmetries, i.e., the quantum mechanical
scenario of the ground state of a quantum system being less symmetric than
the corresponding Hamiltonian. A telltale sign of broken symmetries are
massless bosonic excitations, which generally can pass through each other,
known as Nambu-Goldstone bosons (NGB). The number of such distinct bosonic
excitations are generally taken to be a measure of how many of the original
symmetries are dynamically broken. Despite decades of research on the
subject, a general formula that predicts the number of different NGBs in a
given dynamical system with broken symmetries has eluded theorists \cite%
{brauner, leutwyler}.

\begin{figure}[h]
\centering
\includegraphics[width=5.2555in]{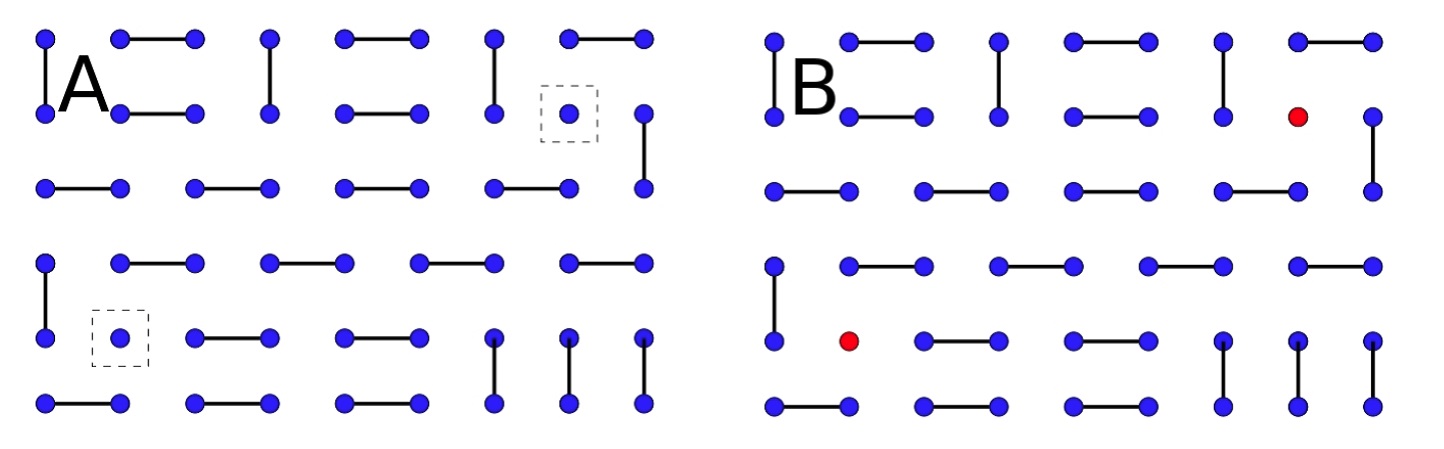}
\caption{An illustration of spin-charge separation in spin singlet dimers.
In figure A is shown a system with two spinons, made by unpairing two of the
electrons and no longer in singlet state. The collective excitations must
not have electric charge because the system still have the same charge. In
figure B the system has two electrons removed resulting in collective
excitations called holons. The system has spin zero but now has charge.
Reproduced from Ref. \protect\cite{schubel}.}
\label{fig5}
\end{figure}

Fermions in one dimension cannot move independently, but only collide with
each other resultng in density fluctuations. An interesting instability of
quasiparticles in condensed matter physics is illustrated by the
transformation of an electron into spinon, holon, and orbiton in in a
one-dimensional sample of strontium cuprate \cite{spholorbit, discovery}.
Spinons and holons are illustrated in Fig. \ref{fig5}. Spin-charge
separations occurs in one-dimensional Tomonaga-Luttinger liquids.

The following table illustrate the difference between baryons and mesons as
subatomic quasiparticles in matter. The key difference between baryons and
mesons is that baryons consist of a combination of three quark particles,
whereas mesons consist of a pair of quark-antiquark particles.

\begin{center}
\begin{tabular}{|r|r|r|}
\hline
\multicolumn{3}{|r|}{%
\begin{tabular}{lll}
\ \ \ \ \ \ \ \ \ \ \ \ \ \ \ \ \ \ \ \ \ \ \ \ \ \ \ \ \ \ \ \ \ \  & 
{\small Baryons versus Mesons} & \ \ \ \ \ \ \ \ \ \ \ \ \ \ \ \ \ \ \ \ \ \
\ \ \ \ \ \ \ \ \ \ \ 
\end{tabular}%
} \\ \hline
& 
\begin{tabular}{lll}
\ \ \ \ \ \ \ \ \ \  & {\small Baryons} & \ \ \ \ \ \ \ \ \ \ \ \ \ 
\end{tabular}
& 
\begin{tabular}{lll}
\ \ \ \ \ \ \ \ \ \ \ \ \  & {\small Mesons} & \ \ \ \ \ \ \ \ \ \ \ \ \ \ 
\end{tabular}
\\ \hline
{\small Definitions} & 
\begin{tabular}{l}
{\small Baryons are subatomic particles,} \\ 
{\small which have three quark particles}%
\end{tabular}%
{\small \ } & 
\begin{tabular}{l}
{\small Mesons are hadronic particles} \\ 
{\small that have a pair} \\ 
{\small of quark and anti-quark}%
\end{tabular}
\ \ \ \ {\small \ \ } \\ \hline
{\small Category} & {\small Fermions \ \ \ \ \ \ \ \ \ \ \ \ \ \ \ \ \ \ \ \ 
} & {\small Bosons \ \ \ \ \ \ \ \ \ \ \ \ \ \ \ \ \ \ \ \ \ \ \ \ \ } \\ 
\hline
{\small Spin \ \ \ } & {\small Half-integer spin \ \ \ \ \ \ \ \ \ \ \ \ \ \
\ } & {\small Integer spin \ \ \ \ \ \ \ \ \ \ \ \ \ \ \ \ \ \ \ \ \ \ \ }
\\ \hline
{\small Quarks \ } & {\small Has three quarks \ \ \ \ \ \ \ \ \ \ \ \ \ \ }
& {\small Has a pair of quarks anti-quarks \ \ \ \ \ \ \ \ \ } \\ \hline
{\small Interactions} & {\small Participate in strong interactions} & 
\begin{tabular}{l}
{\small Participate in both strong} \\ 
{\small and weak interactions}%
\end{tabular}
\ \ \ \ \ \ \ \ {\small \ } \\ \hline
{\small Examples } & {\small Protons and Neutrons \ \ \ \ \ \ \ \ \ } & 
\begin{tabular}{l}
{\small Heavier mesons decay to} \\ 
{\small lighter mesons and} \\ 
{\small ultimately to stable electrons} \\ 
{\small neutrinos and photons, etc.}%
\end{tabular}
\ \ \ \ \ \ \  \\ \hline
\end{tabular}
\end{center}

\subsection{The Tomonaga-Luttinger liquids and Hubbard models}

A prominent bosonization of a \textit{one-dimensional} interacting fermi
system occurs in the so-called Tomonaga-Luttinger liquids and Hubbard
models. Although the bare system of electrons in a quantum wire consists of
fermions, the low-energy elementary excitations in a Luttinger liquid are
bosons, these are sound waves of the 1-D electron. In one dimension, a
fermion or electron that tries to propagate has to push its neighbours
because of electron-electron interactions and Pauli exclusion principle. No
individual motion in possible. Any individual excitation has to become a
collective one, e.g., density fluctuation quasiparticles. The peculiarities
of a 1-D system of interacting fermions makes bosonization of low-energy
excitation spectrum is so powerful. The low-energy is supposed to be
universal and well described by the Tomonaga-Luttinger liquid (LL) theory.

Moreover, the one-dimensional Hubbard model in a magnetic field is
equivalent under renormalization-group transformation to a multicomponent
Tomonaga-Luttinger model. The numerical evaluations of the correlation
functions and the analytic results \cite{tomo,lutt,haldane} indicated
clearly that the 1-D Hubbard Model is a Tomonaga-Luttinger liquid (TLL).

\subsubsection{"Decay" of fermion into spinon, holon and orbiton in CMP}

Remarkably, the Tomonaga-Luttinger model exhibits spin-charge separation, in
which the spin and charge of the fermions possess independent dynamics, so
the velocities of the spin and charge excitations of a Tomonaga-Luttinger
liquid are distinct. These quasiparticles are the so-called spinons, holons,
and in some cases orbitons, e.g. in \textit{quasi} one-dimensional sample of
strontium cuprate \cite{brink,schlappa}. Since spinons are bosons, we expect
spinons move faster than holons and separates.

\section{The Expansion of Any Operators in CMP}

Any operator, $\hat{A}$, can be can be expanded in terms of mixed-space
projector by Eq. (\ref{any_op}), which we rewrite as,

\begin{eqnarray}
A &=&\sum\limits_{p^{\prime \prime },q^{\prime }}\left\vert p^{\prime \prime
}\right\rangle \left\langle p^{\prime \prime }\right\vert A\left\vert
q^{\prime }\right\rangle \left\langle q^{\prime }\right\vert  \notag \\
A &=&\sum\limits_{p^{\prime \prime },q^{\prime }}\left\langle p^{\prime
\prime }\right\vert A\left\vert q^{\prime }\right\rangle \ \left\vert
p^{\prime \prime }\right\rangle \left\langle q^{\prime }\right\vert
\label{any_op2}
\end{eqnarray}%
We wish to express $\left\langle p^{\prime \prime }\right\vert A\left\vert
q^{\prime }\right\rangle $and $\left\vert p^{\prime \prime }\right\rangle
\left\langle q^{\prime }\right\vert $in terms of the position eigenstate
matrix elements and momentum space projectors or \textit{vice versa}.

\subsubsection{Change of units for a unification}

In preparations of our discussions using creation and annihilation operators
on the unification with quantum field theory, we let $\hbar =1$in what
follows. Using, Eqs. (\ref{eq_p})-(\ref{p2q}), we write%
\begin{eqnarray}
\left\langle p^{\prime \prime }\right\vert A\left\vert q^{\prime
}\right\rangle &=&\frac{1}{\sqrt{N}}\sum\limits_{q^{\prime \prime
}}e^{-ip^{\prime \prime }\cdot q^{\prime \prime }}\left\langle q^{\prime
\prime }\right\vert A\left\vert q^{\prime }\right\rangle  \notag \\
&=&\frac{1}{\sqrt{N}}\sum\limits_{q^{\prime \prime }}\frac{\left\langle
q^{\prime \prime }\right\vert A\left\vert q^{\prime }\right\rangle }{%
\left\langle q^{\prime \prime }\right\vert \left\vert p^{\prime \prime
}\right\rangle }  \notag \\
\left\vert p^{\prime \prime }\right\rangle \left\langle q^{\prime
}\right\vert &=&\frac{1}{\sqrt{N}}\sum\limits_{p^{\prime }}e^{ip^{\prime
}\cdot q^{\prime }}\left\vert p^{\prime \prime }\right\rangle \left\langle
p^{\prime }\right\vert  \notag \\
&=&\frac{1}{\sqrt{N}}\sum\limits_{p^{\prime }}\frac{\left\vert p^{\prime
\prime }\right\rangle \left\langle p^{\prime }\right\vert }{\left\langle
p^{\prime }\right\vert \left\vert q^{\prime }\right\rangle }
\label{matrix_project}
\end{eqnarray}%
with completeness relation, characteristic of dual Hilbert spaces,%
\begin{eqnarray*}
\frac{1}{\sqrt{N}}\sum\limits_{p,q}e^{-ip\cdot q}\left\vert p\right\rangle
\left\langle q\right\vert &=&1 \\
\frac{1}{\sqrt{N}}\sum\limits_{p,q}\frac{\left\vert p\right\rangle
\left\langle q\right\vert }{\left\langle q\right\vert \left\vert
p\right\rangle } &=&1
\end{eqnarray*}%
Introducing the notation in Eq. (\ref{matrix_project}),%
\begin{eqnarray*}
\vec{p}^{\prime } &=&\vec{p}+\vec{u}\text{, \ \ \ \ \ \ \ }\vec{q}^{\prime }=%
\vec{q}+\vec{v}, \\
\vec{p}^{\prime \prime } &=&\vec{p}-\vec{u}\text{, \ \ \ \ \ \ \ }\vec{q}%
^{\prime \prime }=\vec{q}-\vec{v},
\end{eqnarray*}%
we have, upon substituting in Eq. (\ref{any_op2}), 
\begin{eqnarray*}
A &=&\sum\limits_{p,q,u,v}\left\langle p^{\prime \prime }\right\vert
A\left\vert q^{\prime }\right\rangle \ \left\vert p^{\prime \prime
}\right\rangle \left\langle q^{\prime }\right\vert \\
&=&\frac{1}{N}\sum\limits_{p,q,u,v}\left( e^{i2\vec{p}\cdot \vec{v}%
}\left\langle \vec{q}-\vec{v}\right\vert A\left\vert \vec{q}+\vec{v}%
\right\rangle \right) \left( \ e^{i2\vec{u}\cdot \vec{q}}\left\vert \vec{p}-%
\vec{u}\right\rangle \left\langle \vec{p}+\vec{u}\right\vert \right)
\end{eqnarray*}

\subsubsection{Mixed space operator basis}

We write the last result as an expansion in terms of \textit{phase-space
point projector}, $\hat{\Delta}\left( p,q\right) $, defined as the \textit{%
lattice Weyl transform} of a projector, by%
\begin{equation}
\hat{\Delta}\left( p,q\right) =\sum\limits_{u}e^{i2\vec{u}\cdot \vec{q}%
}\left\vert \vec{p}-\vec{u}\right\rangle \left\langle \vec{p}+\vec{u}%
\right\vert  \label{delta_ps}
\end{equation}%
and the coefficient of expansion, the so-called \textit{lattice Weyl
transform} of matrix element of operator, $A\left( p,q\right) $, defined by%
\begin{equation*}
A\left( p,q\right) =\sum\limits_{v}e^{i2\vec{p}\cdot \vec{v}}\left\langle 
\vec{q}-\vec{v}\right\vert A\left\vert \vec{q}+\vec{v}\right\rangle \text{.}
\end{equation*}%
Clearly, for a density matrix operator $\hat{\rho}$, the lattice Weyl
transform obeys,%
\begin{equation*}
\sum\limits_{p,q}\rho \left( p,q\right) =1
\end{equation*}%
If one accounts for other extra discrete quantum labels like spin and
energy-band indices, we can incorporate this in the summation in a form of a
trace.

Thus. we eventually have any operator expanded in terms of $\hat{\Delta}%
_{\lambda ^{\prime }\lambda }\left( p,q\right) $,%
\begin{equation}
\hat{A}=\sum\limits_{p,q,\lambda ,\lambda ^{\prime }}A_{\lambda \lambda
^{\prime }}\left( p,q\right) \ \hat{\Delta}_{\lambda ^{\prime }\lambda
}\left( p,q\right)  \label{LWT1}
\end{equation}%
and hence it follows,%
\begin{equation*}
A_{\lambda \lambda ^{\prime }}\left( p,q\right) =Tr\left( \hat{A}\hat{\Delta}%
_{\lambda ^{\prime }\lambda }\left( p,q\right) \right)
\end{equation*}%
Upon similar procedure based on Eq. (\ref{any_op2}), an equivalent
expression can be obtain for $A_{\lambda \lambda ^{\prime }}\left(
p,q\right) $and $\hat{\Delta}_{\lambda ^{\prime }\lambda }\left( p,q\right) $%
, namely,%
\begin{eqnarray}
A_{\lambda \lambda ^{\prime }}\left( p,q\right) &=&\sum\limits_{\bar{u}}e^{2i%
\bar{q}\cdot \bar{u}}\left\langle \bar{p}+\bar{u},\lambda \right\vert \hat{A}%
\left\vert \bar{p}-\bar{u},\lambda ^{\prime }\right\rangle  \label{alt1} \\
\hat{\Delta}_{\lambda ^{\prime }\lambda }\left( p,q\right) &=&\sum\limits_{%
\bar{v}}e^{2i\bar{p}\cdot \bar{v}}\left\vert \bar{q}+\bar{v},\lambda
\right\rangle \left\langle \bar{q}-\bar{v},\lambda ^{\prime }\right\vert
\label{alt2}
\end{eqnarray}

\subsubsection{Fully symmetric translation operators}

The state 
\begin{equation*}
\left[ \exp \left\{ -iq\cdot P\right\} \left\vert 0\right\rangle \right]
=\left\vert q\right\rangle
\end{equation*}%
is an eigenstate of the position operator with displaced eigenvalue by $q$.
However, if the limit $q^{\prime }\Rightarrow 0$is not taken then the state $%
\left[ \exp \left\{ -iq\cdot P\right\} \left\vert q^{\prime }\right\rangle %
\right] =\left\vert q^{\prime }+q\right\rangle $is an eigenstate of the
position operator with eigenvalue $q^{\prime }+q$.

We can symmetrize the translation operator by inserting $\exp \left\{
ip\cdot Q\right\} $in front of $\left\vert 0\right\rangle $which effectively
insert unity. We thus have%
\begin{equation}
\left\vert q\right\rangle =\exp \left\{ -iq\cdot P\right\} \exp \left\{
ip\cdot Q\right\} \left\vert 0\right\rangle .  \label{gen_q-p}
\end{equation}%
By the use of the Campbell-Baker-Hausdorff operator identity, we obtained%
\begin{eqnarray}
&&\exp \left\{ -iq\cdot P\right\} \exp \left\{ ip\cdot Q\right\}  \notag \\
&=&\exp \left\{ -i\left( q\cdot P-p\cdot Q\right) \right\} \exp \left\{ 
\frac{\left[ -iq\cdot P,ip\cdot Q\right] }{2}\right\}  \notag \\
&=&\exp \left\{ -i\left( q\cdot P-p\cdot Q\right) \right\} \exp \left\{ -i%
\frac{p\cdot q}{2}\right\} .  \label{displacement-op}
\end{eqnarray}%
Therefore we have the symmetric form for the displacement operator
generating the state $\left\vert q\right\rangle $from $\left\vert
0\right\rangle $given by 
\begin{equation}
T\left( q\right) ^{sym}=\exp \left\{ -i\frac{p\cdot q}{2}\right\} \exp
\left\{ -i\left( q\cdot P-p\cdot Q\right) \right\} .
\label{eingenvector-displacement-op}
\end{equation}%
The displacement operator may also be interpreted as an operator for the
preparation of the quantum eigenstate out of the 'vacuum', $\left\vert
0\right\rangle \footnote{%
Here, the concept of a vacuum state does not have a special \ meaning since $%
\left\vert 0\right\rangle $represent arbitrary reference position. It is
introduced simply to bring analogy with zero-eigenvalue of non-Hermitian
operators in later chapters, there the state $\left\vert 0\right\rangle $
has a distinguished position.}$, basis eigenstate.

\subsubsection{The symmetric operator basis for mixed representations}

The symmetric operator factor,%
\begin{eqnarray}
Y_{p,q} &=&\left( \exp \left\{ -i\left( q\cdot P-p\cdot Q\right) \right\}
\right)  \notag \\
&=&\exp \left\{ i\frac{p\cdot q}{2}\right\} \exp \left\{ -iq\cdot P\right\}
\exp \left\{ ip\cdot Q\right\} ,  \label{gen_pauli-op}
\end{eqnarray}%
is referred to here as the generalized mixed-space projector or sometimes
referred to as the generalized Pauli-matrix operator \cite{buot8}. It can be
considered the universal form of projector in mixed representations of
condensed matter physics. We shall see that this also holds with dual space
of non-Hermitian operators in quantum field theory.

\paragraph{Symmetric form of $\hat{\Delta}_{\protect\lambda ^{\prime }%
\protect\lambda }\left( p,q\right) $}

Consider Eq. (\ref{alt2}). The following identities can be verified,%
\begin{equation}
\left\vert \bar{q}+\bar{v},\lambda \right\rangle =\exp \left[ -2i\hat{P}%
\cdot \bar{v}\right] \left\vert \bar{q}-\bar{v},\lambda \right\rangle
\label{subs1}
\end{equation}%
Then 
\begin{eqnarray}
\left\vert \bar{q}-\bar{v},\lambda \right\rangle \left\langle \bar{q}-\bar{v}%
,\lambda ^{\prime }\right\vert &=&\left( N\right) ^{-1}\sum\limits_{\bar{u}%
}\exp \left[ -2i\left( \bar{q}-\bar{v}-\hat{Q}\right) \cdot \bar{u}\right]
\sum\limits_{q}\left\vert q,\lambda \right\rangle \left\langle q,\lambda
^{\prime }\right\vert  \label{subs2} \\
&=&\sum\limits_{q}\delta \left( \bar{q}-\bar{v}-\hat{Q}\right) \left\vert
q,\lambda \right\rangle \left\langle q,\lambda ^{\prime }\right\vert
=\sum\limits_{q}\delta \left( \bar{q}-\bar{v}-q\right) \ \left\vert
q,\lambda \right\rangle \left\langle q,\lambda ^{\prime }\right\vert  \notag
\\
\Omega _{\lambda \lambda ^{\prime }} &=&\sum\limits_{q}\left\vert q,\lambda
\right\rangle \left\langle q,\lambda ^{\prime }\right\vert  \notag
\end{eqnarray}%
Substituting the expressions, Eqs (\ref{subs1}) and (\ref{subs2}) in Eq. (%
\ref{alt2}), we obtain a completely symmetric expression of $\hat{\Delta}%
_{\lambda ^{\prime }\lambda }\left( p,q\right) $, 
\begin{eqnarray*}
\hat{\Delta}_{\lambda ^{\prime }\lambda }\left( p,q\right) &=&\sum\limits_{%
\bar{v}}e^{2i\bar{p}\cdot \bar{v}}\left\vert \bar{q}+\bar{v},\lambda
\right\rangle \left\langle \bar{q}-\bar{v},\lambda ^{\prime }\right\vert \\
&=&\left( N\right) ^{-1}\sum\limits_{\bar{v},\bar{u}}e^{\frac{2i}{\hbar }%
\bar{p}\cdot \bar{v}}\exp \left[ -2i\hat{P}\cdot \vec{v}\right] \exp \left[
-2i\left( \bar{q}-\bar{v}-\hat{Q}\right) \cdot \bar{u}\right]
\sum\limits_{q}\left\vert q,\lambda \right\rangle \left\langle q,\lambda
^{\prime }\right\vert \\
&=&\exp \left\{ -2i\left( P\cdot v-Q\cdot u\right) \right\} \exp 2i\left\{
p\cdot v-q\cdot u\right\}
\end{eqnarray*}

Thus, we have changed the seemingly asymmetric expression of the first line
into a symmetric form of the last line. We can combine the exponential
operators to obtain%
\begin{equation*}
\hat{\Delta}_{\lambda ^{\prime }\lambda }\left( p,q\right) =\left( N\right)
^{-1}\sum\limits_{v,u}\exp -2i\left[ \left( \hat{P}-p\right) \cdot v-\left(
Q-q\right) \cdot u\right] \Omega _{\lambda \lambda ^{\prime }}
\end{equation*}%
where now we can write, in general, 
\begin{eqnarray*}
\Omega _{\lambda \lambda ^{\prime }} &=&\sum\limits_{q^{\prime }}\left\vert
q^{\prime },\lambda \right\rangle \left\langle q^{\prime },\lambda ^{\prime
}\right\vert \\
&=&\sum\limits_{p^{\prime }}\left\vert p^{\prime },\lambda \right\rangle
\left\langle p^{\prime },\lambda ^{\prime }\right\vert
\end{eqnarray*}%
We therefore have,%
\begin{eqnarray*}
A\left( p,q\right) &=&Tr\left( \hat{A}\hat{\Delta}\right) \\
&=&\left( N\right) ^{-1}\left( 
\begin{array}{c}
\sum\limits_{v,u}\exp 2i\left[ p\cdot v-q\cdot u\right] \  \\ 
\times Tr\left\{ \hat{A}\exp -2i\left[ \hat{P}\cdot v-Q\cdot u\right] \Omega
_{\lambda \lambda ^{\prime }}\right\}%
\end{array}%
\right)
\end{eqnarray*}%
Therefore, the charactetic distribution of $A\left( p,q\right) $is
identically, 
\begin{equation*}
A_{\lambda \lambda ^{\prime }}\left( u,v\right) =Tr\left\{ \hat{A}\exp
\left( -2i\left[ \hat{P}\cdot v-Q\cdot u\right] \right) \Omega _{\lambda
\lambda ^{\prime }}\right\}
\end{equation*}%
By using this characteristic distribution function for $A_{\lambda \lambda
^{\prime }}\left( p,q\right) $ 
\begin{equation}
A_{\lambda \lambda ^{\prime }}\left( u,v\right) =\left( \frac{1}{N}\right) ^{%
\frac{1}{2}}\sum\limits_{p,q}A_{\lambda \lambda ^{\prime }}\left( p,q\right)
e^{2i\left( p\cdot v-q\cdot u\right) }  \label{char1}
\end{equation}%
with inverse%
\begin{equation*}
A_{\lambda \lambda ^{\prime }}\left( p,q\right) =\left( \frac{1}{N}\right) ^{%
\frac{1}{2}}\sum\limits_{u,v}A_{\lambda \lambda ^{\prime }}\left( u,v\right)
e^{-2i\left( p\cdot v-q\cdot u\right) }
\end{equation*}

Then we can write Eq. (\ref{LWT1}) simply like a Fourier transform (caveat:
Fourier transform to\textit{\ operator space}) of the characteristic
function of the lattice Weyl transform of the operator $\hat{A}$, 
\begin{equation}
\hat{A}=\sum\limits_{u,v,\lambda ,\lambda ^{\prime }}A_{\lambda \lambda
^{\prime }}\left( u,v\right) \exp \left[ \left( -2i\left( \hat{P}\cdot v-%
\hat{Q}\cdot u\right) \right) \right] \Omega _{\lambda \lambda ^{\prime }}
\label{caveat}
\end{equation}%
In continuum approximation, we have,%
\begin{eqnarray}
\hat{\Delta}\left( p,q\right) &=&\left( 2\pi \right) ^{-3}\int du\ dv\ e^{-i%
\left[ \left( \hat{P}-\bar{p}\right) \cdot \vec{v}-\left( Q-\bar{q}\right)
\cdot \bar{u}\right] .u\ }\Omega _{\lambda \lambda ^{\prime }}  \notag \\
&=&\left( 2\pi \right) ^{-3}\int du\ dv\ e^{\left( i\right) \left( p\cdot
v-q\cdot u\right) }e^{\left( -i\right) \left( P.v-Q.u\right) }\Omega
_{\lambda \lambda ^{\prime }},  \label{sym_delta}
\end{eqnarray}%
From%
\begin{eqnarray*}
A_{\lambda \lambda ^{\prime }}\left( p,q\right) &=&Tr\left( \hat{A}\hat{%
\Delta}\right) =\left( 2\pi \right) ^{-3}\left( \int du\ dv\ e^{i\left(
p\cdot v-q\cdot u\right) }Tr\left( \hat{A}e^{\left( -i\right) \left(
P.v-Q.u\right) }\right) \right) \\
&=&\left( 2\pi \right) ^{-3}\left( \int du\ dv\ e^{\left( i\right) \left(
p\cdot v-q\cdot u\right) }A_{\lambda \lambda ^{\prime }}\left( u,v\right)
\right)
\end{eqnarray*}

\subsection{Characteristic distribution of lattice Weyl transform}

In general, we can have different expression for the characteristic function
depending on the use of the often referred to in corresponding CS formalism 
\footnote{%
For creation and annihilation operators in many-body quantum physics, the
proof relies on the use of normal or anti-normal ordering of canonical
operators, which can then be treated like $%
\mathbb{C}
$-numbers in expansion of exponentials. The exponential in Eq. (\ref%
{characteristics}) is sometimes referred to as the generalized Pauli-spin
operator.} as the normal and anti-normal expessions,%
\begin{eqnarray*}
e^{\left( -i\right) \left( \ P.v-Q.u\right) } &=&\exp \left\{ -i\frac{u\cdot
v}{2}\right\} \exp \left\{ -iv\cdot P\right\} \exp \left\{ iu\cdot Q\right\}
\\
&=&\exp \left\{ i\frac{u\cdot v}{2}\right\} \exp \left\{ iu\cdot Q\right\}
\exp \left\{ -iv\cdot P\right\}
\end{eqnarray*}%
so that with $\exp \left\{ i\frac{u\cdot v}{2}\right\} e^{\left( -i\right)
\left( P.v-Q.u\right) }$as our reference, we have the so-called normal and
anti-normal order expressions given by, 
\begin{eqnarray}
\exp \left\{ -iv\cdot P\right\} \exp \left\{ iu\cdot Q\right\} &=&\exp
\left\{ i\frac{u\cdot v}{2}\right\} e^{\left( -i\right) \left(
P.v-Q.u\right) },  \label{smooth1} \\
\exp \left\{ iu\cdot Q\right\} \exp \left\{ -iv\cdot P\right\} &=&\exp
\left\{ -i\frac{u\cdot v}{2}\right\} e^{\left( -i\right) \left(
P.v-Q.u\right) },  \label{smooth2}
\end{eqnarray}%
respectively, yielding the following differrent expressions for $A_{\lambda
\lambda ^{\prime }}\left( u,v\right) $, namely, 
\begin{equation}
A_{\lambda \lambda ^{\prime }}^{w}\left( u,v\right) =Tr\left( \hat{A}\exp
\left( -i\right) \left( \ P.v-Q.u\right) \right)  \label{symmetricchardef}
\end{equation}%
\begin{eqnarray}
A_{\lambda \lambda ^{\prime }}^{n}\left( u,v\right) &=&Tr\ \left\{ \hat{A}\
\exp \left\{ -iv\cdot P\right\} \exp \left\{ iu\cdot Q\right\} \right\} , 
\notag \\
&=&\exp \left\{ i\frac{u\cdot v}{2}\right\} Tr\left( \hat{A}\exp \left(
-i\right) \left( \ P.v-Q.u\right) \right)  \label{normalchardef}
\end{eqnarray}%
\begin{eqnarray}
A_{\lambda \lambda ^{\prime }}^{a}\left( u,v\right) &=&Tr\ \left\{ \hat{A}\
\exp \left\{ ip\cdot Q\right\} \exp \left\{ -iq\cdot P\right\} \right\} , 
\notag \\
&=&\exp \left\{ -i\frac{u\cdot v}{2}\right\} Tr\left( \hat{A}\exp \left(
-i\right) \left( \ P.v-Q.u\right) \right)  \label{antinormalchardef}
\end{eqnarray}

Although, Eqs. (\ref{normalchardef}) and (\ref{antinormalchardef}) only
amounts to difference in the phase factors, the corresponding quantities in
non-Hermitian dual spaces gives a very different distributions often
referred \ to as smooth-out distribtuions. Examining Eqs. (\ref{smooth1})
and (\ref{smooth2}), and the fact that in non-Hermitian mixed representation,%
$\left\{ i\frac{u\cdot v}{2}\right\} $is a real quantity that resembles a
Guassian function, Eqs. (\ref{smooth1}) clearly represent some smoothing of
the Wigner distribution characteristic function and hence the Wigner
distribution itself. In Eqs. (\ref{normalchardef}) and (\ref%
{antinormalchardef}) no informations are lost.

Indeed, more \textit{general} phase-space distribution functions, $f^{\left(
g\right) }\left( p,q,t\right) $, can be obtained from the expression%
\begin{eqnarray}
f^{\left( g\right) }\left( p,q,t\right) &=&\frac{1}{2\pi \hbar }\int
dudve^{-i\left( qu+pv\right) }\left[ \left\{ Tr\ \hat{\rho}\ \exp \left(
-i\right) \left( \ P.v-Q.u\right) \right\} g\left( u,v\right) \right]  \notag
\\
&=&\frac{1}{2\pi \hbar }\int dudve^{-\frac{i}{\hbar }\left( qu+pv\right) }%
\left[ C^{\left( w\right) }\left( u,v,t\right) \ g\left( u,v\right) \right] ,
\end{eqnarray}%
where $g\left( u,v\right) $is some choosen phase function.

\subsubsection{Alternative symmetrization of Eq. (\protect\ref{delta_ps})
for $\hat{\Delta}_{\protect\lambda ^{\prime }\protect\lambda }\left(
p,q\right) $}

We can also take up Eq. (\ref{delta_ps}) to symmetrize $\hat{\Delta}%
_{\lambda ^{\prime }\lambda }\left( p,q\right) $. We have%
\begin{eqnarray*}
\hat{\Delta}\left( p,q\right) &=&\sum\limits_{u}e^{i2\vec{u}\cdot \vec{q}%
}\left\vert \vec{p}-\vec{u}\right\rangle \left\langle \vec{p}+\vec{u}%
\right\vert \\
&=&\sum\limits_{u}e^{-i2\vec{u}\cdot \vec{q}}\left\vert \vec{p}+\vec{u}%
\right\rangle \left\langle \vec{p}-\vec{u}\right\vert \\
&=&\sum\limits_{u}e^{i2\left( Q-q\right) \cdot u}\left\vert \vec{p}-\vec{u}%
\right\rangle \left\langle \vec{p}-\vec{u}\right\vert
\end{eqnarray*}%
where%
\begin{equation*}
\left\vert \vec{p}-\vec{u}\right\rangle \left\langle \vec{p}-\vec{u}%
\right\vert =\sum\limits_{\vec{v},\vec{p}^{\prime }}\exp \left[ 2i\left( 
\vec{p}-\vec{u}-\hat{P}\right) \cdot \vec{v}\right] \ \left\vert \vec{p}%
^{\prime }\right\rangle \left\langle \vec{p}^{\prime }\right\vert
\end{equation*}%
Therefore,%
\begin{eqnarray*}
\hat{\Delta}\left( p,q\right) &=&\sum\limits_{u}e^{i2\vec{u}\cdot \vec{q}%
}\sum\limits_{v}\exp \left[ 2i\left( \vec{p}-\vec{u}-P\right) \cdot v\right]
\ \left\vert \vec{p}^{\prime }\right\rangle \left\langle \vec{p}^{\prime
}\right\vert \exp \left[ i\left( \vec{p}-\vec{u}-P\right) \cdot v\right]
e^{-i2Q\cdot u} \\
&=&\sum\limits_{u,v}e^{i2\left( Q-q\right) \cdot u}\exp \left[ -i2\left(
P-p\right) \cdot v\right] \ \exp \left[ -i2\vec{u}\cdot v\right] \left\vert 
\vec{p}^{\prime }\right\rangle \left\langle \vec{p}^{\prime }\right\vert
e^{-i2\left( Q-q\right) \cdot u}
\end{eqnarray*}%
\begin{eqnarray*}
e^{-i2Q\cdot u}e^{-i2P\cdot v} &=&e^{-2i\left( P\cdot v-Q\cdot u\right) }e^{%
\frac{4}{2}\left[ Q,P\right] \vec{u}\cdot v} \\
&=&e^{-2i\left( P\cdot v-Q\cdot u\right) }e^{2i\vec{u}\cdot v}
\end{eqnarray*}%
Therefore,%
\begin{equation*}
\hat{\Delta}\left( p,q\right) =\exp 2i\left( p\cdot v-q\cdot u\right) \exp %
\left[ -2i\left( P\cdot v-Q\cdot u\right) \right] \sum\limits_{\vec{p}%
^{\prime }}\left\vert \vec{p}^{\prime }\right\rangle \left\langle \vec{p}%
^{\prime }\right\vert
\end{equation*}%
Thus, 
\begin{eqnarray*}
\hat{A} &=&\sum\limits_{p,q,\lambda ,\lambda ^{\prime }}A_{\lambda \lambda
^{\prime }}\left( p,q\right) \ \hat{\Delta}_{\lambda ^{\prime }\lambda
}\left( p,q\right) \\
&=&\sum\limits_{p,q,\lambda ,\lambda ^{\prime },u,v}A_{\lambda \lambda
^{\prime }}\left( p,q\right) \ \exp 2i\left( p\cdot v-q\cdot u\right) \exp 
\left[ -2i\left( P\cdot v-Q\cdot u\right) \right] \sum\limits_{\vec{p}%
^{\prime }}\left\vert \vec{p}^{\prime }\right\rangle \left\langle \vec{p}%
^{\prime }\right\vert
\end{eqnarray*}%
with characteristic function given by,%
\begin{equation*}
A\left( u,v\right) =\sum\limits_{p,q}A_{\lambda \lambda ^{\prime }}\left(
p,q\right) \ \exp 2i\left( p\cdot v-q\cdot u\right)
\end{equation*}%
and%
\begin{equation*}
\hat{A}=\sum\limits_{\lambda ,\lambda ^{\prime },u,v}A\left( u,v\right) \exp %
\left[ -2i\left( P\cdot v-Q\cdot u\right) \right] \ \left\vert \vec{p}%
_{0}\right\rangle \left\langle \vec{p}_{0}\right\vert
\end{equation*}%
The algebra of $\exp \left\{ -i\left( q^{\prime }\cdot P-p^{\prime }\cdot
Q\right) \right\} $as well as its relevance to the physics of two-state
systems, spin systems, quantum computing, entanglements \cite{mechanical}
and teleportation, are discussed in one of the author's book \cite{buot8}.

\subsection{Construction of path integral in CMP: Mixed space representation}

We are here interested in expressing the transition amplitude as a lattice
path integral in solid-state physics. A well-known procedure to implement
this is to decomposed the evolution operator, $\hat{U}\left( t,t_{0}\right)
\equiv \exp \frac{-i}{\hbar }\left( t-t_{0}\right) \mathcal{H}$, in terms of
freely wandering paths during infinitisimal time increments, $\epsilon $.
First, we decomposed in time increments as%
\begin{equation*}
\hat{U}\left( t,t_{0}\right) =Lim_{n\Longrightarrow \infty
}\prod\limits_{j=1}^{n+1}\hat{U}\left( t_{j},t_{j-1}\right)
\end{equation*}%
where $t_{n+1}=t$. As a wandering incremental paths between $t_{n+1}$and $%
t_{0}$in the crystal lattice, we make use of the Wannier function wavevector
and its completeness (wanderer) as,%
\begin{eqnarray}
&&\left\langle q\left( t_{n+1}\right) \right\vert \hat{U}\left(
t,t_{0}\right) \left\vert q\left( t_{0}\right) \right\rangle  \notag \\
&=&\left\langle q\left( t_{n+1}\right) \right\vert \hat{U}\left(
t_{n+1},t_{n}\right) \sum\limits_{q_{n}}\left\vert q\left( t_{n}\right)
\right\rangle \left\langle q\left( t_{n}\right) \right\vert \hat{U}\left(
t_{n},t_{n-1}\right) \sum\limits_{q_{n-1}}\left\vert q\left( t_{n-1}\right)
\right\rangle \left\langle q\left( t_{n-1}\right) \right\vert .  \notag \\
&&\times ....\sum\limits_{q_{1}}\left\vert q\left( t_{1}\right)
\right\rangle \left\langle q\left( t_{1}\right) \right\vert \hat{U}\left(
t_{1},t_{0}\right) \left\vert q\left( t_{0}\right) \right\rangle  \notag \\
&=&\sum\limits_{q_{_{1}}}........\sum\limits_{q_{_{n}}}\prod%
\limits_{j=1}^{n+1}\left\langle q\left( t_{j}\right) \right\vert \hat{U}%
\left( t_{j},t_{j-1}\right) \left\vert q\left( t_{j-1}\right) \right\rangle
\label{path1}
\end{eqnarray}%
where the end coordinates, namely, $q\left( t_{n+1}\right) $and $q\left(
t_{0}\right) $are fixed. By making the time intervals infinitely small or by
letting $Lim_{n\Longrightarrow \infty }$, we can take advantage of the
linearity of the small time evolution operators to calculate the matrix
elements after which it can be recomposed in exponential form but now cast
as $%
\mathbb{C}
$-numbers. We have%
\begin{equation*}
\hat{U}\left( t_{j},t_{j-1}\right) \simeq 1-\frac{i}{\hbar }\left(
t_{j}-t_{j-1}\right) \mathcal{H}+O\left( t_{j}-t_{j-1}\right) ^{2}+..
\end{equation*}

Now comes our big advantage in calculating the matrix elements in
mixed-space representation. We now expand the operator $\hat{U}\left(
t_{j},t_{j-1}\right) $in terms of the phase-space point projector $\hat{%
\Delta}\left( p,q\right) $. We have%
\begin{equation*}
\mathcal{H=}\sum\limits_{p,q}H\left( p,q\right) \hat{\Delta}\left( p,q\right)
\end{equation*}%
Then%
\begin{eqnarray*}
&&\left\langle q\left( t_{j}\right) \right\vert \hat{U}\left(
t_{j},t_{j-1}\right) \left\vert q\left( t_{j-1}\right) \right\rangle \\
&=&\left\langle q\left( t_{j}\right) \right\vert \left( 1-\frac{i}{\hbar }%
\left( t_{j}-t_{j-1}\right) \mathcal{H}\right) \left\vert q\left(
t_{j-1}\right) \right\rangle \\
&=&1-\frac{i}{\hbar }\left( t_{j}-t_{j-1}\right) \sum\limits_{p,q}H\left(
p,q\right) \ \left\{ \left\langle q\left( t_{j}\right) \right\vert \hat{%
\Delta}\left( p,q\right) \left\vert q\left( t_{j-1}\right) \right\rangle
\right\}
\end{eqnarray*}%
where for intermediate points,%
\begin{equation*}
\left\{ \left\langle q\left( t_{j}\right) \right\vert \hat{\Delta}\left(
p,q\right) \left\vert q\left( t_{j-1}\right) \right\rangle \right\} =\exp
\left( \frac{i}{\hbar }p_{j}\cdot \left( q_{j}-q_{j-1}\right) \right) \delta
\left( 2q-\left[ q_{j}+q_{j-1}\right] \right)
\end{equation*}%
and for the end points, we have%
\begin{equation*}
\left\{ \left\langle q\left( t_{n+1}\right) \right\vert \hat{\Delta}\left(
p,q\right) \left\vert q\left( t_{n}\right) \right\rangle \right\} =\exp
\left( \frac{i}{\hbar }p_{n}\cdot \left( q_{n+1}-q_{n}\right) \right) \delta
\left( 2q-\left[ q_{n+1}+q_{n}\right] \right)
\end{equation*}%
\begin{equation*}
\left\{ \left\langle q\left( t_{1}\right) \right\vert \hat{\Delta}\left(
p,q\right) \left\vert q\left( t_{0}\right) \right\rangle \right\} =\exp
\left( \frac{i}{\hbar }p_{1}\cdot \left( q_{1}-q_{0}\right) \right) \delta
\left( 2q-\left[ q_{1}+q_{0}\right] \right)
\end{equation*}%
so that we can write to first order in time increment, $\left(
t_{j}-t_{j-1}\right) $, as,%
\begin{eqnarray*}
&&\left\langle q\left( t_{j}\right) \right\vert \hat{U}\left(
t_{j},t_{j-1}\right) \left\vert q\left( t_{j-1}\right) \right\rangle \\
&=&\left( N\hbar ^{3}\right) ^{-1}\sum\limits_{p_{j},q,q_{j}}\exp \left( 
\frac{i}{\hbar }p_{j}\cdot \left( q_{j}-q_{j-1}\right) \right) \\
&&\times \exp \left( \frac{-i}{\hbar }\left( t_{j}-t_{j-1}\right) H\left(
p,q\right) \right) \delta \left( 2q-\left[ q_{j}+q_{j-1}\right] \right) \\
&=&\left( N\hbar ^{3}\right) ^{-1}\sum\limits_{j=1}^{n}\exp \left[ \frac{i}{%
\hbar }\left( t_{j}-t_{j-1}\right) \left( p_{j}\cdot \frac{\left(
q_{j}-q_{j-1}\right) }{\left( t_{j}-t_{j-1}\right) }-H\left( p,\frac{%
q_{j}+q_{j-1}}{2}\right) \right) \right]
\end{eqnarray*}%
Substituting in Eq. (\ref{path1}), we obtained%
\begin{eqnarray*}
&&\left\langle q\left( t_{n+1}\right) \right\vert \hat{U}\left(
t,t_{0}\right) \left\vert q\left( t_{0}\right) \right\rangle \\
&=&\left( N\hbar ^{3}\right)
^{-1}\sum\limits_{q_{_{1}}}....\sum\limits_{q_{_{n}}}\sum%
\limits_{p_{_{1}}}....\sum\limits_{p_{_{n}}}\prod\limits_{j=1}^{n} \\
&&\times \exp \left[ \frac{i}{\hbar }\left( t_{j}-t_{j-1}\right) \left(
p\cdot \frac{\left( q_{j}-q_{j-1}\right) }{\left( t_{j}-t_{j-1}\right) }%
-H\left( p,\frac{q_{j}+q_{j-1}}{2}\right) \right) \right] \\
&=&\left( N\hbar ^{3}\right)
^{-1}\prod\limits_{j=1}^{n}\sum\limits_{q_{_{j}}}\prod\limits_{j=1}^{n}\sum%
\limits_{p}\exp \left[ \frac{i}{\hbar }\left( t_{j}-t_{j-1}\right) \left( 
\begin{array}{c}
p\cdot \frac{\left( q_{j}-q_{j-1}\right) }{\left( t_{j}-t_{j-1}\right) } \\ 
-H\left( p,\frac{q_{j}+q_{j-1}}{2}\right)%
\end{array}%
\right) \right]
\end{eqnarray*}%
Now since $\sum\limits_{p}$is performed for each term laveled by $j=1.....n $%
, we can also label summation over $p$by summation over $p_{j}$, i.e., $%
\sum\limits_{p_{j}}$giving us a more symmetrical form,%
\begin{eqnarray*}
&&\left\langle q\left( t_{n+1}\right) \right\vert \hat{U}\left(
t,t_{0}\right) \left\vert q\left( t_{0}\right) \right\rangle \\
&=&\left( N\hbar ^{3}\right)
^{-1}\prod\limits_{j=1}^{n}\sum\limits_{q_{_{j}}}\prod\limits_{j=1}^{n}\sum%
\limits_{p_{j}} \\
&&\times \exp \sum\limits_{j=1}^{n}\left[ \frac{i}{\hbar }\left(
t_{j}-t_{j-1}\right) \left( p_{j}\cdot \frac{\left( q_{j}-q_{j-1}\right) }{%
\left( t_{j}-t_{j-1}\right) }-H\left( p_{j},\frac{q_{j}+q_{j-1}}{2}\right)
\right) \right]
\end{eqnarray*}

\subsubsection{Continuum limit}

If we take the conituum limit, i.e., $N\Longrightarrow \infty ,$and $\Delta
t=t_{j}-t_{j-1}\Longrightarrow 0$, and defining%
\begin{eqnarray*}
\int \mathcal{D}p &\equiv &\prod\limits_{j=1}^{n}\sum\limits_{p_{j}} \\
\int\limits_{q_{i}=q\left( t_{0}\right) }^{q_{f}=q\left( t_{n}\right) }%
\mathcal{D}q &\equiv &\prod\limits_{j=1}^{n}\sum\limits_{q_{j}}
\end{eqnarray*}%
We have in the exponential, 
\begin{eqnarray*}
S &=&\sum\limits_{j=1}^{n}\left[ \frac{i}{\hbar }\left( t_{j}-t_{j-1}\right)
\left( p_{j}\cdot \frac{\left( q_{j}-q_{j-1}\right) }{\left(
t_{j}-t_{j-1}\right) }-H\left( p_{j},\frac{q_{j}+q_{j-1}}{2}\right) \right) %
\right] \\
&\Longrightarrow &\frac{i}{\hbar }\int\limits_{t_{i}}^{t_{f}}dt\ \left[
p\left( t\right) \cdot \dot{q}\left( t\right) -H\left( p\left( t\right)
,q\left( t\right) \right) \right]
\end{eqnarray*}%
where the the integral in $S$ is identified as the Hamiltonian action. Then
we end up with%
\begin{equation*}
\left\langle q\left( t_{f}\right) \right\vert \hat{U}\left( t,t_{0}\right)
\left\vert q\left( t_{i}\right) \right\rangle =\left( 2\pi \hbar ^{3}\right)
^{-1}\iint \mathcal{D}q\mathcal{D}p\ \exp \left( S\right)
\end{equation*}%
This last result is often called functional integral since now the $p\left(
t\right) $,$q\left( t\right) $ defines phase-space path functions.

\section{General Mixed Space Representation in QFT}

Let the operators $\hat{\psi}$ and $\hat{\psi}^{\dagger }$be non-Hermitian
operators. \ These operators obey either communtation or anti-commutation
relations, i.e.,%
\begin{equation}
\left[ \hat{\psi},\hat{\psi}^{\dagger }\right] _{\eta }=1
\label{anticom_com}
\end{equation}%
where the subscript $\eta =$$+$ is for the anti-commutation and $\eta =$$-$
stands for commutation relation. These non-Hermitian operators have
distinguished left and right eigenvectors. We have%
\begin{eqnarray}
\hat{\psi}\left\vert \alpha \right\rangle &=&\alpha \left\vert \alpha
\right\rangle  \label{ann-eigen-eq1} \\
\left\langle \alpha \right\vert \hat{\psi}^{\dagger } &=&\left\langle \alpha
\right\vert \alpha ^{\ast }  \label{ann-eigen-eq2}
\end{eqnarray}%
\begin{eqnarray}
\hat{\psi}^{\dagger }\left\vert \beta \right\rangle &=&\beta \left\vert
\beta \right\rangle  \label{ann-eigen-eq3} \\
\left\langle \beta \right\vert \hat{\psi} &=&\left\langle \beta \right\vert
\beta ^{\ast }  \label{ann-eigen-eq4}
\end{eqnarray}%
This means the left eigenvector of $\hat{\psi}$ is $\left\langle \beta
\right\vert $, with eigenvalue $\beta ^{\ast }$, whereas the left
eigenvector of $\hat{\psi}^{\dagger }$is $\left\langle \alpha \right\vert $,
with eigenvalue $\alpha ^{\ast }$. We have,%
\begin{eqnarray}
\left\langle \beta \right\vert \hat{\psi}\left\vert \alpha \right\rangle
&=&\alpha \left\langle \beta \right\vert \left\vert \alpha \right\rangle 
\notag \\
&=&\beta ^{\ast }\left\langle \beta \right\vert \left\vert \alpha
\right\rangle  \notag \\
\left( \alpha -\beta ^{\ast }\right) \left\langle \beta \right\vert
\left\vert \alpha \right\rangle &=&0  \label{biortho1}
\end{eqnarray}%
and similarly%
\begin{eqnarray}
\left\langle \alpha \right\vert \hat{\psi}^{\dagger }\left\vert \beta
\right\rangle &=&\alpha ^{\ast }\left\langle \alpha \right\vert \left\vert
\beta \right\rangle  \notag \\
&=&\beta \left\langle \alpha \right\vert \left\vert \beta \right\rangle 
\notag \\
\left( \alpha ^{\ast }-\beta \right) \left\langle \alpha \right\vert
\left\vert \beta \right\rangle &=&0  \label{biortho2}
\end{eqnarray}%
From Eqs. (\ref{biortho1}) and (\ref{biortho2}), if $\alpha _{n}\neq \beta
_{m}^{\ast }$, $\left\langle \beta _{m}\right\vert \left\vert \alpha
_{n}\right\rangle =0$, imitating orthogonal Hermitian Hilbert space.
However, if $\alpha _{n}=\beta _{m}^{\ast }$, then $\left\langle \beta
_{m}\right\vert \left\vert \alpha _{n}\right\rangle \neq 0$, amenable to
probabilistic interpretation. Moreover, with $\left\vert \alpha
\right\rangle $ and $\left\langle \beta \right\vert $ we have the general
projection given by%
\begin{equation}
\sum\limits_{n}\frac{\left\vert \alpha _{n}\right\rangle \left\langle \beta
_{n}\right\vert }{\left\langle \beta _{n}\right\vert \left\vert \alpha
_{n}\right\rangle }=1  \label{gen_complete}
\end{equation}%
which project only within the paired eigenspace of $\left\{ \left\vert
\alpha \right\rangle ,\left\langle \beta \right\vert \right\} $. Similarly, 
\begin{equation*}
\sum\limits_{n}\frac{\left\vert \beta _{n}\right\rangle \left\langle \alpha
_{n}\right\vert }{\left\langle \alpha _{n}\right\vert \left\vert \beta
_{n}\right\rangle }=1
\end{equation*}%
only projects within the paired space, $\left\{ \left\langle \alpha
\right\vert ,\left\vert \beta \right\rangle \right\} $.

The general proof of Eq. (\ref{gen_complete}) lies in the following
expansion of $\left\vert \Psi \right\rangle $, where $\left\vert \Psi
\right\rangle $ is a complete orthonormal eigenvector. Let, 
\begin{equation*}
\left\vert \Psi \right\rangle =\sum\limits_{n^{\prime }}c^{n^{\prime
}}\left\vert \alpha _{n^{\prime }}\right\rangle .
\end{equation*}%
Then we have%
\begin{eqnarray*}
\sum\limits_{n}\frac{\left\vert \alpha _{n}\right\rangle \left\langle \beta
_{n}\right\vert \left\vert \Psi \right\rangle }{\left\langle \beta
_{n}\right\vert \left\vert \alpha _{n}\right\rangle } &=&\sum\limits_{n,n^{%
\prime }}\frac{\left\vert \alpha _{n}\right\rangle \left\langle \beta
_{n}\right\vert \left\vert \alpha _{n^{\prime }}\right\rangle c^{n^{\prime }}%
}{\left\langle \beta _{n}\right\vert \left\vert \alpha _{n}\right\rangle } \\
&=&\sum\limits_{n}c^{n}\left\vert \alpha _{n}\right\rangle =\left\vert \Psi
\right\rangle
\end{eqnarray*}

\subsection{\label{dual_H}Mixed Hilbert-space construction}

Thus, from Eqs. (\ref{biortho1}) and (\ref{biortho2}), we have a
well-defined Hermitian-like operation in terms of the paired eigenvector set 
$\left\{ \left\langle \alpha \right\vert ,\left\vert \beta \right\rangle
\right\} $ and$\left\{ \left\langle \beta \right\vert ,\left\vert \alpha
\right\rangle \right\} $ as dual eigenspaces. Indeed, assuming
nondegenerates countable states, we can pair the states so that $\alpha
_{n}=\beta _{m}^{\ast }$ and rewrite the pairs with same quantum subscript,
i.e., $\alpha _{n}=\beta _{n}^{\ast }$, which would be consistent with the
commutation relation, $\left[ \hat{\psi}_{n},\hat{\psi}_{n}^{\dagger }\right]
_{\eta }=1$.

Just as the dot product $\left\langle p\right\vert \left\vert q\right\rangle
\neq 0$ in condensed matter physics discussions, so $\left\langle \alpha
_{n}\right\vert \left\vert \alpha _{n}\right\rangle \neq 0$, which means
that $\left\langle \alpha _{n}\right\vert \left\vert \alpha
_{n}\right\rangle $ cannot be made equal to zero, since these belongs to
separate paired set, $\left\{ \left\langle \alpha \right\vert ,\left\vert
\beta \right\rangle \right\} $ and$\left\{ \left\langle \beta \right\vert
,\left\vert \alpha \right\rangle \right\} $, respectively . In some sense,
the $\left\vert \alpha \right\rangle $,$\left\vert \beta \right\rangle $
respective spaces are reminiscent of the $q$-$p$ phase space, where $%
\left\langle p\right\vert \left\vert q\right\rangle \neq 0$ is the
transition function. This conclusion can be made quite general in what
follows.

\subsection{\label{pairing}Pairing algorithm and Hermitianization}

Thus, in order to work with Hermitian-like operators, we want the
eigenvalues $\alpha =\beta ^{\ast }$ and $\alpha ^{\ast }=\beta $. For
nondegenerate countable finite system, this type of pairing of the different 
$\left\vert \alpha \right\rangle $and $\left\vert \beta \right\rangle $
spaces is well defined. For convenience in what follows, we relabel the $%
\left\vert \alpha \right\rangle $ and $\left\vert \beta \right\rangle $
notations and their adjoints to reflect the Hermitian-like new spaces, sort
of \textit{renormalize} new dynamical vector spaces.

From Eqs. (\ref{biortho1}) and (\ref{biortho2}), it seems trivial just like
for the Hermitian operators to prove orthogonality or more appropriately,
biorthogonality, in terms of the $\left\{ \left\langle \alpha \right\vert
,\left\vert \beta \right\rangle \right\} $ and $\left\{ \left\langle \beta
\right\vert ,\left\vert \alpha \right\rangle \right\} $ 'dual' eigenspaces.
This seems to suggest the following Hermitian-like relations%
\begin{eqnarray*}
\left\langle \alpha \right\vert \hat{\psi}^{\dagger }\left\vert \beta
\right\rangle &\equiv &\alpha ^{\ast }\left\langle \alpha \right\vert
\left\vert \beta \right\rangle \\
\left\langle \beta \right\vert \hat{\psi}\left\vert \alpha \right\rangle
&\equiv &\alpha \left\langle \beta \right\vert \left\vert \alpha
\right\rangle
\end{eqnarray*}%
where $\left\langle \alpha \right\vert $ and $\left\vert \beta \right\rangle 
$ are the left and right eigenvectors, respectively of $\hat{\psi}^{\dagger
} $, whereas, $\left\langle \beta \right\vert $ and $\left\vert \alpha
\right\rangle $ are the left and right eigenvectors, respectively, of $\hat{%
\psi}$. The above pairing allows us to form dual eigenvectors to simulate
the Hilbert-spaces of Hermitian operators.

We now denote the $\left\langle \alpha \right\vert $ and $\left\vert \beta
\right\rangle $ left and right eigenvectors, respectively of $\hat{\psi}%
^{\dagger }$as making up the $\alpha ^{\ast }$-Hilbert space, and we will
adopt a new consistent labels, $\left\langle \alpha \right\vert
\Longrightarrow \left\langle \mathfrak{p}\right\vert $ and $\left\vert \beta
\right\rangle \Longrightarrow \left\vert \mathfrak{p}\right\rangle $.
Similary, we relabel the $\left\langle \beta \right\vert $ and $\left\vert
\alpha \right\rangle $ left and right eigenvectors, respectively, of $\hat{%
\psi}$ as making up the $\alpha $-Hilbert space, with $\left\langle \beta
\right\vert \Longrightarrow \left\langle \mathfrak{q}\right\vert $ and $%
\left\vert \alpha \right\rangle \Longrightarrow \left\vert \mathfrak{q}%
\right\rangle $. The $\mathfrak{q}$ and $\mathfrak{p}$- eigenspaces
constitute our newly-formed quantum label for Hilbert spaces for $\hat{\psi}$
and $\hat{\psi}^{\dagger }$, respectively.

\subsection{Completeness relations}

In the new dual space representation, Eq. (\ref{gen_complete}) becomes
simply a completeness relation,%
\begin{eqnarray}
\sum\limits_{n}\frac{\left\vert \alpha _{n}\right\rangle \left\langle \beta
_{n}\right\vert }{\left\langle \beta _{n}\right\vert \left\vert \alpha
_{n}\right\rangle } &=&1\Longrightarrow \sum\limits_{\mathfrak{q}}\frac{%
\left\vert \mathfrak{q}\right\rangle \left\langle \mathfrak{q}\right\vert }{%
\left\langle \mathfrak{q}\right\vert \left\vert \mathfrak{q}\right\rangle }=1
\notag \\
&\Longrightarrow &\sum\limits_{\mathfrak{q}}\left\vert \mathfrak{q}%
\right\rangle \left\langle \mathfrak{q}\right\vert =1  \label{derived_compl}
\end{eqnarray}%
which yields the completeness relation, with normalized $\left\langle 
\mathfrak{q}\right\vert \left\vert \mathfrak{q}\right\rangle \equiv 1$.
Thus, the the $\left\vert \mathfrak{q}\right\rangle $and $\left\vert 
\mathfrak{p}\right\rangle $eigenstates obey the completenes relations,%
\begin{eqnarray}
\sum\limits_{\mathfrak{q}}\left\vert \mathfrak{q}\right\rangle \left\langle 
\mathfrak{q}\right\vert &=&1  \notag \\
\sum\limits_{\mathfrak{p}}\left\vert \mathfrak{p}\right\rangle \left\langle 
\mathfrak{p}\right\vert &=&1  \label{complete theta_phi}
\end{eqnarray}%
Then it becomes trivial to see the transformation between the dual spaces, $%
\left\vert \mathfrak{q}\right\rangle $and $\left\vert \mathfrak{p}%
\right\rangle $eigenstates,%
\begin{eqnarray}
\left\vert \mathfrak{q}\right\rangle &=&\sum\limits_{\phi }\left\vert 
\mathfrak{p}\right\rangle \left\langle \mathfrak{p}\right\vert \left\vert 
\mathfrak{q}\right\rangle  \label{trans1} \\
\left\vert \mathfrak{p}\right\rangle &=&\sum\limits_{\theta }\left\vert 
\mathfrak{q}\right\rangle \left\langle \mathfrak{q}\right\vert \left\vert 
\mathfrak{p}\right\rangle  \label{trans2}
\end{eqnarray}%
with transformation function between elements of new dual spaces given by $%
\left\langle \mathfrak{p}\right\vert \left\vert \mathfrak{q}\right\rangle $
and $\left\langle \mathfrak{q}\right\vert \left\vert \mathfrak{p}%
\right\rangle $, respectively. From Eqs. (\ref{biortho1}) and (\ref{biortho2}%
), we have 
\begin{eqnarray*}
\left\langle \mathfrak{q}_{m}\right\vert \left\vert \mathfrak{q}%
_{n}\right\rangle &=&\delta _{m,n} \\
\left\langle \mathfrak{p}_{m}\right\vert \left\vert \mathfrak{p}%
_{n}\right\rangle &=&\delta _{m,n}
\end{eqnarray*}%
firmly defining the complete and orthogonal dual Hilbert spaces, $\left\{
\left\vert \mathfrak{q}\right\rangle \right\} $ and$\left\{ \left\vert 
\mathfrak{p}\right\rangle \right\} $.

\subsection{Generation of states}

Equation (\ref{anticom_com}), defines the generation of state $\left\vert 
\mathfrak{q}\right\rangle $, 
\begin{equation}
\left\vert \mathfrak{q}\right\rangle =C_{o}\exp \alpha \hat{\psi}^{\dagger
}\left\vert 0\right\rangle  \label{qft_gen}
\end{equation}%
\begin{equation*}
\hat{\psi}^{\dagger }\left\vert \mathfrak{q}\right\rangle =\hat{\psi}%
^{\dagger }\exp \left( q\hat{\psi}^{\dagger }\left\vert 0\right\rangle
\right) =\frac{\partial }{\partial \mathfrak{q}}\mathcal{\ }\left\vert 
\mathfrak{q}\right\rangle
\end{equation*}%
Inserting the term $\exp \left\{ -\alpha ^{\ast }\hat{\psi}\right\} $ right
in front of $\left\vert 0\right\rangle $\footnote{%
There is arbitrarines in incorporating $\exp \left\{ \phi \hat{a}\right\} $,
either positive or negative exponent, operating on vacuum state. To be
symmetric we should use positive exponent, $\exp \left\{ \phi \hat{a}%
\right\} $. For convenience, we want the generation os state unitary, so is
advisable to use the negative exponent. We will follow this convention is
what follows.}, which has the effect of multiplying by unity, we obtain a
fully symmetric form as 
\begin{equation}
\left\vert \mathfrak{q}\right\rangle =C_{o}\exp \alpha \hat{\psi}^{\dagger
}\exp \left\{ -\alpha ^{\ast }\hat{\psi}\right\} \left\vert \psi
_{0}\right\rangle  \label{qft_sym_gen}
\end{equation}%
To avoid confusion, we set the eigenvalues $\alpha =\mathfrak{q}$ and $%
\alpha ^{\ast }=\mathfrak{p}$. We also set $\left\vert \mathfrak{q}%
\right\rangle =\exp \left( -i\mathfrak{q}\cdot \mathcal{\hat{P}}\right)
\left\vert 0\right\rangle $, i.e., we have,%
\begin{eqnarray}
\hat{\psi}^{\dagger }\left\vert \mathfrak{q}\right\rangle &=&\frac{\partial 
}{\partial \mathfrak{q}}\mathcal{\ }\left\vert \mathfrak{q}\right\rangle =-i%
\mathcal{\hat{P}\ }\left\vert \mathfrak{q}\right\rangle  \label{qft_p} \\
\mathcal{\hat{P}}\left\vert \mathfrak{p}\right\rangle &=&\ \mathfrak{p}%
\mathcal{\ }\left\vert \mathfrak{p}\right\rangle  \label{qft_p1}
\end{eqnarray}%
To calculate $\left\langle \mathfrak{p}\right\vert \left\vert \mathfrak{q}%
\right\rangle $, we proceed as follows.%
\begin{eqnarray}
\left\langle \mathfrak{p}\right\vert \hat{\psi}^{\dagger }\left\vert 
\mathfrak{q}\right\rangle &=&\left\langle \mathfrak{p}\right\vert \frac{%
\partial }{\partial \mathfrak{q}}\mathcal{\ }\left\vert \mathfrak{q}%
\right\rangle  \label{qft_q} \\
-i\mathfrak{p}\left\langle \mathfrak{p}\right\vert \left\vert \mathfrak{q}%
\right\rangle &=&\frac{\partial }{\partial \mathfrak{q}}\left\langle 
\mathfrak{p}\right\vert \left\vert \mathfrak{q}\right\rangle  \label{qft_q2}
\\
\frac{\frac{\partial }{\partial \mathfrak{q}}\left\langle \mathfrak{p}%
\right\vert \left\vert \mathfrak{q}\right\rangle }{\left\langle \mathfrak{p}%
\right\vert \left\vert \mathfrak{q}\right\rangle } &=&-i\mathfrak{p}
\label{qft_q3} \\
\frac{\partial }{\partial \mathfrak{q}}\ln \left\langle \mathfrak{p}%
\right\vert \left\vert \mathfrak{q}\right\rangle &=&-i\mathfrak{p}
\label{qft_q4} \\
\left\langle \mathfrak{p}\right\vert \left\vert \mathfrak{q}\right\rangle
&=&\exp \left( -i\mathfrak{q}\cdot \mathfrak{p}\right)  \label{qft_q5}
\end{eqnarray}%
Similarly, we have 
\begin{eqnarray}
\hat{\psi}\left\vert \mathfrak{p}\right\rangle &=&\frac{\partial }{\partial 
\mathfrak{p}}\mathfrak{\ }\left\vert \mathfrak{p}\right\rangle =i\mathfrak{%
\mathcal{\hat{Q}\ }}\left\vert \mathfrak{p}\right\rangle  \label{diff_Q} \\
\mathfrak{\mathcal{\hat{Q}}}\ \left\vert \mathfrak{q}\right\rangle &=&%
\mathfrak{q\ }\left\vert \mathfrak{q}\right\rangle  \label{eigen_Q1}
\end{eqnarray}%
and%
\begin{eqnarray}
\left\langle \mathfrak{q}\right\vert \hat{\psi}\left\vert \mathfrak{p}%
\right\rangle &=&\left\langle \mathfrak{q}\right\vert \frac{\partial }{%
\partial \mathfrak{p}}\mathfrak{\ }\left\vert \mathfrak{p}\right\rangle 
\notag \\
i\mathfrak{q}\left\langle \mathfrak{q}\right\vert \left\vert \mathfrak{p}%
\right\rangle &=&\frac{\partial }{\partial \mathfrak{p}}\mathfrak{\ }%
\left\langle \mathfrak{q}\right\vert \left\vert \mathfrak{p}\right\rangle 
\notag \\
\left\langle \mathfrak{q}\right\vert \left\vert \mathfrak{p}\right\rangle
&=&\exp i\mathfrak{p\cdot q}  \label{qp_qft}
\end{eqnarray}%
So far all the above developments holds for fermions and bosons. However,
note that for fermions the eigenvalues corresponding to $\mathfrak{q}$ and $%
\mathfrak{p}$ are elements of the Grassmann algebra. First we will treat the
case of mixed space representation of boson quantum field theory.

\section{Mixed Space Representation in QFT: Bosons}

Since the commutator for bosons, $\left[ \hat{\psi},\hat{\psi}^{\dagger }%
\right] \Longrightarrow \left[ \mathcal{\hat{Q}},-i\mathcal{\hat{P}}\right]
=1_{-}$, is a $%
\mathbb{C}
$-number (i.e., not an operator), we can readily make use of the
Campbell-Baker-Hausdorff operator identity, namely, \footnote{%
Equation (\ref{sym_translation-op2}) readily follows from the symmetric form
of translation operator in Eq. (\ref{displacement-op}), involving the
universal canonical operators, $\hat{Q}$and $\hat{P}$, 
\begin{eqnarray*}
&&\exp \left\{ -\frac{i}{\hbar }q\cdot \hat{P}\right\} \exp \left\{ \frac{i}{%
\hbar }p\cdot \hat{Q}\right\} \\
&=&\exp \left\{ -\frac{i}{\hbar }\left( q\cdot \hat{P}-p\cdot \hat{Q}\right)
\right\} \exp \left\{ -\frac{i}{\hbar }\frac{p\cdot q}{2}\right\}
\end{eqnarray*}%
if one substitute the following relations%
\begin{eqnarray*}
P &=&i\hbar \hat{a}^{\dagger } \\
p &=&i\hbar \alpha ^{\ast } \\
Q &=&\hat{a} \\
q &=&\alpha
\end{eqnarray*}%
}%
\begin{eqnarray}
C_{o}\exp \left\{ -i\mathfrak{q\cdot }\mathcal{\hat{P}}\right\} \exp \left\{
i\mathfrak{p\cdot }\mathcal{\hat{Q}}\right\} &=&C_{o}\exp \left( -i\left\{ 
\mathfrak{q\cdot }\mathcal{\hat{P}}-\mathfrak{p\cdot }\mathcal{\hat{Q}}%
\right\} \right) \exp \left\{ -\frac{\left[ \mathfrak{q\cdot }\mathcal{\hat{P%
}},\mathfrak{p\cdot }\mathcal{\hat{Q}}\right] }{2}\right\} _{\pm }  \notag \\
&=&C_{o}\exp \left( -i\left\{ \mathfrak{q\cdot }\mathcal{\hat{P}}-\mathfrak{%
p\cdot }\mathcal{\hat{Q}}\right\} \right) \exp \left\{ i\frac{\mathfrak{q}%
\cdot \mathfrak{p}}{2}\right\} .  \label{sym_translation-op2}
\end{eqnarray}%
Therefore, the generator of $\left\vert \mathfrak{q}\right\rangle $ states
as,%
\begin{eqnarray}
\left\vert \mathfrak{q}\right\rangle &=&C_{o}\exp \left\{ i\frac{\mathfrak{q}%
\cdot \mathfrak{p}}{2}\right\} \exp \left( -i\left\{ \mathfrak{q\cdot }%
\mathcal{\hat{P}}-\mathfrak{p\cdot }\mathcal{\hat{Q}}\right\} \right)
\left\vert 0\right\rangle  \notag \\
&=&\mathcal{D}\left( \mathfrak{q}\right) \left\vert 0\right\rangle .
\label{renorm_gen}
\end{eqnarray}%
Thus $\mathcal{D}\left( \mathfrak{q}\right) $ is the displacement operator
that generates the state $\left\vert \mathfrak{q}\right\rangle $ from the
vacuum, $\left\vert 0\right\rangle $. Note that $\exp \left( i\mathfrak{p}%
^{\prime }\mathfrak{\cdot }\mathcal{\hat{Q}}\right) \left\vert \mathfrak{p}%
\right\rangle =\exp \left( \mathfrak{p}^{\prime }\cdot \frac{\partial }{%
\partial \mathfrak{p}}\right) \left\vert \mathfrak{p}\right\rangle $ when
operating on $\left\vert \mathfrak{p}\right\rangle $ states.

\subsection{Reformulation as mixed Space Representation}

We can now formulate the non-Hermitian operators to Hermitian-like dual
space representation, similar to the $q$-$p$ dual space representation of
Sec. \ref{q-p_represent}.

\subsubsection{Isomorphism between $\mathfrak{q}$-$\mathfrak{p}$ and $q$-$p$
dual spaces}

The development that follows basically demonstrate the remarkable
isomorphism of annihilation and creation operators, $\mathfrak{q}$-$%
\mathfrak{p}$ or $\alpha $-$\alpha ^{\ast }\left( \psi \text{-}\psi ^{\ast
}\right) $ and their dual Hilbert spaces with the position and momentum
operators and their dual Hilbert spaces, $q$-$p$. This is clearly seen by
the following bijective and homomorphic or isomorphic transformation, i.e.,
the harmonic oscillator transformation from position and momentum operators
to annihilation and creation operators,%
\begin{eqnarray}
\left( 
\begin{array}{c}
\hat{\psi} \\ 
\hat{\psi}^{\dagger }%
\end{array}%
\right) &=&\left( 
\begin{array}{c}
\mathcal{Q} \\ 
-i\mathcal{P}%
\end{array}%
\right) =\frac{1}{\sqrt{2}}\left( 
\begin{array}{cc}
1 & 1 \\ 
1 & -1%
\end{array}%
\right) \left( 
\begin{array}{c}
\hat{Q} \\ 
i\hat{P}%
\end{array}%
\right)  \label{harmonic_map} \\
&=&\frac{1}{\sqrt{2}}\left( 
\begin{array}{c}
\hat{Q}+i\hat{P} \\ 
\hat{Q}-i\hat{P}%
\end{array}%
\right)
\end{eqnarray}%
where $\hat{Q}$ and $\hat{P}$ are dimensionless canonical operators, with
eigenvalues, $q$ and $p$, respectively. Then the transformation of Eq. (\ref%
{harmonic_map}) induces a transformation of Eq. (\ref{PQ_trans2}) as%
\begin{eqnarray}
\mathcal{D}_{sym}\left( \mathfrak{q,p}\right) &=&\exp \left\{ -i\left( 
\mathcal{P\cdot }\mathfrak{q-}\mathcal{Q\cdot }\mathfrak{p}\right) \right\} 
\notag \\
&=&\exp \left\{ -i\mathcal{P\cdot }\mathfrak{q+i}\mathcal{Q\cdot }\mathfrak{p%
}\right\}  \notag \\
&=&\exp \left\{ \hat{\psi}^{\dagger }\mathcal{\cdot }\mathfrak{q+}\hat{\psi}%
\mathcal{\cdot }\mathfrak{ip}\right\}  \label{PQ_trans2}
\end{eqnarray}%
We have%
\begin{eqnarray*}
\hat{\psi}^{\dagger }\mathcal{\cdot }\mathfrak{q+}\hat{\psi}\mathcal{\cdot }%
\mathfrak{ip} &\mathfrak{=}&\frac{1}{2}\left[ \left( \hat{Q}-i\hat{P}\right)
\cdot \left( v+iu\right) +\left( \hat{Q}+i\hat{P}\right) \cdot \left(
v-iu\right) \right] \\
&=&\frac{1}{2}\left[ \left( \hat{Q}-i\hat{P}\right) \cdot \left( v+iu\right)
+\left( \hat{Q}+i\hat{P}\right) \cdot \left( -v+iu\right) \right] \\
&=&-i\left( \hat{P}\cdot v-\hat{Q}\cdot u\right)
\end{eqnarray*}%
which upon substituting in Eq. (\ref{PQ_trans2} gives 
\begin{eqnarray}
Y\left( \mathfrak{q,p}\right) &=&\exp \left\{ -i\mathcal{P\cdot }\mathfrak{%
q+i}\mathcal{Q\cdot }\mathfrak{p}\right\}  \notag \\
&=&\exp \left[ -i\left( \hat{P}\cdot v-\hat{Q}\cdot u\right) \right]  \notag
\end{eqnarray}%
where $Y\left( \mathfrak{q,p}\right) $ is the symmetric translation operator
in position-momentum, $q$-$p$ dual space. Therefore the dual space $\left\{ 
\mathfrak{q,p}\right\} $ or $\left\{ \alpha ,\alpha ^{\ast }\right\} $is
isomorphic to $q$-$p$ dual space. \ \ \ \ \ \ \ \ \ \ \ \ \ \ \ \ \ \ \ \ \
\ \ \ \ \ 

\subsection{The $\mathfrak{q}$-$\mathfrak{p}$ representations and lattice
Weyl Transform}

The mixed $\mathfrak{q}$-$\mathfrak{p}$ representation basically start by
expanding any quantum operator, $\hat{A}$, in terms of mutually unbiased
basis states, namely the eigenvector of annihilation operator, $\hat{\psi}$
or $\mathcal{Q}$, and the eigenvector of creation operator, $\hat{\psi}%
^{\dagger }$or $\mathcal{P}$. We have%
\begin{eqnarray}
\hat{A} &=&\sum\limits_{\mathfrak{p},\mathfrak{q}}\left\vert \mathfrak{q}%
\right\rangle \left\langle \mathfrak{q}\right\vert \hat{A}\left\vert 
\mathfrak{p}\right\rangle \left\langle \mathfrak{p}\right\vert  \notag \\
&=&\sum\limits_{\mathfrak{p},\mathfrak{q}}\left\langle \mathfrak{q}%
\right\vert \hat{A}\left\vert \mathfrak{p}\right\rangle \ \left\vert 
\mathfrak{q}\right\rangle \left\langle \mathfrak{p}\right\vert
\label{any_op_qft}
\end{eqnarray}

\subsection{The completeness of dual spaces}

The set $\left\{ \left\vert \mathfrak{q}\right\rangle \left\langle \mathfrak{%
p}\right\vert \right\} $ is the basis operators for the mixed $\mathfrak{q}$-%
$\mathfrak{p}$ representation. From the completeness relations of the
unbiased basis states, $\left\{ \left\vert \mathfrak{q}\right\rangle
\right\} $ and $\left\{ \left\vert \mathfrak{p}\right\rangle \right\} $, the
set $\left\{ \left\vert \mathfrak{q}\right\rangle \left\langle \mathfrak{p}%
\right\vert \right\} $ obeys the completeness relation%
\begin{eqnarray}
\sum\limits_{\mathfrak{q},\mathfrak{p}}\left\vert \mathfrak{q}\right\rangle
\left\langle \mathfrak{q}\right\vert \left\vert \mathfrak{p}\right\rangle
\left\langle \mathfrak{p}\right\vert &=&1  \label{complete1_qft} \\
\sum\limits_{\mathfrak{q},\mathfrak{p}}\left\langle \mathfrak{q}\right\vert
\left\vert \mathfrak{p}\right\rangle \ \left\vert \mathfrak{q}\right\rangle
\left\langle \mathfrak{p}\right\vert &=&1  \label{complete2_qft}
\end{eqnarray}%
Substituting the expression for $\left\langle \mathfrak{q}\right\vert
\left\vert \mathfrak{p}\right\rangle $, 
\begin{eqnarray}
\left\langle \mathfrak{q}\right\vert \left\vert \mathfrak{p}\right\rangle
&=&\exp \left( i\mathfrak{p}\cdot \mathfrak{q}\right)  \label{qp_eq_qft} \\
\left\langle \mathfrak{p}\right\vert \left\vert \mathfrak{q}\right\rangle
&=&\exp \left( -i\mathfrak{p}\cdot \mathfrak{q}\right)  \label{pq_eq_qft}
\end{eqnarray}%
we obtained, for the completeness relation, 
\begin{equation}
C_{0}\sum\limits_{\mathfrak{q},\mathfrak{p}}\exp \left( i\mathfrak{p}\cdot 
\mathfrak{q}\right) \left\vert \mathfrak{q}\right\rangle \left\langle 
\mathfrak{p}\right\vert =1  \label{completeness_qft}
\end{equation}%
where $C_{o}$can be choosen as 
\begin{equation*}
C_{o}=\left( N\right) ^{-\frac{1}{2}}
\end{equation*}%
Equation (\ref{completeness}) can be rewritten as%
\begin{equation}
\left( N\right) ^{-\frac{1}{2}}\sum\limits_{\mathfrak{q},\mathfrak{p}}\frac{%
\left\vert \mathfrak{q}\right\rangle \left\langle \mathfrak{p}\right\vert }{%
\left\langle \mathfrak{p}\right\vert \left\vert \mathfrak{q}\right\rangle }=1
\label{completeness2_qft}
\end{equation}%
and similarly,%
\begin{equation*}
\left( N\right) ^{-\frac{1}{2}}\sum\limits_{\mathfrak{q},\mathfrak{p}}\frac{%
\left\vert \mathfrak{p}\right\rangle \left\langle \mathfrak{q}\right\vert }{%
\left\langle \mathfrak{q}\right\vert \left\vert \mathfrak{p}\right\rangle }=1
\end{equation*}%
Here we use the transformation identities in the mixed $\mathfrak{q}$-$%
\mathfrak{p}$ representation, 
\begin{eqnarray}
\left\vert \mathfrak{p}\right\rangle &=&\sum\limits_{\mathfrak{q}%
}\left\langle \mathfrak{q}\right\vert \left\vert \mathfrak{p}\right\rangle \
\left\vert \mathfrak{q}\right\rangle  \label{eq_p_qft} \\
\left\langle \mathfrak{p}\right\vert &=&\sum\limits_{\theta }\left\langle 
\mathfrak{p}\right\vert \left\vert \mathfrak{q}\right\rangle \ \left\langle 
\mathfrak{q}\right\vert  \label{eq_p1_qft} \\
\left\vert \mathfrak{q}\right\rangle &=&\sum\limits_{\phi }\left\langle 
\mathfrak{p}\right\vert \left\vert \mathfrak{q}\right\rangle \ \left\vert 
\mathfrak{p}\right\rangle  \label{eq_q_qft} \\
\left\langle \mathfrak{q}\right\vert &=&\sum\limits_{\phi }\left\langle 
\mathfrak{q}\right\vert \left\vert \mathfrak{p}\right\rangle \ \left\langle 
\mathfrak{p}\right\vert  \label{eq_q1_qft}
\end{eqnarray}%
with transformation functions given by Eqs. (\ref{qp_eq})-(\ref{pq_eq}).

\subsubsection{The expansion of any operators in dual space}

Any operator, $\hat{A}$, can be expressed in terms of the unbiased
eigenvector spaces, $\left\vert \mathfrak{q}^{\prime }\right\rangle $and $%
\left\vert \mathfrak{p}^{\prime \prime }\right\rangle $, respectively, in a
mixed representation by Eq. (\ref{any_op}), which we rewrite as,

\begin{eqnarray}
\hat{A} &=&\sum\limits_{\mathfrak{p}^{\prime \prime },\mathfrak{q}^{\prime
}}\left\vert \mathfrak{p}^{\prime \prime }\right\rangle \left\langle 
\mathfrak{p}^{\prime \prime }\right\vert A\left\vert \mathfrak{q}^{\prime
}\right\rangle \left\langle \mathfrak{q}^{\prime }\right\vert  \notag \\
&=&\sum\limits_{\mathfrak{p}^{\prime \prime },\mathfrak{q}^{\prime
}}\left\langle \mathfrak{p}^{\prime \prime }\right\vert A\left\vert 
\mathfrak{q}^{\prime }\right\rangle \ \left\vert \mathfrak{p}^{\prime \prime
}\right\rangle \left\langle \mathfrak{q}^{\prime }\right\vert
\label{any_op2_qft}
\end{eqnarray}%
We wish to express $\left\langle \mathfrak{p}^{\prime \prime }\right\vert
A\left\vert \mathfrak{q}^{\prime }\right\rangle $ and $\ \left\vert 
\mathfrak{p}^{\prime \prime }\right\rangle \left\langle \mathfrak{q}^{\prime
}\right\vert $ in terms of the $\left\vert \mathfrak{q}\right\rangle $%
-eigenstate matrix elements and $\left\vert \mathfrak{p}\right\rangle $%
-space projectors, respectively. Using, Eqs. (\ref{eq_p})-(\ref{p2q}), we
write%
\begin{eqnarray}
\left\langle \mathfrak{p}^{\prime \prime }\right\vert A\left\vert \mathfrak{q%
}^{\prime }\right\rangle &=&\frac{1}{\sqrt{N}}\sum\limits_{\mathfrak{q}%
^{\prime \prime }}e^{-i\mathfrak{p}^{\prime \prime }\cdot \mathfrak{q}%
^{\prime \prime }}\left\langle \mathfrak{q}^{\prime \prime }\right\vert
A\left\vert \mathfrak{q}^{\prime }\right\rangle  \notag \\
\left\vert \mathfrak{p}^{\prime \prime }\right\rangle \left\langle \mathfrak{%
q}^{\prime }\right\vert &=&\frac{1}{\sqrt{N}}\sum\limits_{\mathfrak{p}%
^{\prime }}e^{i\mathfrak{p}^{\prime }\cdot \mathfrak{q}^{\prime }}\left\vert 
\mathfrak{p}^{\prime \prime }\right\rangle \left\langle \mathfrak{p}^{\prime
}\right\vert  \label{matrix_project_qft}
\end{eqnarray}%
with completeness relation, using the transformation function characteristic
of dual spaces,%
\begin{equation*}
\frac{1}{\sqrt{N}}\sum\limits_{\mathfrak{q},\mathfrak{p}}\exp \left( i%
\mathfrak{p}\cdot \mathfrak{q}\right) \left\vert \mathfrak{q}\right\rangle
\left\langle \mathfrak{p}\right\vert =1=\sum\limits_{\mathfrak{p}}\left\vert 
\mathfrak{p}\right\rangle \left\langle \mathfrak{p}\right\vert
\end{equation*}%
Introducing the notation in Eq. (\ref{matrix_project}),%
\begin{eqnarray*}
\mathfrak{p}^{\prime } &=&\mathfrak{p}+\mathfrak{u}\text{, \ \ \ \ \ \ \ }%
\mathfrak{q}^{\prime }=\mathfrak{q}+\mathfrak{v}, \\
\mathfrak{p}^{\prime \prime } &=&\mathfrak{p}-\mathfrak{u}\text{, \ \ \ \ \
\ \ }\mathfrak{q}^{\prime \prime }=\mathfrak{q}-\mathfrak{v}\text{.}
\end{eqnarray*}%
Then, upon substituting in Eq. (\ref{any_op2}), we end up with%
\begin{eqnarray*}
A &=&\sum\limits_{\mathfrak{p}^{\prime \prime },\mathfrak{q}^{\prime
}}\left\vert \mathfrak{p}^{\prime \prime }\right\rangle \left\langle 
\mathfrak{p}^{\prime \prime }\right\vert A\left\vert \mathfrak{q}^{\prime
}\right\rangle \left\langle \mathfrak{q}^{\prime }\right\vert \\
&=&\frac{1}{N}\sum\limits_{\mathfrak{p},\mathfrak{q},\mathfrak{u},\mathfrak{v%
}}e^{i2\left( \mathfrak{p}\cdot \mathfrak{v}+\mathfrak{u}\cdot \mathfrak{q}%
\right) }\left\langle \mathfrak{q}-\mathfrak{v}\right\vert A\left\vert 
\mathfrak{q}+\mathfrak{v}\right\rangle \ \left\vert \mathfrak{p}-\mathfrak{u}%
\right\rangle \left\langle \mathfrak{p}+\mathfrak{u}\right\vert
\end{eqnarray*}

\subsubsection{Mixed space operator basis, $\hat{\Delta}\left( \mathfrak{p},%
\mathfrak{q}\right) $}

We write the last result as an expansion in terms of \textit{mixed}-\textit{%
phase point projector}, $\hat{\Delta}\left( \mathfrak{p},\mathfrak{q}\right) 
$, defined as the\textit{\ Weyl transform} of a projector, by%
\begin{eqnarray}
\hat{\Delta}\left( \mathfrak{p},\mathfrak{q}\right) &=&\sum\limits_{%
\mathfrak{u}}e^{-i2\mathfrak{u}\cdot \mathfrak{q}}\left\vert \mathfrak{p}+%
\mathfrak{u}\right\rangle \left\langle \mathfrak{p}-\mathfrak{u}\right\vert 
\notag \\
&=&\sum\limits_{\mathfrak{u}}e^{-i2\mathfrak{u}\cdot \mathfrak{q}%
}e^{2iQ\cdot u}\left\vert \mathfrak{p}-\mathfrak{u}\right\rangle
\left\langle \mathfrak{p}-\mathfrak{u}\right\vert  \notag \\
&=&\sum\limits_{\mathfrak{u}}e^{-i2\mathfrak{u\cdot q}}e^{2i\mathcal{Q}\cdot 
\mathfrak{u}}\sum\limits_{\mathfrak{v}}e^{2i\left( \mathfrak{p-u}-\mathcal{P}%
\right) \cdot \mathfrak{v}}\left\vert \mathfrak{p}_{0}\right\rangle
\left\langle \mathfrak{p}_{0}\right\vert  \notag \\
&=&\sum\limits_{\mathfrak{u,v}}e^{2i\left( \mathcal{Q-}\mathfrak{q}\right)
\cdot \mathfrak{u}}e^{-2i\left( \mathcal{P-}\mathfrak{p}\right) \cdot 
\mathfrak{v}}e^{-2i\mathfrak{u}\cdot \mathfrak{v}}\left\vert \mathfrak{p}%
_{0}\right\rangle \left\langle \mathfrak{p}_{0}\right\vert  \notag \\
&=&\sum\limits_{\mathfrak{u,v}}e^{2i\left( \mathfrak{p}\cdot \mathfrak{%
v-q\cdot u}\right) }e^{-2i\left( \mathcal{P}\cdot \mathfrak{v-}\mathcal{%
Q\cdot }\mathfrak{u}\right) }\sum\limits_{\mathfrak{p}_{0}}\left\vert 
\mathfrak{p}_{0}\right\rangle \left\langle \mathfrak{p}_{0}\right\vert
\label{delta_sym}
\end{eqnarray}%
and the coefficient of expansion, the so-called \textit{Weyl transform} of
matrix element of operator, $A\left( \mathfrak{p},\mathfrak{q}\right) $,
defined by%
\begin{equation*}
A\left( \mathfrak{p},\mathfrak{q}\right) =\sum\limits_{\mathfrak{v}}e^{i2%
\mathfrak{p}\cdot \mathfrak{v}}\left\langle \mathfrak{q}-\mathfrak{v}%
\right\vert A\left\vert \mathfrak{q}+\mathfrak{v}\right\rangle \text{.}
\end{equation*}%
Clearly, for a density matrix operator $\hat{\rho}$, the Weyl transform
obeys,%
\begin{equation*}
\sum\limits_{\mathfrak{p},\mathfrak{q}}\rho \left( \mathfrak{p},\mathfrak{q}%
\right) =1
\end{equation*}%
If one accounts for other extra discrete quantum labels like spin and
energy-band indices, we can incorporate this in the summation in a form of a
trace.

Thus, we eventually have any operator expanded in terms of mixed space
operator basis, $\hat{\Delta}\left( \mathfrak{p},\mathfrak{q}\right) $,%
\begin{eqnarray}
\hat{A} &=&\sum\limits_{\mathfrak{p},\mathfrak{q}}A\left( \mathfrak{p},%
\mathfrak{q}\right) \ \hat{\Delta}\left( \mathfrak{p},\mathfrak{q}\right)
\label{LWT1_qft} \\
&=&\sum\limits_{\mathfrak{u,v}}\left( \sum\limits_{\mathfrak{p},\mathfrak{q}%
}A\left( \mathfrak{p},\mathfrak{q}\right) e^{2i\left( \mathfrak{p}\cdot 
\mathfrak{v-q\cdot u}\right) }\right) e^{-2i\left( \mathcal{P}\cdot 
\mathfrak{v-}\mathcal{Q\cdot }\mathfrak{u}\right) }\left\vert \mathfrak{p}%
_{0}\right\rangle \left\langle \mathfrak{p}_{0}\right\vert
\end{eqnarray}%
We have%
\begin{eqnarray*}
A\left( \mathfrak{p},\mathfrak{q}\right) &=&Tr\left( \hat{A}\ \hat{\Delta}%
\left( \mathfrak{p},\mathfrak{q}\right) \right) \\
&=&\ \sum\limits_{\mathfrak{u,v}}e^{2i\left( \mathfrak{p}\cdot \mathfrak{%
v-q\cdot u}\right) }Tr\left( \hat{A}e^{-2i\left( \mathcal{P}\cdot \mathfrak{%
v-}\mathcal{Q\cdot }\mathfrak{u}\right) }\left\vert \mathfrak{p}%
_{0}\right\rangle \left\langle \mathfrak{p}_{0}\right\vert \right) \\
&=&\sum\limits_{\mathfrak{u,v}}e^{2i\left( \mathfrak{p}\cdot \mathfrak{%
v-q\cdot u}\right) }A\left( \mathfrak{u,v}\right)
\end{eqnarray*}%
where $A\left( \mathfrak{u,v}\right) $ is the characteristic function of $%
A\left( \mathfrak{p},\mathfrak{q}\right) $ distribution. Upon similar
procedure based on Eq. (\ref{any_op2}), an equivalent expression can be
obtain for $A\left( \mathfrak{p},\mathfrak{q}\right) $ and $\hat{\Delta}%
\left( \mathfrak{p},\mathfrak{q}\right) $, namely,%
\begin{eqnarray}
A\left( \mathfrak{p},\mathfrak{q}\right) &=&\sum\limits_{\mathfrak{u}}e^{i2%
\mathfrak{u}\cdot \mathfrak{q}}\left\langle \mathfrak{p}+\mathfrak{u}%
\right\vert \hat{A}\left\vert \mathfrak{p}-\mathfrak{u}\right\rangle
\label{alt1_qft} \\
\hat{\Delta}\left( \mathfrak{p},\mathfrak{q}\right) &=&\sum\limits_{%
\mathfrak{v}}e^{i2\mathfrak{p}\cdot \mathfrak{v}}\left\vert \mathfrak{q}+%
\mathfrak{v}\right\rangle \left\langle \mathfrak{q}-\mathfrak{v}\right\vert
\label{alt2_qft}
\end{eqnarray}

\subsubsection{Symmetrization of the generator of eigenstates}

Generator of eigenstates occupy a central role in non-Hermitian quantum
mechanics. We can also define this in the Hermitian mixed $q$-$p$
representation. We have for $\mathfrak{q}$,%
\begin{equation}
\left\vert \mathfrak{q}\right\rangle =\exp \left\{ -i\mathfrak{q}\cdot 
\mathcal{P}\right\} \left\vert 0\right\rangle ,
\label{translate-op-position}
\end{equation}%
where the operator $\mathcal{P}=i\nabla _{\mathfrak{q}}$is operating on the
basis eigenvector $\left\vert \mathfrak{q}\right\rangle $, since $\mathcal{P}%
=-i\nabla _{x}$acting on the $x$-components behaves contravariantly. Here,
we basically made an assumption in the above expression that there exist
continuous \ function of $\mathfrak{q}$ having an infinite radius of
convergence, which are equal to $\left\vert \mathfrak{q}\right\rangle $ at
the lattice points, i.e., we expand the exponential in Eq. (\ref%
{translate-op-position}) as a well-defined Taylor series.

\subsubsection{Fully symmetric translation operators}

The state 
\begin{equation}
\left\vert \mathfrak{q}\right\rangle =\exp \left\{ -i\mathfrak{q}\cdot 
\mathcal{P}\right\} \left\vert 0\right\rangle  \label{PQ_trans}
\end{equation}%
is an eigenstate with displaced eigenvalue by $\mathfrak{q}$. However, if
the limit $\mathfrak{q}^{\prime }\Rightarrow 0$ is not taken then the state $%
\left[ \exp \left\{ -i\mathfrak{q}\cdot \mathcal{P}\right\} \left\vert 
\mathfrak{q}^{\prime }\right\rangle \right] =\left\vert \mathfrak{q}^{\prime
}+\mathfrak{q}\right\rangle $ is an eigenstate of the position operator with
eigenvalue $\mathfrak{q}^{\prime }+\mathfrak{q}$.

We can symmetrize the translation operator by inserting $\exp \left\{ i%
\mathfrak{p}\cdot \mathcal{Q}\right\} $ in front of $\left\vert
0\right\rangle $, in Eq. (\ref{PQ_trans}), which effectively insert unity.
We thus have%
\begin{equation}
\left\vert \mathfrak{q}\right\rangle =\exp \left\{ -i\mathfrak{q}\cdot 
\mathcal{P}\right\} \exp \left\{ i\mathfrak{p}\cdot \mathcal{Q}\right\}
\left\vert 0\right\rangle .  \label{gen_q-p_qft}
\end{equation}%
By the use of the Campbell-Baker-Hausdorff operator identity, we obtained%
\begin{eqnarray}
&&\exp \left\{ -i\mathfrak{q}\cdot \mathcal{P}\right\} \exp \left\{ i%
\mathfrak{p}\cdot \mathcal{Q}\right\}  \notag \\
&=&\exp \left\{ -i\left( \mathfrak{q}\cdot \mathcal{P}-\mathfrak{p}\cdot 
\mathcal{Q}\right) \right\} \exp \left\{ \frac{\left[ -i\mathfrak{q}\cdot 
\mathcal{P},i\mathfrak{p}\cdot \mathcal{Q}\right] }{2}\right\}  \notag \\
&=&\exp \left\{ -i\left( q\cdot P-p\cdot Q\right) \right\} \exp \left\{ -i%
\frac{\mathfrak{p}\cdot \mathfrak{q}}{2}\right\} .
\label{displacement-op_qft}
\end{eqnarray}%
Therefore we have the symmetric form for the displacement operator
generating the state $\left\vert \mathfrak{q}\right\rangle $ from $%
\left\vert 0\right\rangle $given by 
\begin{equation}
T\left( \mathfrak{q,p}\right) _{sym}=\exp \left\{ -i\frac{\mathfrak{p}\cdot 
\mathfrak{q}}{2}\right\} \exp \left\{ -\frac{i}{\hbar }\left( \mathfrak{q}%
\cdot \mathcal{P}-\mathfrak{p}\cdot \mathcal{Q}\right) \right\} .
\label{eingenvect-displace-op}
\end{equation}%
where we used the symbol $T\left( \mathfrak{q,p}\right) _{sym}$because of
isomorphism demonstrated in Eq. (\ref{eingenvector-displacement-op}). The
displacement operator may also be interpreted as an operator for the
preparation of the quantum eigenstate out of the 'vacuum', $\left\vert
0\right\rangle \footnote{%
Here, the concept of a vacuum state does not have a special \ meaning since $%
\left\vert 0\right\rangle $represent arbitrary reference position. It is
introduced simply to bring analogy with zero-eigenvalue of non-Hermitian
operators in later chapters, there the state $\left\vert 0\right\rangle $
has a distinguished position.}$, basis eigenstate.

\subsubsection{The symmetric operator basis for mixed representations}

The ubiquitous appearance of the symmetric operator factor,%
\begin{eqnarray}
Y_{\mathfrak{p},\mathfrak{q}} &=&\exp \left( -i\left( \mathfrak{q}\cdot 
\mathcal{P}-\mathfrak{p}\cdot \mathcal{Q}\right) \right)  \notag \\
&=&\exp \left\{ i\frac{\mathfrak{p}\cdot \mathfrak{q}}{2}\right\} \exp
\left\{ -i\mathfrak{q}\cdot \mathcal{P}\right\} \exp \left\{ i\mathfrak{p}%
\cdot \mathcal{Q}\right\} ,  \label{gen_pauli-op_qft}
\end{eqnarray}%
suggest that this operator is the basis operator for mixed space
representation theiry. This is referred to here as the generalized projector
or as generalized Pauli-matrix operator \cite{buot8}. It can be considered
the universal form of projector in mixed representations of quantum physics,
either dealing with Hermitian or with non-Hermitian operators.

\paragraph{Symmetric form of $\hat{\Delta}_{\protect\lambda ^{\prime }%
\protect\lambda }\left( \mathfrak{p},\mathfrak{q}\right) $}

If we consider Eq. (\ref{alt2}) as the expression for $\hat{\Delta}_{\lambda
^{\prime }\lambda }\left( \mathfrak{p},\mathfrak{q}\right) $, the following
identities can be verified,%
\begin{equation}
\left\vert \mathfrak{q+v},\lambda \right\rangle =\exp \left[ -2i\mathcal{P}%
\cdot \mathfrak{v}\right] \left\vert \mathfrak{q-v},\lambda \right\rangle
\label{subs1_qft}
\end{equation}%
Then 
\begin{eqnarray}
\left\vert \mathfrak{q-v},\lambda \right\rangle \left\langle \mathfrak{q-v}%
,\lambda ^{\prime }\right\vert &=&\left( N\right) ^{-1}\sum\limits_{%
\mathfrak{u,q}_{0}}\exp \left[ -2i\left( \mathfrak{q-v}-\mathcal{Q}\right)
\cdot \mathfrak{u}\right] \left\vert \mathfrak{q}_{0},\lambda \right\rangle
\left\langle \mathfrak{q}_{0},\lambda ^{\prime }\right\vert
\label{subs2_qft} \\
&=&\sum\limits_{\mathfrak{q}_{0}}\delta \left( \mathfrak{q-v}-\mathcal{\hat{Q%
}}\right) \left\vert \mathfrak{q}_{0},\lambda \right\rangle \left\langle 
\mathfrak{q}_{0},\lambda ^{\prime }\right\vert =\sum\limits_{\mathfrak{q}%
_{0}}\delta \left( \mathfrak{q-v-q}_{0}\right) \ \left\vert \mathfrak{q}%
_{0},\lambda \right\rangle \left\langle \mathfrak{q}_{0},\lambda ^{\prime
}\right\vert  \notag \\
\Omega _{\lambda \lambda ^{\prime }} &=&\left\vert \mathfrak{q}_{0},\lambda
\right\rangle \left\langle \mathfrak{q}_{0},\lambda ^{\prime }\right\vert 
\notag
\end{eqnarray}%
Substituting the expressions, Eqs (\ref{subs1}) and (\ref{subs2}) in Eq. (%
\ref{alt2}), we obtain a completely symmetric expression of $\hat{\Delta}%
\left( \mathfrak{p},\mathfrak{q}\right) $, 
\begin{eqnarray*}
\hat{\Delta}\left( \mathfrak{p},\mathfrak{q}\right) &=&\sum\limits_{%
\mathfrak{v}}e^{i2\mathfrak{p}\cdot \mathfrak{v}}\left\vert \mathfrak{q}+%
\mathfrak{v}\right\rangle \left\langle \mathfrak{q}-\mathfrak{v}\right\vert
\\
&=&\left( N\right) ^{-1}\sum\limits_{\mathfrak{\bar{v}},\mathfrak{\bar{u}}%
}e^{i2\mathfrak{p}\cdot \mathfrak{v}}\exp \left[ -2i\mathcal{P}\cdot 
\mathfrak{u}\right] \exp \left[ -2i\left( \mathfrak{q}-\mathfrak{v}-\mathcal{%
\hat{Q}}\right) \cdot \mathfrak{u}\right] \sum\limits_{\mathfrak{q}%
_{0}}\left\vert \mathfrak{q}_{0},\lambda \right\rangle \left\langle 
\mathfrak{q}_{0},\lambda ^{\prime }\right\vert
\end{eqnarray*}%
Then we have%
\begin{eqnarray}
&&\exp \left\{ -i\left( \mathcal{P}-\mathfrak{p}\right) \cdot 2\mathfrak{v}%
\right\} \exp \left\{ 2i\left( \mathcal{\hat{Q}-}\mathfrak{q}\right) \cdot 
\mathfrak{\bar{u}}\right\} \exp 2i\mathfrak{v}\cdot \mathfrak{u}  \notag \\
&=&\exp \left\{ -i\left( \mathcal{P}\cdot \mathfrak{v}-\mathcal{Q}\cdot 
\mathfrak{u}\right) \right\} \exp \left\{ 2i\left( \mathfrak{p}\cdot 
\mathfrak{v}-\mathfrak{q}\cdot \mathfrak{u}\right) \right\}
\label{delta_sym2}
\end{eqnarray}%
which is the same as Eq. (\ref{delta_sym}). Thus, we have changed the
seemingly asymmetric expression of the first line into a symmetric form of
the last line.

We can combine the exponential operators to obtain%
\begin{eqnarray*}
&&\left( N\right) ^{-1}\sum\limits_{\bar{v},\bar{u}}e^{2i\mathfrak{p}\cdot 
\mathfrak{v}}\exp \left[ -2i\mathcal{P}\cdot \mathfrak{v}\right] \exp \left[
2i\left( \mathfrak{q}-\mathfrak{v}-\mathcal{Q}\right) \cdot \mathfrak{\bar{u}%
}\right] \sum\limits_{\mathfrak{q}_{0}}\left\vert \mathfrak{q}_{0},\lambda
\right\rangle \left\langle \mathfrak{q}_{0},\lambda ^{\prime }\right\vert \\
&=&\left( N\right) ^{-1}\sum\limits_{\bar{v},\bar{u}}\exp -2i\left[ \left( 
\mathcal{P}-\mathfrak{p}\right) \cdot \mathfrak{v}+\left( \mathcal{Q}-%
\mathfrak{q}\right) \cdot \mathfrak{u}\right] \Omega _{\lambda \lambda
^{\prime }}
\end{eqnarray*}%
where, 
\begin{equation*}
\Omega _{\lambda \lambda ^{\prime }}=\sum\limits_{\mathfrak{q}%
_{0}}\left\vert \mathfrak{q}_{0},\lambda \right\rangle \left\langle 
\mathfrak{q}_{0},\lambda ^{\prime }\right\vert =\sum\limits_{\mathfrak{p}%
_{0}}\left\vert \mathfrak{p}_{0},\lambda \right\rangle \left\langle 
\mathfrak{p}_{0},\lambda ^{\prime }\right\vert
\end{equation*}%
\begin{eqnarray*}
A\left( \mathfrak{p},\mathfrak{q}\right) &=&Tr\left( \hat{A}\hat{\Delta}%
\right) \\
&=&\left( N\right) ^{-1}\left( 
\begin{array}{c}
\sum\limits_{\bar{v},\bar{u}}\exp 2i\left[ \mathfrak{p}\cdot \mathfrak{v}-%
\mathfrak{q}\cdot \mathfrak{u}\right] \  \\ 
\times Tr\left\{ \hat{A}\exp \left\{ -2i\left[ \mathcal{P}\cdot \mathfrak{v}-%
\mathcal{Q}\cdot \mathfrak{u}\right] \right\} \Omega _{\lambda \lambda
^{\prime }}\right\}%
\end{array}%
\right)
\end{eqnarray*}%
Therefore, the charactetic distribution for $A\left( p,q\right) $is
identically, 
\begin{equation*}
A_{\lambda \lambda ^{\prime }}\left( \mathfrak{u},\mathfrak{v}\right)
=Tr\left\{ \hat{A}\exp \left\{ -2i\left[ \mathcal{P}\cdot \mathfrak{v}-%
\mathcal{Q}\cdot \mathfrak{u}\right] \right\} \Omega _{\lambda \lambda
^{\prime }}\right\}
\end{equation*}%
as before. By using the characteristic distribution function for $A_{\lambda
\lambda ^{\prime }}\left( \mathfrak{p},\mathfrak{q}\right) $ 
\begin{equation}
A_{\lambda \lambda ^{\prime }}\left( \mathfrak{u},\mathfrak{v}\right)
=\left( \frac{1}{N}\right) ^{\frac{1}{2}}\sum\limits_{p,q}A_{\lambda \lambda
^{\prime }}\left( \mathfrak{p},\mathfrak{q}\right) \exp 2i\left[ \mathfrak{p}%
\cdot \mathfrak{v}-\mathfrak{q}\cdot \mathfrak{u}\right]  \label{char1_qft}
\end{equation}%
with inverse%
\begin{equation*}
A_{\lambda \lambda ^{\prime }}\left( \mathfrak{p},\mathfrak{q}\right)
=\left( \frac{1}{N}\right) ^{\frac{1}{2}}\sum\limits_{u,v}A_{\lambda \lambda
^{\prime }}\left( \mathfrak{u},\mathfrak{v}\right) \exp \left\{ -2i\left[ 
\mathfrak{p}\cdot \mathfrak{v}-\mathfrak{q}\cdot \mathfrak{u}\right] \right\}
\end{equation*}%
Then we can write Eq. (\ref{LWT1}) simply like a Fourier transform (caveat:
Fourier transform to\textit{\ operator space}) of the characteristic
function of the lattice Weyl transform of the operator $\hat{A}$, 
\begin{equation}
\hat{A}=\sum\limits_{\bar{u},\bar{v},\lambda ,\lambda ^{\prime }}A_{\lambda
\lambda ^{\prime }}\left( \mathfrak{u},\mathfrak{v}\right) \exp \left\{ -2i%
\left[ \mathcal{P}\cdot \mathfrak{v}-\mathcal{Q}\cdot \mathfrak{u}\right]
\right\} \Omega _{\lambda ^{\prime }\lambda }  \label{caveat_qft}
\end{equation}%
where the inverse can be written as%
\begin{equation*}
A_{\lambda \lambda ^{\prime }}\left( \mathfrak{u},\mathfrak{v}\right)
=Tr\left\{ \hat{A}\exp \left\{ -2i\left[ \mathcal{P}\cdot \mathfrak{v}-%
\mathcal{Q}\cdot \mathfrak{u}\right] \right\} \right\} \Omega _{\lambda
\lambda ^{\prime }}
\end{equation*}%
In continuum approximation, we have,%
\begin{eqnarray}
\hat{\Delta}\left( \mathfrak{p,q}\right) &=&\left( 2\pi \right) ^{-1}\int d%
\mathfrak{u}\wedge d\mathfrak{v}\ e^{2i\left[ \left( \ \left( \mathfrak{p}-%
\mathcal{P}\right) .v-\mathfrak{q}-\mathcal{Q}\right) .\mathfrak{u}\ \right]
}  \notag \\
&=&\left( 2\pi \right) ^{-1}\int d\mathfrak{u}\wedge d\mathfrak{v}\ e^{2i%
\left[ \mathfrak{p}\cdot \mathfrak{v}-\mathfrak{q}\cdot \mathfrak{u}\right]
}e^{\left( -2i\right) \left( \mathcal{P}\cdot \mathfrak{v}-\mathcal{Q}\cdot 
\mathfrak{u}\right) },  \label{sym_delta_qft}
\end{eqnarray}

\subsubsection{On the counting of states: Coherent state formulation}

The measure of counting of states in Eq. (\ref{sym_delta_qft}) can be shown
to be related to original annihilation and creation operator in the case of
harmonic oscillator. We have 
\begin{eqnarray*}
\left( 2\pi \right) ^{-1}\int d\mathfrak{u}\wedge d\mathfrak{v} &\mathfrak{%
\Longrightarrow }&\left( 2\pi \right) ^{-1}\int -id\mathfrak{u}\wedge d%
\mathfrak{v} \\
&=&\left( 2\pi \right) ^{-1}\int \frac{1}{2}d\left( \mathfrak{\bar{q}-i\bar{p%
}}\right) \wedge d\left( \mathfrak{\bar{q}+i\bar{p}}\right) \\
&=&\left( 2\pi \right) ^{-1}\int \left( d\mathfrak{\bar{q}\wedge }d\mathfrak{%
\bar{p}}\right) \\
&=&\frac{1}{\pi }\int \left( d\func{Re}\alpha \wedge d\func{Im}\alpha \right)
\end{eqnarray*}%
since $d\func{Re}\alpha =\frac{d\mathfrak{\bar{q}}}{\sqrt{2}}$ and $d\func{Im%
}\alpha =\frac{d\mathfrak{\bar{p}}}{\sqrt{2}}$.

\subsection{Characteristic distribution of lattice Weyl transform}

Therefore, we have the identity for the characteristic function $A_{\lambda
\lambda ^{\prime }}\left( \mathfrak{u},\mathfrak{v}\right) $%
\begin{equation}
A_{\lambda \lambda ^{\prime }}\left( \mathfrak{u},\mathfrak{v}\right)
=Tr\left\{ \hat{A}\exp \left\{ -2i\left[ \mathcal{P}\cdot \mathfrak{v}-%
\mathcal{Q}\cdot \mathfrak{u}\right] \right\} \right\} \Omega _{\lambda
\lambda ^{\prime }}  \label{characteristics}
\end{equation}%
The characteristic function exist for all function of canonical quantum
operators, either Hermitian or non-Hermitian, spinor (fermions) or boson
operators \footnote{%
For creation and annihilation operators in many-body quantum physics, the
proof relies on the use of normal or anti-normal ordering of canonical
operators, which can then be treated like $%
\mathbb{C}
$-numbers in expansion of exponentials. The exponential in Eq. (\ref%
{characteristics}) is sometimes referred to as the generalized Pauli-spin
operator.}.

\subsubsection{Implications on coherent states (CS) formulation}

In what follows, we will drop the discrete indices $\lambda $ and $\lambda
^{\prime }$ to make contact with CS formulation of quantum physics. In
general, we can have different expression for the characteristic function
depending on the use of, what is often referred to in corresponding CS
formalism as the normal and anti-normal expessions,%
\begin{eqnarray*}
\exp \left\{ -2i\left[ \mathcal{P}\cdot \mathfrak{v}-\mathcal{Q}\cdot 
\mathfrak{u}\right] \right\} &=&\exp \left\{ i\mathfrak{u}\cdot \mathfrak{v}%
\right\} \exp \left\{ -2i\mathfrak{v}\cdot \mathcal{P}\right\} \exp \left\{
2i\mathfrak{u}\cdot \mathcal{Q}\right\} \\
&=&\exp \left\{ -i\mathfrak{u}\cdot \mathfrak{v}\right\} \exp \left\{ 2i%
\mathfrak{u}\cdot \mathcal{Q}\right\} \exp \left\{ -2i\mathfrak{v}\cdot 
\mathcal{P}\right\}
\end{eqnarray*}%
so that 
\begin{eqnarray}
\exp \left\{ -2i\mathfrak{v}\cdot \mathcal{P}\right\} \exp \left\{ 2i%
\mathfrak{u}\cdot \mathcal{Q}\right\} &=&\exp \left\{ -i\mathfrak{u}\cdot 
\mathfrak{v}\right\} \exp \left\{ -2i\left[ \mathcal{P}\cdot \mathfrak{v}-%
\mathcal{Q}\cdot \mathfrak{u}\right] \right\}  \label{smooth1_qft} \\
\exp \left\{ 2i\mathfrak{u}\cdot \mathcal{Q}\right\} \exp \left\{ -2i%
\mathfrak{v}\cdot \mathcal{P}\right\} &=&\exp \left\{ i\mathfrak{u}\cdot 
\mathfrak{v}\right\} \exp \left\{ -2i\left[ \mathcal{P}\cdot \mathfrak{v}-%
\mathcal{Q}\cdot \mathfrak{u}\right] \right\}  \label{smooth2_qft}
\end{eqnarray}%
yielding the following differrent expressions for $A_{\lambda \lambda
^{\prime }}\left( u,v\right) $, namely, 
\begin{equation}
A_{\lambda \lambda ^{\prime }}^{w}\left( u,v\right) =Tr\left( \hat{A}\exp
\left( -2i\right) \left( \mathcal{P}.\mathfrak{v-}\mathcal{Q}.\mathfrak{u}\
\right) \right)  \label{symmchardef}
\end{equation}%
which is the characteristic function for the Wigner distribution function.
We also have the so-called normal characteristic distribution function, 
\begin{eqnarray}
A_{\lambda \lambda ^{\prime }}^{n}\left( u,v\right) &=&Tr\left[ \ \hat{A}%
\exp \left\{ -2i\mathfrak{v}\cdot \mathcal{P}\right\} \exp \left\{ 2i%
\mathfrak{u}\cdot \mathcal{Q}\right\} \right] ,  \notag \\
&=&\exp \left\{ -i\mathfrak{u}\cdot \mathfrak{v}\right\} Tr\left( \hat{A}%
\exp \left( -2i\right) \left( \mathcal{P}.\mathfrak{v-}\mathcal{Q}.\mathfrak{%
u}\ \right) \right)  \label{normalchardef_qft}
\end{eqnarray}%
and the anti-normal characteristic distribution function given by%
\begin{eqnarray}
A_{\lambda \lambda ^{\prime }}^{a}\left( u,v\right) &=&Tr\ \left\{ \hat{A}\
\exp \left\{ 2i\mathfrak{u}\cdot \mathcal{Q}\right\} \exp \left\{ -2i%
\mathfrak{v}\cdot \mathcal{P}\right\} \right\} ,  \notag \\
&=&\exp \left\{ i\mathfrak{u}\cdot \mathfrak{v}\right\} Tr\left( \hat{A}\exp
\left( -2i\right) \left( \mathcal{P}.\mathfrak{v-}\mathcal{Q}.\mathfrak{u}%
\right) \right)  \label{antinormchardef}
\end{eqnarray}%
Although, Eqs. (\ref{normalchardef}) and (\ref{antinormalchardef}) only
amounts to difference in the phase factors in the canonical
position-momentum $q$-$p$ ordinary mixed space representation, similar
quantities in non-Hermitian dual spaces gives a real exponents giving very
different distributions often referred \ to as smooth-out distributions.
Examining Eqs. (\ref{smooth1}) and (\ref{smooth2}), and the fact that in
non-Hermitian mixed representation,%
\begin{eqnarray}
\exp \left\{ -i\mathfrak{u}\cdot \mathfrak{v}\right\} &\equiv &\bar{\alpha}%
^{\ast }\bar{\alpha}=\frac{1}{2}\left( \bar{q}-i\bar{p}\right) \left( \bar{q}%
+i\bar{p}\right)  \notag \\
&=&\frac{1}{2}\left( \func{Re}\bar{\alpha}^{2}+\func{Im}\bar{\alpha}%
^{2}\right) \text{,}  \label{P_func} \\
\exp \left\{ i\mathfrak{u}\cdot \mathfrak{v}\right\} &\equiv &-\bar{\alpha}%
^{\ast }\bar{\alpha}=-\frac{1}{2}\left( \bar{q}-i\bar{p}\right) \left( \bar{q%
}+i\bar{p}\right)  \notag \\
&=&-\frac{1}{2}\left( \func{Re}\bar{\alpha}^{2}+\func{Im}\bar{\alpha}%
^{2}\right) \text{,}  \label{Q_func}
\end{eqnarray}%
in original notation is a real quantity that resembles a Guassian function,
Eqs. (\ref{smooth2_qft}) clearly represent some smoothing of the Wigner
distribution characteristic function and hence the Wigner distribution
itself. Indeed, in Eqs. (\ref{normalchardef}) and (\ref{antinormalchardef})
no informations are lost.

Indeed, more \textit{general} phase-space distribution functions, $f^{\left(
g\right) }\left( \mathfrak{p,q},t\right) $, can be obtained from the
expression%
\begin{eqnarray}
f^{\left( g\right) }\left( \mathfrak{p,q},t\right) &=&\left( \frac{1}{N}%
\right) ^{\frac{1}{2}}\sum\limits_{u,v}A_{\lambda \lambda ^{\prime }}\left( 
\mathfrak{u},\mathfrak{v}\right) \exp \left\{ -2i\left[ \mathfrak{p}\cdot 
\mathfrak{v}-\mathfrak{q}\cdot \mathfrak{u}\right] \right\} g\left( 
\mathfrak{u,v}\right)  \notag \\
&=&\frac{1}{2\pi }\int dudve^{-i\left( \left[ \mathfrak{p}\cdot \mathfrak{v}-%
\mathfrak{q}\cdot \mathfrak{u}\right] \right) }\left[ C^{\left( w\right)
}\left( \mathfrak{u,v},t\right) \ g\left( \mathfrak{u,v}\right) \right] ,
\end{eqnarray}%
where $g\left( \mathfrak{u,v}\right) $is some choosen smoothing function.

\subsubsection{P- and Q-function or Husimi distribution: smoothing}

Generally all distribution function will become meaningful under the
integral sign, thus these have the properties of generalized distribution
functions. The distribution function $f^{a}\left( \mathfrak{p,q},t\right) $
that one obtain from Eq. (\ref{P_func}) is known as the $P$-function and
that obtain from Eq. (\ref{Q_func}) is also known as the $Q$-function or the
Husimi distributionin quantum optics. A detailed discussion of these two
distribution is given by the author's book on the topic of coherent state
formulation, and will not be repeated here.

The identical algebra of $\exp \left\{ -i\left( q^{\prime }\cdot P-p^{\prime
}\cdot Q\right) \right\} $ as well as its relevance to the physics of
two-state systems, spin systems, quantum computing, entanglements \cite%
{mechanical} and teleportation, are discussed in one of the author's book 
\cite{buot8}

\subsection{Path integral for bosons}

The path integral for bosons straightforwardly follows from the above $%
\mathfrak{q}$-$\mathfrak{p}$ mixed space representation. This is given by
the author \cite{direct} and will not be repeated here. We will just give
the result as,%
\begin{eqnarray*}
&&\left\langle \mathfrak{q}\right\vert U\left( t,t_{o}\right) \left\vert 
\mathfrak{q}_{o}\right\rangle \\
&=&\lim_{n\Longrightarrow \infty }\int .....\int
\prod\limits_{i=1}^{n}d^{N\lambda }\mathfrak{q}_{i}\prod\limits_{i=1}^{n+1}%
\frac{d^{N\lambda }\mathfrak{p}_{i}}{\hbar ^{N\lambda }}\exp \left\{ \frac{i%
}{\hbar }\sum\limits_{j=1}^{n}\epsilon \left[ \mathfrak{p}_{j}\cdot \frac{%
\mathfrak{q}_{j}-\mathfrak{q}_{j-1}}{\epsilon }-H\left( \mathfrak{p}_{j},%
\frac{\mathfrak{q}_{j}+\mathfrak{q}_{j-1}}{2}\right) \right] \right\}
\end{eqnarray*}%
where $\epsilon =\left( t_{j}-t_{j-1}\right) $ and $H\left( \mathfrak{p}_{j},%
\frac{\mathfrak{q}_{j}+\mathfrak{q}_{j-1}}{2}\right) $ is the Weyl transform
of the Hamiltonian.

\section{Mixed Space Representation in QFT: Fermions}

The canonical field operators satisfy the following anticommutation
relations,%
\begin{eqnarray*}
\left[ \psi _{\mu }^{\dagger },\psi _{\nu }^{\dagger }\right] _{+} &\equiv
&\left\{ \psi _{\mu }^{\dagger },\psi _{\nu }^{\dagger }\right\} =0 \\
\left[ \psi _{\mu },\psi _{\nu }\right] _{+} &\equiv &\left\{ \psi _{\mu
},\psi _{\nu }\right\} =0 \\
\left[ \hat{\psi},\hat{\psi}^{\dagger }\right] _{+} &\equiv &\left\{ \hat{%
\psi},\hat{\psi}^{\dagger }\right\} =1
\end{eqnarray*}%
We have, as before write, $\hat{\psi}^{\dagger }=-i\mathcal{\hat{P}}$ as
before, 
\begin{eqnarray}
\left[ \hat{\psi},\hat{\psi}^{\dagger }\right] &\Longrightarrow &\left[ 
\mathcal{\hat{Q}},-i\mathcal{\hat{P}}\right] _{+}=1  \notag \\
\left[ \mathcal{\hat{Q}},\mathcal{\hat{P}}\right] _{+} &=&i  \label{anti_com}
\end{eqnarray}%
To avoid confusion, we set the respective eigenvalues of $\mathcal{\hat{Q}}$
and $\mathcal{\hat{P}}$ as $\mathfrak{q}$ and $\mathfrak{p}$, respectively.
As before, we have the left and right eigenvector defined by%
\begin{eqnarray}
\mathcal{\hat{P}}\left\vert \mathfrak{p}\right\rangle &=&\mathfrak{p}%
\left\vert \mathfrak{p}\right\rangle =\left\vert \mathfrak{p}\right\rangle 
\mathfrak{p,}  \notag \\
\left\langle \mathfrak{p}\right\vert \mathcal{\hat{P}} &\mathcal{=}%
&\left\langle \mathfrak{p}\right\vert \mathfrak{p=p}\left\langle \mathfrak{p}%
\right\vert \text{.}  \label{commute_vect_grass1}
\end{eqnarray}%
Likewise we have%
\begin{eqnarray}
\mathcal{\hat{Q}}\left\vert \mathfrak{q}\right\rangle &=&\mathfrak{q}%
\left\vert \mathfrak{q}\right\rangle =\left\vert \mathfrak{q}\right\rangle 
\mathfrak{q}\text{,}  \notag \\
\left\langle \mathfrak{q}\right\vert \mathcal{\hat{Q}} &\mathcal{=}%
&\left\langle \mathfrak{q}\right\vert \mathfrak{q=q}\left\langle \mathfrak{q}%
\right\vert \text{.}  \label{commute_vect_grass2}
\end{eqnarray}%
Thus, although the eigenvalues and fermion operators are elements of
Grassmann algebra, the eigenvectors are $%
\mathbb{C}
$-vectors and commutes with Grassmann variables.

For fermions, we can with Eqs. (\ref{qft_gen}) - (\ref{qp_qft}) which are
common to both bosons and fermions. However, for fermions, the eigenvalues
are elements of the Grassmann algebra. The following eigenvector generation
holds for bosons and fermions, namely,%
\begin{eqnarray*}
\hat{\psi}^{\dagger }\left\vert \mathfrak{q}\right\rangle &=&-i\mathcal{\hat{%
P}}\left\vert \mathfrak{q}\right\rangle =\overrightarrow{\frac{\partial }{%
\partial \mathfrak{q}}}\left\vert \mathfrak{q}\right\rangle \Longrightarrow 
\mathcal{\hat{P}}\left\vert \mathfrak{q}\right\rangle =\overrightarrow{i%
\frac{\partial }{\partial \mathfrak{q}}}\left\vert \mathfrak{q}\right\rangle
\\
\left\vert \mathfrak{q}\right\rangle \overleftarrow{\hat{\psi}^{\dagger }}
&=&\left\langle \mathfrak{q}\right\vert \left( i\mathcal{\hat{P}}\right)
=\left\langle \mathfrak{q}\right\vert \overleftarrow{\left( -\frac{\partial 
}{\partial \mathfrak{q}}\right) }\Longrightarrow \left\langle \mathfrak{q}%
\right\vert \mathcal{\hat{P}=}\left\langle \mathfrak{q}\right\vert 
\overleftarrow{\left( i\frac{\partial }{\partial \mathfrak{q}}\right) } \\
\hat{\psi}\left\vert \mathfrak{p}\right\rangle &=&i\mathcal{\hat{Q}}%
\left\vert \mathfrak{p}\right\rangle =\frac{\partial }{\partial \mathfrak{p}}%
\left\vert \mathfrak{p}\right\rangle \Longrightarrow \mathcal{\hat{Q}}%
\left\vert \mathfrak{p}\right\rangle =\overrightarrow{-i\frac{\partial }{%
\partial \mathfrak{p}}}\left\vert \mathfrak{p}\right\rangle \\
\left\langle \mathfrak{p}\right\vert \overleftarrow{\left( -i\mathcal{\hat{Q}%
}\right) } &=&\left\langle \mathfrak{p}\right\vert \overleftarrow{\left( -%
\frac{\partial }{\partial \mathfrak{p}}\right) }\Longrightarrow \left\langle 
\mathfrak{p}\right\vert \mathcal{\hat{Q}}=\left\langle \mathfrak{p}%
\right\vert \overleftarrow{\left( -i\frac{\partial }{\partial \mathfrak{p}}%
\right) }
\end{eqnarray*}%
which is compatible with Eqs. (\ref{commute_vect_grass1}) -(\ref%
{commute_vect_grass2}). We also have%
\begin{eqnarray*}
\left\vert \mathfrak{q}\right\rangle &=&\exp \left( -i\mathfrak{q}.\mathcal{%
\hat{P}}\right) \left\vert 0\right\rangle \Longrightarrow \mathcal{\hat{P}}%
\left\vert \mathfrak{q}\right\rangle =\overrightarrow{i\frac{\partial }{%
\partial \mathfrak{q}}}\left\vert \mathfrak{q}\right\rangle \\
\left\langle \mathfrak{q}\right\vert &=&\left\langle 0\right\vert \exp
\left( i\mathfrak{q}.\mathcal{\hat{P}}\right) \Longrightarrow \left\langle 
\mathfrak{q}\right\vert \mathcal{\hat{P}=}\left\langle \mathfrak{q}%
\right\vert \overleftarrow{\left( i\frac{\partial }{\partial \mathfrak{q}}%
\right) } \\
\left\vert \mathfrak{p}\right\rangle &=&\exp \left( i\mathfrak{p}.\mathcal{%
\hat{Q}}\right) \left\vert 0\right\rangle \Longrightarrow \mathcal{\hat{Q}}%
\left\vert \mathfrak{p}\right\rangle =\overrightarrow{-i\frac{\partial }{%
\partial \mathfrak{p}}}\left\vert \mathfrak{p}\right\rangle \\
\left\langle \mathfrak{p}\right\vert &=&\left\langle 0\right\vert \exp
\left( -i\mathfrak{p}.\mathcal{\hat{Q}}\right) \Longrightarrow \left\langle 
\mathfrak{p}\right\vert \left( \mathcal{\hat{Q}}\right) =\left\langle 
\mathfrak{p}\right\vert \overleftarrow{\left( -i\frac{\partial }{\partial 
\mathfrak{p}}\right) }
\end{eqnarray*}%
which is again compatible with Eqs. (\ref{commute_vect_grass1}) -(\ref%
{commute_vect_grass2}), where the arrows denote the left and right
derivatives and the dot product is defined by 
\begin{equation*}
\psi ^{\dagger }\cdot \mathfrak{q}=\sum\limits_{\mu }\mathfrak{q}_{\mu }\psi
_{\mu }^{\dagger }\equiv \sum\limits_{\mu }\mathfrak{q}_{\mu }\left( -i%
\mathcal{\hat{P}}_{\mu }\right)
\end{equation*}%
Note that we retained the notation, $\mathfrak{p}$ and $\mathfrak{q}$ for
eigenvalues, with the understanding that for fermions, $\mathfrak{p}$ and $%
\mathfrak{q}$ are elements of Grassmann algebra, not ordinary $%
\mathbb{C}
$-numbers.

\subsection{Transformation functions}

Using the identity: $e^{A}e^{B}=e^{B}e^{A}e^{\left[ A,B\right] }$ for the
case where the commutator $\left[ A,B\right] $ is a $%
\mathbb{C}
$-number, we have the following expression for the transformation functions%
\begin{eqnarray*}
\left\langle \mathfrak{p}\right\vert \left\vert \mathfrak{q}\right\rangle
&=&\left\langle \mathfrak{p=}0\right\vert \exp \left( -i\mathfrak{p}.%
\mathcal{\hat{Q}}\right) \exp \left( -i\mathfrak{q}.\mathcal{\hat{P}}\right)
\left\vert \mathfrak{q=}0\right\rangle \\
&=&\left\langle 0\right\vert \left\vert 0\right\rangle \exp \left[ i%
\mathfrak{p}.\mathfrak{q}\right] =\exp \left[ i\mathfrak{p}.\mathfrak{q}%
\right]
\end{eqnarray*}%
since, we have%
\begin{eqnarray*}
\left[ -i\mathfrak{p}.\mathcal{\hat{Q}},-i\mathfrak{q}.\mathcal{\hat{P}}%
\right] _{-} &=&-\left[ \left( \mathfrak{p}.\mathcal{\hat{Q}}\right) \left( 
\mathfrak{q}.\mathcal{\hat{P}}\right) -\left( \mathfrak{q}.\mathcal{\hat{P}}%
\right) \left( \mathfrak{p}.\mathcal{\hat{Q}}\right) \right] \\
&=&\left[ \sum\limits_{\mu ,v}\mathfrak{p}_{\mu }\mathfrak{q}_{\nu }\left\{ 
\mathcal{\hat{Q}}_{\mu }\mathcal{\hat{P}}_{\nu }+\mathcal{\hat{P}}_{\nu }%
\mathcal{\hat{Q}}_{\mu }\right\} \right] =i\sum\limits_{\mu ,v}\mathfrak{p}%
_{\mu }\mathfrak{q}_{\nu }\delta _{\mu \nu } \\
&=&i\sum\limits_{\mu }\mathfrak{p}_{\mu }\mathfrak{q}_{\mu }=i\mathfrak{%
p\cdot q}
\end{eqnarray*}

Equation (\ref{anti_com}), defines the generation of state $\left\vert 
\mathfrak{q}\right\rangle $, 
\begin{equation}
\left\vert \mathfrak{q}\right\rangle =C_{o}\exp \left( -i\mathfrak{q}%
\mathcal{\hat{P}}\right) \left\vert 0\right\rangle
\end{equation}%
Inserting the term $\exp \left\{ ip\cdot \mathcal{\hat{Q}}\right\} $ right
in front of $\left\vert 0\right\rangle $\footnote{%
There is arbitrarines in incorporating $\exp \left\{ \phi \hat{a}\right\} $,
either positive or negative exponent, operating on vacuum state. To be
symmetric we should use positive exponent, $\exp \left\{ \phi \hat{a}%
\right\} $. For convenience, we want the generation os state unitary, so is
advisable to use the negative exponent. We will follow this convention is
what follows.}, which has the effect of multiplying by unity, we obtain a
fully symmetric form as 
\begin{equation}
\left\vert \mathfrak{q}\right\rangle =C_{o}\exp \left( -i\mathfrak{q\cdot }%
\mathcal{\hat{P}}\right) \exp \left\{ ip\cdot \mathcal{\hat{Q}}\right\}
\left\vert 0\right\rangle
\end{equation}%
We have,%
\begin{eqnarray}
\hat{\psi}^{\dagger }\left\vert \mathfrak{q}\right\rangle &=&\frac{\partial 
}{\partial \mathfrak{q}}\mathcal{\ }\left\vert \mathfrak{q}\right\rangle =-i%
\mathcal{\hat{P}\ }\left\vert \mathfrak{q}\right\rangle \\
\mathcal{\hat{P}}\left\vert \mathfrak{p}\right\rangle &=&\ \mathfrak{p}%
\mathcal{\ }\left\vert \mathfrak{p}\right\rangle
\end{eqnarray}%
To calculate $\left\langle \mathfrak{p}\right\vert \left\vert \mathfrak{q}%
\right\rangle $, on the same manner as for boson but accounting for
Grassmann variables, we proceed as follows.%
\begin{eqnarray}
\left\langle \mathfrak{p}\right\vert \hat{\psi}^{\dagger }\left\vert 
\mathfrak{q}\right\rangle &=&\left\langle \mathfrak{p}\right\vert \frac{%
\partial }{\partial \mathfrak{q}}\mathcal{\ }\left\vert \mathfrak{q}%
\right\rangle \\
-i\mathfrak{p}\left\langle \mathfrak{p}\right\vert \left\vert \mathfrak{q}%
\right\rangle &=&\frac{\partial }{\partial \mathfrak{q}}\left\langle 
\mathfrak{p}\right\vert \left\vert \mathfrak{q}\right\rangle \\
\frac{\frac{\partial }{\partial \mathfrak{q}}\left\langle \mathfrak{p}%
\right\vert \left\vert \mathfrak{q}\right\rangle }{\left\langle \mathfrak{p}%
\right\vert \left\vert \mathfrak{q}\right\rangle } &=&-i\mathfrak{p} \\
\frac{\partial }{\partial \mathfrak{q}}\ln \left\langle \mathfrak{p}%
\right\vert \left\vert \mathfrak{q}\right\rangle &=&-i\mathfrak{p} \\
\left\langle \mathfrak{p}\right\vert \left\vert \mathfrak{q}\right\rangle
&=&\exp \left( -i\mathfrak{q}\cdot \mathfrak{p}\right) \\
&=&\exp \left( i\mathfrak{p}\cdot \mathfrak{q}\right)
\end{eqnarray}%
The last line occurs by virtue of Grassmann algebra for the eigenvalues.
Similarly, we have 
\begin{eqnarray}
\hat{\psi}\left\vert \mathfrak{p}\right\rangle &=&\frac{\partial }{\partial 
\mathfrak{p}}\mathfrak{\ }\left\vert \mathfrak{p}\right\rangle =i\mathfrak{%
\mathcal{\hat{Q}\ }}\left\vert \mathfrak{p}\right\rangle \\
\mathfrak{\mathcal{\hat{Q}}}\ \left\vert \mathfrak{q}\right\rangle &=&%
\mathfrak{q\ }\left\vert \mathfrak{q}\right\rangle
\end{eqnarray}%
and%
\begin{eqnarray}
\left\langle \mathfrak{q}\right\vert i\mathfrak{\mathcal{\hat{Q}}}\left\vert 
\mathfrak{p}\right\rangle &=&i\mathfrak{q}\left\langle \mathfrak{q}%
\right\vert \left\vert \mathfrak{p}\right\rangle  \notag \\
&=&\frac{\partial }{\partial \mathfrak{p}}\left\langle \mathfrak{q}%
\right\vert \left\vert \mathfrak{p}\right\rangle \\
i\mathfrak{q}\left\langle \mathfrak{q}\right\vert \left\vert \mathfrak{p}%
\right\rangle &=&\frac{\partial }{\partial \mathfrak{p}}\mathfrak{\ }%
\left\langle \mathfrak{q}\right\vert \left\vert \mathfrak{p}\right\rangle 
\notag \\
\left\langle \mathfrak{q}\right\vert \left\vert \mathfrak{p}\right\rangle
&=&\exp i\mathfrak{p\cdot q} \\
&=&\exp \left( -i\mathfrak{q\cdot p}\right)
\end{eqnarray}%
Note that for fermions the eigenvalues corresponding to $\mathfrak{q}$ and $%
\mathfrak{p}$ are elements of the Grassmann algebra.

From 
\begin{eqnarray}
\left\langle \mathfrak{p}\right\vert \left\vert \mathfrak{q}\right\rangle
&=&\left\langle \mathfrak{p=}0\right\vert \exp \left( -i\mathfrak{p}.%
\mathcal{\hat{Q}}\right) \exp \left( -i\mathfrak{q}.\mathcal{\hat{P}}\right)
\left\vert \mathfrak{q=}0\right\rangle  \notag \\
&=&\left\langle 0\right\vert \left\vert 0\right\rangle \exp \left[ i%
\mathfrak{p}.\mathfrak{q}\right] =\exp \left[ i\mathfrak{p}.\mathfrak{q}%
\right]  \label{fermion_trans}
\end{eqnarray}%
We can write%
\begin{equation*}
\left\langle \mathfrak{p=}0\right\vert \exp \left( -i\mathfrak{p}.\mathcal{%
\hat{Q}}\right) \exp \left[ -i\mathfrak{p}.\mathfrak{q}\right] \exp \left( -i%
\mathfrak{q}.\mathcal{\hat{P}}\right) \left\vert \mathfrak{q=}0\right\rangle
\equiv 1
\end{equation*}%
We have from Eqs. (\ref{complete1_qft}) - (\ref{complete2_qft}) 
\begin{eqnarray}
\sum\limits_{\mathfrak{q},\mathfrak{p}}\left\vert \mathfrak{q}\right\rangle
\left\langle \mathfrak{q}\right\vert \left\vert \mathfrak{p}\right\rangle
\left\langle \mathfrak{p}\right\vert &=&1  \label{unity1} \\
\sum\limits_{\mathfrak{q},\mathfrak{p}}\left\vert \mathfrak{q}\right\rangle
\exp \left( -i\mathfrak{q\cdot p}\right) \left\langle \mathfrak{p}%
\right\vert &=&1  \label{unity2} \\
\sum\limits_{\mathfrak{q},\mathfrak{p}}\exp \left( i\mathfrak{p\cdot q}%
\right) \ \left\vert \mathfrak{q}\right\rangle \left\langle \mathfrak{p}%
\right\vert &=&1  \label{unity3}
\end{eqnarray}%
and fro Eqs. (\ref{eq_p_qft})-(\ref{eq_q1_qft})%
\begin{eqnarray}
\left\vert \mathfrak{p}\right\rangle &=&\sum\limits_{\mathfrak{q}}\exp
\left( i\mathfrak{p\cdot q}\right) \ \left\vert \mathfrak{q}\right\rangle
\label{fermion_fourier1} \\
\left\langle \mathfrak{p}\right\vert &=&\sum\limits_{\mathfrak{q}%
}\left\langle \mathfrak{q}\right\vert \exp \left( -i\mathfrak{p\cdot q}%
\right)  \label{fermion_fourier2} \\
\left\vert \mathfrak{q}\right\rangle &=&\sum\limits_{\mathfrak{p}}\exp
\left( i\mathfrak{p\cdot q}\right) \ \left\vert \mathfrak{p}\right\rangle
\label{fermion_fourier3} \\
\left\langle \mathfrak{q}\right\vert &=&\sum\limits_{\mathfrak{p}%
}\left\langle \mathfrak{p}\right\vert \exp \left( -i\mathfrak{p\cdot q}%
\right)  \label{fermion_fourier4}
\end{eqnarray}%
Therefore, we have also the completeness relation expressed as%
\begin{eqnarray}
\sum\limits_{\mathfrak{q}}\left\vert \mathfrak{q}\right\rangle \left\langle 
\mathfrak{q}\right\vert &=&1  \notag \\
\sum\limits_{\mathfrak{p}}\left\vert \mathfrak{p}\right\rangle \left\langle 
\mathfrak{p}\right\vert &=&1  \notag \\
\sum\limits_{\mathfrak{q,p}}\exp \left( i\mathfrak{p\cdot q}\right) \
\left\vert \mathfrak{p}\right\rangle \left\langle q\right\vert &=&1
\label{complete_fermion1} \\
\sum\limits_{\mathfrak{q,p}}\exp \left( i\mathfrak{p\cdot q}\right) \
\left\vert \mathfrak{q}\right\rangle \left\langle p\right\vert &=&1
\label{complete_fermion2}
\end{eqnarray}

Equations (\ref{complete_fermion1})-(\ref{complete_fermion2}) are the ones
used by Halperin et al \cite{halperin} in following the coherent state
formulation.

\subsection{"Lattice" Weyl transform for fermions}

First we will formulate the Weyl transformation based on finite set of
eigenvalues discussed in Sec. \ref{dual_H}. On the other hand, for
continuous eigenvalues or fields, one need to use the Berezin \cite{berezin}
Grassman calculus involving Berezin integration of anticommuting variables.

We have for finite set of eigenvalues, expand any operator as%
\begin{eqnarray}
\hat{A} &=&\sum\limits_{\mathfrak{p}^{\prime \prime },\mathfrak{q}^{\prime
}}\left\vert \mathfrak{p}^{\prime \prime }\right\rangle \left\langle 
\mathfrak{p}^{\prime \prime }\right\vert A\left\vert \mathfrak{q}^{\prime
}\right\rangle \left\langle \mathfrak{q}^{\prime }\right\vert  \notag \\
&=&\sum\limits_{\mathfrak{p}^{\prime \prime },\mathfrak{q}^{\prime
}}\left\langle \mathfrak{p}^{\prime \prime }\right\vert A\left\vert 
\mathfrak{q}^{\prime }\right\rangle \ \left\vert \mathfrak{p}^{\prime \prime
}\right\rangle \left\langle \mathfrak{q}^{\prime }\right\vert
\label{resolve_operator}
\end{eqnarray}%
or 
\begin{eqnarray}
\hat{A} &=&\sum\limits_{\mathfrak{p}^{\prime \prime },\mathfrak{q}^{\prime
}}\left\vert \mathfrak{q}^{\prime \prime }\right\rangle \left\langle 
\mathfrak{q}^{\prime \prime }\right\vert A\left\vert \mathfrak{p}^{\prime
}\right\rangle \left\langle \mathfrak{p}^{\prime }\right\vert  \notag \\
&=&\sum\limits_{\mathfrak{p}^{\prime \prime },\mathfrak{q}^{\prime
}}\left\langle \mathfrak{q}^{\prime \prime }\right\vert A\left\vert 
\mathfrak{p}^{\prime }\right\rangle \ \left\vert \mathfrak{q}^{\prime \prime
}\right\rangle \left\langle \mathfrak{p}^{\prime }\right\vert
\end{eqnarray}%
Using, Eqs. (\ref{eq_p})-(\ref{p2q}), we write%
\begin{eqnarray}
\left\langle \mathfrak{p}^{\prime \prime }\right\vert A\left\vert \mathfrak{q%
}^{\prime }\right\rangle &=&\frac{1}{\sqrt{N}}\sum\limits_{\mathfrak{q}%
^{\prime \prime }}e^{-i\mathfrak{p}^{\prime \prime }\cdot \mathfrak{q}%
^{\prime \prime }}\left\langle \mathfrak{q}^{\prime \prime }\right\vert
A\left\vert \mathfrak{q}^{\prime }\right\rangle  \notag \\
\left\vert \mathfrak{p}^{\prime \prime }\right\rangle \left\langle \mathfrak{%
q}^{\prime }\right\vert &=&\frac{1}{\sqrt{N}}\sum\limits_{\mathfrak{p}%
^{\prime }}e^{i\mathfrak{p}^{\prime }\cdot \mathfrak{q}^{\prime }}\left\vert 
\mathfrak{p}^{\prime \prime }\right\rangle \left\langle \mathfrak{p}^{\prime
}\right\vert
\end{eqnarray}%
Introducing the notation in Eq. (\ref{matrix_project}),%
\begin{eqnarray*}
\mathfrak{p}^{\prime } &=&\mathfrak{p}+\mathfrak{u}\text{, \ \ \ \ \ \ \ }%
\mathfrak{q}^{\prime }=\mathfrak{q}+\mathfrak{v}, \\
\mathfrak{p}^{\prime \prime } &=&\mathfrak{p}-\mathfrak{u}\text{, \ \ \ \ \
\ \ }\mathfrak{q}^{\prime \prime }=\mathfrak{q}-\mathfrak{v}\text{.}
\end{eqnarray*}%
Then%
\begin{eqnarray*}
p^{\prime }\cdot q^{\prime } &=&\left( \vec{p}+\vec{u}\right) \cdot \left( 
\vec{q}+\vec{v}\right) =\vec{p}\cdot \vec{q}+\vec{u}\cdot \vec{v}+\vec{p}%
\cdot \vec{v}+\vec{u}\cdot \vec{q}\text{,} \\
-\vec{p}^{\prime \prime }\cdot \text{\ }\vec{q}^{\prime \prime } &=&-\left( 
\vec{p}-\vec{u}\right) \cdot \left( \vec{q}-\vec{v}\right) =-\vec{p}\cdot 
\vec{q}-\vec{u}\cdot \vec{v}+\vec{p}\cdot \vec{v}+\vec{u}\cdot \vec{q}
\end{eqnarray*}%
Then, upon substituting in Eq. (\ref{any_op2}), we end up with%
\begin{eqnarray*}
A &=&\sum\limits_{\mathfrak{p}^{\prime \prime },\mathfrak{q}^{\prime
}}\left\vert \mathfrak{p}^{\prime \prime }\right\rangle \left\langle 
\mathfrak{p}^{\prime \prime }\right\vert A\left\vert \mathfrak{q}^{\prime
}\right\rangle \left\langle \mathfrak{q}^{\prime }\right\vert \\
&=&\frac{1}{N}\sum\limits_{\mathfrak{p},\mathfrak{q},\mathfrak{u},\mathfrak{v%
}}e^{i2\left( \mathfrak{p}\cdot \mathfrak{v}+\mathfrak{u}\cdot \mathfrak{q}%
\right) }\left\langle \mathfrak{q}-\mathfrak{v}\right\vert A\left\vert 
\mathfrak{q}+\mathfrak{v}\right\rangle \ \left\vert \mathfrak{p}-\mathfrak{u}%
\right\rangle \left\langle \mathfrak{p}+\mathfrak{u}\right\vert
\end{eqnarray*}%
Likewise as in previous formulations, we write the last result as an
expansion in terms of "\textit{phase-space" point projector}, $\hat{\Delta}%
\left( p,q\right) $, defined as the fermion "\textit{lattice" Weyl transform}
of a projector, by%
\begin{equation}
\hat{\Delta}\left( p,q\right) =\sum\limits_{u}e^{i2\vec{u}\cdot \vec{q}%
}\left\vert \vec{p}-\vec{u}\right\rangle \left\langle \vec{p}+\vec{u}%
\right\vert
\end{equation}%
and the coefficient of expansion, the fermion "\textit{lattice" Weyl
transform} of matrix element of operator, $A\left( p,q\right) $, defined by%
\begin{equation*}
A\left( p,q\right) =\sum\limits_{v}e^{i2\vec{p}\cdot \vec{v}}\left\langle 
\vec{q}-\vec{v}\right\vert A\left\vert \vec{q}+\vec{v}\right\rangle \text{.}
\end{equation*}%
An equivalent expression can also be obtain for $A\left( \mathfrak{p},%
\mathfrak{q}\right) $and $\hat{\Delta}\left( \mathfrak{p},\mathfrak{q}%
\right) $, namely,%
\begin{eqnarray}
A\left( \mathfrak{p},\mathfrak{q}\right) &=&\sum\limits_{\mathfrak{u}}e^{i2%
\mathfrak{u}\cdot \mathfrak{q}}\left\langle \mathfrak{p}+\mathfrak{u}%
\right\vert \hat{A}\left\vert \mathfrak{p}-\mathfrak{u}\right\rangle \\
\hat{\Delta}\left( \mathfrak{p},\mathfrak{q}\right) &=&\sum\limits_{%
\mathfrak{v}}e^{i2\mathfrak{p}\cdot \mathfrak{v}}\left\vert \mathfrak{q}+%
\mathfrak{v}\right\rangle \left\langle \mathfrak{q}-\mathfrak{v}\right\vert
\label{symmetrize}
\end{eqnarray}

\subsection{Symmetric form of $\hat{\Delta}_{\protect\lambda ^{\prime }%
\protect\lambda }\left( \mathfrak{p},\mathfrak{q}\right) $}

In the expression, Eq. (\ref{symmetrize}) for $\hat{\Delta}_{\lambda
^{\prime }\lambda }\left( \mathfrak{p},\mathfrak{q}\right) $, we can use the
following identities,%
\begin{equation}
\left\vert \mathfrak{q+v},\lambda \right\rangle =\exp \left[ -2i\mathcal{P}%
\cdot \mathfrak{v}\right] \left\vert \mathfrak{q-v},\lambda \right\rangle
\end{equation}%
Then 
\begin{eqnarray}
\left\vert \mathfrak{q-v},\lambda \right\rangle \left\langle \mathfrak{q-v}%
,\lambda ^{\prime }\right\vert &=&\left( N\right) ^{-1}\sum\limits_{%
\mathfrak{u,q}_{0}}\exp \left[ -2i\left( \mathfrak{q-v}-\mathcal{Q}\right)
\cdot \mathfrak{u}\right] \left\vert \mathfrak{q}_{0},\lambda \right\rangle
\left\langle \mathfrak{q}_{0},\lambda ^{\prime }\right\vert \\
&=&\sum\limits_{\mathfrak{q}_{0}}\delta \left( \mathfrak{q-v}-\mathcal{\hat{Q%
}}\right) \left\vert \mathfrak{q}_{0},\lambda \right\rangle \left\langle 
\mathfrak{q}_{0},\lambda ^{\prime }\right\vert =\sum\limits_{\mathfrak{q}%
_{0}}\delta \left( \mathfrak{q-v-q}_{0}\right) \ \left\vert \mathfrak{q}%
_{0},\lambda \right\rangle \left\langle \mathfrak{q}_{0},\lambda ^{\prime
}\right\vert  \notag \\
\Omega _{\lambda \lambda ^{\prime }} &=&\left\vert \mathfrak{q}_{0},\lambda
\right\rangle \left\langle \mathfrak{q}_{0},\lambda ^{\prime }\right\vert 
\notag
\end{eqnarray}%
Substituting the expressions, Eqs (\ref{subs1}) and (\ref{subs2}) in Eq. (%
\ref{alt2}), we obtain a completely symmetric expression of $\hat{\Delta}%
\left( \mathfrak{p},\mathfrak{q}\right) $, 
\begin{eqnarray*}
\hat{\Delta}\left( \mathfrak{p},\mathfrak{q}\right) &=&\sum\limits_{%
\mathfrak{v}}e^{i2\mathfrak{p}\cdot \mathfrak{v}}\left\vert \mathfrak{q}+%
\mathfrak{v}\right\rangle \left\langle \mathfrak{q}-\mathfrak{v}\right\vert
\\
&=&\left( N\right) ^{-1}\sum\limits_{\mathfrak{\bar{v}},\mathfrak{\bar{u}}%
}e^{i2\mathfrak{p}\cdot \mathfrak{v}}\exp \left[ -2i\mathcal{P}\cdot 
\mathfrak{v}\right] \exp \left[ -2i\left( \mathfrak{q}-\mathfrak{v}-\mathcal{%
\hat{Q}}\right) \cdot \mathfrak{u}\right] \sum\limits_{\mathfrak{q}%
_{0}}\left\vert \mathfrak{q}_{0},\lambda \right\rangle \left\langle 
\mathfrak{q}_{0},\lambda ^{\prime }\right\vert
\end{eqnarray*}%
Then we have, even accounting for the anticommutating Grassman variables,%
\begin{eqnarray}
&&\exp \left\{ -i\left( \mathcal{P}-\mathfrak{p}\right) \cdot 2\mathfrak{v}%
\right\} \exp \left\{ 2i\left( \mathcal{\hat{Q}-}\mathfrak{q}\right) \cdot 
\mathfrak{\bar{u}}\right\} \exp 2i\mathfrak{v}\cdot \mathfrak{u}  \notag \\
&=&\exp \left\{ -i\left( \mathcal{P}\cdot \mathfrak{v}-\mathcal{Q}\cdot 
\mathfrak{u}\right) \right\} \exp \left\{ 2i\left( \mathfrak{p}\cdot 
\mathfrak{v}-\mathfrak{q}\cdot \mathfrak{u}\right) \right\}
\end{eqnarray}%
which is the same as Eq. (\ref{delta_sym}). Thus, we have changed the
seemingly asymmetric expression of the first line into a symmetric form of
the last line. And again, we have%
\begin{eqnarray}
\hat{\Delta}\left( \mathfrak{p},\mathfrak{q}\right) &=&\left( N\right)
^{-1}\sum\limits_{\mathfrak{v},\mathfrak{u}}\exp \left\{ -i\left( \mathcal{P}%
\cdot \mathfrak{v}-\mathcal{Q}\cdot \mathfrak{u}\right) \right\} \exp
\left\{ 2i\left( \mathfrak{p}\cdot \mathfrak{v}-\mathfrak{q}\cdot \mathfrak{u%
}\right) \right\} \sum\limits_{\mathfrak{q}_{0}}\left\vert \mathfrak{q}%
_{0},\lambda \right\rangle \left\langle \mathfrak{q}_{0},\lambda ^{\prime
}\right\vert  \notag \\
&=&\left( N\right) ^{-1}\sum\limits_{\mathfrak{v},\mathfrak{u}}\exp \left\{
-i\left( \mathcal{P}\cdot \mathfrak{v}-\mathcal{Q}\cdot \mathfrak{u}\right)
\right\} \exp \left\{ 2i\left( \mathfrak{p}\cdot \mathfrak{v}-\mathfrak{q}%
\cdot \mathfrak{u}\right) \right\} \Omega _{\lambda \lambda ^{\prime }}
\label{fermion_delta1}
\end{eqnarray}%
where, 
\begin{equation*}
\Omega _{\lambda \lambda ^{\prime }}=\sum\limits_{\mathfrak{q}%
_{0}}\left\vert \mathfrak{q}_{0},\lambda \right\rangle \left\langle 
\mathfrak{q}_{0},\lambda ^{\prime }\right\vert =\sum\limits_{\mathfrak{p}%
_{0}}\left\vert \mathfrak{p}_{0},\lambda \right\rangle \left\langle 
\mathfrak{p}_{0},\lambda ^{\prime }\right\vert
\end{equation*}%
Moreover, we also have

\begin{eqnarray}
A\left( \mathfrak{p},\mathfrak{q}\right) &=&Tr\left( \hat{A}\hat{\Delta}%
\right)  \notag \\
&=&\left( N\right) ^{-1}\left( 
\begin{array}{c}
\sum\limits_{\bar{v},\bar{u}}\exp 2i\left[ \mathfrak{p}\cdot \mathfrak{v}-%
\mathfrak{q}\cdot \mathfrak{u}\right] \  \\ 
\times Tr\left\{ \hat{A}\exp \left\{ -2i\left[ \mathcal{P}\cdot \mathfrak{v}-%
\mathcal{Q}\cdot \mathfrak{u}\right] \right\} \Omega _{\lambda \lambda
^{\prime }}\right\}%
\end{array}%
\right)  \label{fermion_latrans1}
\end{eqnarray}

\subsection{Characteristic distribution of "lattice" Weyl transform}

Therefore, the charactetic distribution for $A\left( p,q\right) $is
identically, 
\begin{equation}
A_{\lambda \lambda ^{\prime }}\left( \mathfrak{u},\mathfrak{v}\right)
=Tr\left\{ \hat{A}\exp \left\{ -2i\left[ \mathcal{P}\cdot \mathfrak{v}-%
\mathcal{Q}\cdot \mathfrak{u}\right] \right\} \Omega _{\lambda \lambda
^{\prime }}\right\}  \label{fermion_chardist1}
\end{equation}%
which is formally as before. By using the characteristic distribution
function for $A_{\lambda \lambda ^{\prime }}\left( \mathfrak{p},\mathfrak{q}%
\right) $ 
\begin{equation}
A_{\lambda \lambda ^{\prime }}\left( \mathfrak{u},\mathfrak{v}\right)
=\left( \frac{1}{N}\right) ^{\frac{1}{2}}\sum\limits_{p,q}A_{\lambda \lambda
^{\prime }}\left( \mathfrak{p},\mathfrak{q}\right) \exp 2i\left[ \mathfrak{p}%
\cdot \mathfrak{v}-\mathfrak{q}\cdot \mathfrak{u}\right]
\label{fermion_chardist2}
\end{equation}%
with inverse%
\begin{equation}
A_{\lambda \lambda ^{\prime }}\left( \mathfrak{p},\mathfrak{q}\right)
=\left( \frac{1}{N}\right) ^{\frac{1}{2}}\sum\limits_{u,v}A_{\lambda \lambda
^{\prime }}\left( \mathfrak{u},\mathfrak{v}\right) \exp \left\{ -2i\left[ 
\mathfrak{p}\cdot \mathfrak{v}-\mathfrak{q}\cdot \mathfrak{u}\right] \right\}
\label{fermionLWT_chardist1}
\end{equation}%
Then we can write Eq. (\ref{LWT1}) simply like a Fourier transform (caveat:
Fourier transform to\textit{\ operator space}) of the characteristic
function of the lattice Weyl transform of the operator $\hat{A}$, 
\begin{equation}
\hat{A}=\sum\limits_{\bar{u},\bar{v},\lambda ,\lambda ^{\prime }}A_{\lambda
\lambda ^{\prime }}\left( \mathfrak{u},\mathfrak{v}\right) \exp \left\{ -2i%
\left[ \mathcal{P}\cdot \mathfrak{v}-\mathcal{Q}\cdot \mathfrak{u}\right]
\right\} \Omega _{\lambda ^{\prime }\lambda }
\label{fermionOP_from_chardist1}
\end{equation}%
where the inverse can be written as%
\begin{equation*}
A_{\lambda \lambda ^{\prime }}\left( \mathfrak{u},\mathfrak{v}\right)
=Tr\left\{ \hat{A}\exp \left\{ -2i\left[ \mathcal{P}\cdot \mathfrak{v}-%
\mathcal{Q}\cdot \mathfrak{u}\right] \right\} \right\} \Omega _{\lambda
\lambda ^{\prime }}
\end{equation*}

\subsection{Distribution functions}

\begin{equation}
A_{\lambda \lambda ^{\prime }}^{n}\left( u,v\right) =\exp \left\{ -2i%
\mathfrak{u}\cdot \mathfrak{v}\right\} Tr\left( \hat{A}\exp \left(
-2i\right) \left( \mathcal{P}.\mathfrak{v-}\mathcal{Q}.\mathfrak{u}\ \right)
\right)  \notag
\end{equation}%
and the anti-normal characteristic distribution function given by%
\begin{equation}
A_{\lambda \lambda ^{\prime }}^{a}\left( u,v\right) =\exp \left\{ i\mathfrak{%
u}\cdot \mathfrak{v}\right\} Tr\left( \hat{A}\exp \left( -2i\right) \left( 
\mathcal{P}.\mathfrak{v-}\mathcal{Q}.\mathfrak{u}\right) \right)  \notag
\end{equation}

Formally again, a more \textit{general} phase-space distribution functions, $%
f^{\left( g\right) }\left( \mathfrak{p,q},t\right) $, can be obtained from
the expression%
\begin{equation}
f^{\left( g\right) }\left( \mathfrak{p,q},t\right) =\left( \frac{1}{N}%
\right) ^{\frac{1}{2}}\sum\limits_{u,v}A_{\lambda \lambda ^{\prime }}\left( 
\mathfrak{u},\mathfrak{v}\right) \exp \left\{ -2i\left[ \mathfrak{p}\cdot 
\mathfrak{v}-\mathfrak{q}\cdot \mathfrak{u}\right] \right\} g\left( 
\mathfrak{u,v}\right)  \notag
\end{equation}%
where $g\left( \mathfrak{u,v}\right) $is some choosen smoothing function.
Thus, for finite system, the formalism exactly maps to that of CMP since
Bloch electrons are fermions.

\section{Continuous Fields: Calculations of Integrals for Fermions}

With the continuous annihilation and creation field operators for fermions,
our pairing algorithm in Sec. \ref{pairing} is ambigous, but the results are
assumed to be extendable to continuous fields. The major difference occurs
in the functional integration of continuous Grassman variables. The reason
for this is that Berezin integration \cite{berezin} is actually similar to
(standard) functional differentiation. Formally we can still associate a
sort of unification in a symbolic sense for the Grassman integration.

A sort of justification for the readers is by considering translation
invariance of a definite integral and linearity:

\begin{equation*}
\int\limits_{-\infty }^{\infty }f\left( x\right) dx=\int\limits_{-\infty
}^{\infty }f\left( x+y\right) dx
\end{equation*}%
$f\left( x\right) $is at most a linear function of $x$, a Grassamn variable.
Hence, we write%
\begin{equation*}
f\left( x\right) =a+bx=\left( 
\begin{array}{cc}
a & b%
\end{array}%
\right) \left( 
\begin{array}{c}
1 \\ 
x%
\end{array}%
\right)
\end{equation*}%
or more precisely%
\begin{eqnarray*}
f &=&f_{00}+f_{01}\hat{\psi}^{\dagger }+f_{10}\hat{\psi}+f_{11}\hat{\psi}%
^{\dagger }\hat{\psi} \\
&=&\left( 
\begin{array}{cccc}
f_{00} & f_{01} & f_{10} & f_{11}%
\end{array}%
\right) \left( 
\begin{array}{c}
1 \\ 
\hat{\psi}^{\dagger } \\ 
\hat{\psi} \\ 
\hat{\psi}^{\dagger }\hat{\psi}%
\end{array}%
\right)
\end{eqnarray*}%
Linearity says%
\begin{equation*}
\int\limits_{-\infty }^{\infty }\left( a+bf\left( x\right) \right)
dx=\int\limits_{-\infty }^{\infty }adx+\int\limits_{-\infty }^{\infty
}bf\left( x\right) dx
\end{equation*}%
Thus, we have%
\begin{eqnarray*}
\int\limits_{-\infty }^{\infty }f\left( x+y\right) dx
&=&\int\limits_{-\infty }^{\infty }adx+\int\limits_{-\infty }^{\infty
}b\left( x+y\right) dx \\
&=&\int\limits_{-\infty }^{\infty }\left( a+bx\right)
dx+\int\limits_{-\infty }^{\infty }bydx \\
&=&\int\limits_{-\infty }^{\infty }f\left( x\right)
dx+by\int\limits_{-\infty }^{\infty }dx
\end{eqnarray*}%
But by translational invariance%
\begin{equation*}
\int\limits_{-\infty }^{\infty }f\left( x+y\right) dx=\int\limits_{-\infty
}^{\infty }f\left( x\right) dx
\end{equation*}%
Therefore, since $by$is arbitrary, one is forced to equate

\begin{equation*}
\int\limits_{-\infty }^{\infty }dx=0
\end{equation*}%
which does looks like a differentiation of a constant, here equals to $1$.
Thus, integration of Grassman numbers obeys%
\begin{eqnarray*}
\int dx &=&0 \\
\int xdx &=&1
\end{eqnarray*}

The previous relations involving integrals must now be interpreted as
standard differentiation with respect to the integration variable. We
observe that like before we have the eigenvalue equations for the left and
right eigenvector defined by%
\begin{eqnarray}
\mathcal{\hat{P}}\left\vert \mathfrak{p}\right\rangle &=&\mathfrak{p}%
\left\vert \mathfrak{p}\right\rangle =\left\vert \mathfrak{p}\right\rangle 
\mathfrak{p,}  \notag \\
\left\langle \mathfrak{p}\right\vert \mathcal{\hat{P}} &\mathcal{=}%
&\left\langle \mathfrak{p}\right\vert \mathfrak{p=p}\left\langle \mathfrak{p}%
\right\vert \text{.}
\end{eqnarray}%
Likewise we have%
\begin{eqnarray}
\mathcal{\hat{Q}}\left\vert \mathfrak{q}\right\rangle &=&\mathfrak{q}%
\left\vert \mathfrak{q}\right\rangle =\left\vert \mathfrak{q}\right\rangle 
\mathfrak{q}\text{,}  \notag \\
\left\langle \mathfrak{q}\right\vert \mathcal{\hat{Q}} &\mathcal{=}%
&\left\langle \mathfrak{q}\right\vert \mathfrak{q=q}\left\langle \mathfrak{q}%
\right\vert \text{.}
\end{eqnarray}%
where the eigenvalues are now continuous Grassman variables.

The completeness relation is given formally as before by 
\begin{eqnarray}
1 &=&\sum\limits_{\mathfrak{q}}\left\vert \mathfrak{q}\right\rangle
\left\langle \mathfrak{q}\right\vert \Longrightarrow \int d\left[ \mathfrak{q%
}\right] \ \left\vert \mathfrak{q}\right\rangle \left\langle \mathfrak{q}%
\right\vert  \notag \\
&=&\prod \left( \frac{\partial }{\partial \mathfrak{q}^{\prime }}+\frac{%
\partial }{\partial \mathfrak{q}^{^{\prime \prime }}}\right) \left\vert 
\mathfrak{q}^{\prime }\right\rangle \left\langle \mathfrak{q}^{\prime \prime
}\right\vert  \notag \\
&=&\int \mathfrak{Dq\ }\left\vert \mathfrak{q}\right\rangle \left\langle 
\mathfrak{q}\right\vert  \label{resunity1}
\end{eqnarray}%
where the multicomponent integral $\int \mathfrak{D}\mathfrak{q}$is only a
symbolic meaning for continuous fields, so as to formally resembles that of
discrete pairing case before. Likewise%
\begin{equation}
1=\sum\limits_{\mathfrak{p}}\left\vert \mathfrak{p}\right\rangle
\left\langle \mathfrak{p}\right\vert \Longrightarrow \int \mathfrak{Dp}\
\left\vert \mathfrak{p}\right\rangle \left\langle \mathfrak{p}\right\vert
\label{resunity2}
\end{equation}

\subsection{The transformation function and resolution of identity}

The transformation function is the same as Eq. (\ref{fermion_trans}). The
fermion fourier transforms are similar as Eqs. (\ref{fermion_fourier1}) - (%
\ref{fermion_fourier4}) but for continuous case, the summation is now
replaced by functional integral obeying the Berezin fermion field integrals.
Similarly the completeness relations, often referred to as resolutions of
unity, are as given by Eqs. (\ref{unity1}) - (\ref{unity3}) with summation
replaced by Berezin fermion integrals, Eqs. (\ref{resunity1}) and (\ref%
{resunity2}).

\subsection{Continuous Weyl transform for fermions}

The discrete expression for the mixed representation of any operator given
before, Eq. (\ref{resolve_operator}), now reads in the continuous case as

\begin{equation}
\hat{A}=\sum\limits_{\mathfrak{p}^{\prime \prime },\mathfrak{q}^{\prime
}}\left\vert \mathfrak{p}^{\prime \prime }\right\rangle \left\langle 
\mathfrak{p}^{\prime \prime }\right\vert A\left\vert \mathfrak{q}^{\prime
}\right\rangle \left\langle \mathfrak{q}^{\prime }\right\vert
\Longrightarrow \iint \mathfrak{Dp}^{\prime \prime }\mathfrak{Dq}^{\prime
}\left\langle \mathfrak{p}^{\prime \prime }\right\vert A\left\vert \mathfrak{%
q}^{\prime }\right\rangle \left\vert \mathfrak{p}^{\prime \prime
}\right\rangle \left\langle \mathfrak{q}^{\prime }\right\vert
\label{operator}
\end{equation}%
\begin{equation*}
\hat{A}=\iint \mathfrak{DpDq\ }A\left( \mathfrak{p,q}\right) \ \hat{\Delta}%
\left( \mathfrak{p,q}\right)
\end{equation*}%
where%
\begin{equation*}
A\left( \mathfrak{p,q}\right) =\int \mathfrak{Dv\ }e^{-2\mathfrak{p.v}%
}\left\langle \mathfrak{q+v}\right\vert \hat{A}\left\vert \mathfrak{q-v}%
\right\rangle
\end{equation*}%
\begin{equation*}
\hat{\Delta}\left( \mathfrak{p,q}\right) =\int \mathfrak{Du\ }e^{2\mathfrak{%
q.u}}\left\vert \mathfrak{p+u}\right\rangle \left\langle \mathfrak{p-u}%
\right\vert
\end{equation*}%
We also have the equivalent expressions,%
\begin{eqnarray}
A\left( \mathfrak{p},\mathfrak{q}\right) &=&\int \mathfrak{Du\ }e^{-i2%
\mathfrak{q}\cdot \mathfrak{u}}\left\langle \mathfrak{p}+\mathfrak{u}%
\right\vert \hat{A}\left\vert \mathfrak{p}-\mathfrak{u}\right\rangle \\
\hat{\Delta}\left( \mathfrak{p},\mathfrak{q}\right) &=&\int \mathfrak{Dv\ }%
e^{i2\mathfrak{p}\cdot \mathfrak{v}}\left\vert \mathfrak{q}+\mathfrak{v}%
\right\rangle \left\langle \mathfrak{q}-\mathfrak{v}\right\vert
\end{eqnarray}%
in the sense of Berezin fermionic integrals.

Equation (\ref{fermion_delta1}) can now be written in terms of Berezin
fermion integral as%
\begin{equation}
\hat{\Delta}\left( \mathfrak{p},\mathfrak{q}\right) =\iint \mathfrak{DvDu}%
\exp \left\{ -i\left( \mathcal{P}\cdot \mathfrak{v}-\mathcal{Q}\cdot 
\mathfrak{u}\right) \right\} \exp \left\{ 2i\left( \mathfrak{p}\cdot 
\mathfrak{v}-\mathfrak{q}\cdot \mathfrak{u}\right) \right\} \Omega _{\lambda
\lambda ^{\prime }}  \notag
\end{equation}%
and Eq. (\ref{fermion_latrans1}) is now,%
\begin{eqnarray}
A\left( \mathfrak{p},\mathfrak{q}\right) &=&Tr\left( \hat{A}\hat{\Delta}%
\right)  \notag \\
&=&\left( 
\begin{array}{c}
\iint \mathfrak{DvDu\ }\exp 2i\left[ \mathfrak{p}\cdot \mathfrak{v}-%
\mathfrak{q}\cdot \mathfrak{u}\right] \  \\ 
\times Tr\left\{ \hat{A}\exp \left\{ -2i\left[ \mathcal{P}\cdot \mathfrak{v}-%
\mathcal{Q}\cdot \mathfrak{u}\right] \right\} \Omega _{\lambda \lambda
^{\prime }}\right\}%
\end{array}%
\right)
\end{eqnarray}%
We also have for Eq. (\ref{fermionLWT_chardist1}) the expression,%
\begin{equation}
A_{\lambda \lambda ^{\prime }}\left( \mathfrak{u},\mathfrak{v}\right) =\iint 
\mathfrak{D}\mathcal{\mathfrak{p}}\mathfrak{Dq\ }A_{\lambda \lambda ^{\prime
}}\left( \mathfrak{p},\mathfrak{q}\right) \exp 2i\left[ \mathfrak{p}\cdot 
\mathfrak{v}-\mathfrak{q}\cdot \mathfrak{u}\right]
\end{equation}%
with inverse%
\begin{equation}
A_{\lambda \lambda ^{\prime }}\left( \mathfrak{p},\mathfrak{q}\right) =\iint 
\mathfrak{DvDu\ }A_{\lambda \lambda ^{\prime }}\left( \mathfrak{u},\mathfrak{%
v}\right) \exp \left\{ -2i\left[ \mathfrak{p}\cdot \mathfrak{v}-\mathfrak{q}%
\cdot \mathfrak{u}\right] \right\}
\end{equation}

\subsection{Distribution functions}

\begin{equation}
A_{\lambda \lambda ^{\prime }}^{n}\left( u,v\right) =\exp \left\{ -2i%
\mathfrak{u}\cdot \mathfrak{v}\right\} Tr\left( \hat{A}\exp \left(
-2i\right) \left( \mathcal{P}.\mathfrak{v-}\mathcal{Q}.\mathfrak{u}\ \right)
\right)  \notag
\end{equation}%
and the anti-normal characteristic distribution function given by%
\begin{equation}
A_{\lambda \lambda ^{\prime }}^{a}\left( u,v\right) =\exp \left\{ i\mathfrak{%
u}\cdot \mathfrak{v}\right\} Tr\left( \hat{A}\exp \left( -2i\right) \left( 
\mathcal{P}.\mathfrak{v-}\mathcal{Q}.\mathfrak{u}\right) \right)  \notag
\end{equation}%
Formally, again a more \textit{general} phase-space distribution functions, $%
f^{\left( g\right) }\left( \mathfrak{p,q},t\right) $, can be obtained from
the expression%
\begin{equation}
f^{\left( g\right) }\left( \mathfrak{p,q},t\right) =\iint \mathfrak{DuD%
\mathfrak{v}\ }A_{\lambda \lambda ^{\prime }}\left( \mathfrak{u},\mathfrak{v}%
\right) \exp \left\{ -2i\left[ \mathfrak{p}\cdot \mathfrak{v}-\mathfrak{q}%
\cdot \mathfrak{u}\right] \right\} g\left( \mathfrak{u,v}\right)  \notag
\end{equation}%
where $g\left( \mathfrak{u,v}\right) $is some choosen smoothing function.
Finally, we can write Eq. (\ref{fermionOP_from_chardist1}) as%
\begin{equation}
\hat{A}=\sum\limits_{\lambda ,\lambda ^{\prime }}\iint \mathfrak{DuD%
\mathfrak{v\ }}A_{\lambda \lambda ^{\prime }}\left( \mathfrak{u},\mathfrak{v}%
\right) \exp \left\{ -2i\left[ \mathcal{P}\cdot \mathfrak{v}-\mathcal{Q}%
\cdot \mathfrak{u}\right] \right\} \Omega _{\lambda ^{\prime }\lambda }
\end{equation}

\subsection{Fermion path integrals}

We are now equipped to calculate the path integral expression for the
evolution operator, $U\left( t.t_{o}\right) ,$%
\begin{eqnarray}
\hat{U}\left( t.t_{o}\right) &=&\exp \left( \frac{-i}{\hbar }\left[ t-t_{o}%
\right] \mathcal{H}\right)  \notag \\
&=&\prod\limits_{j=1}^{n+1}\hat{U}\left( t_{j}.t_{j-1}\right)
\label{incremental}
\end{eqnarray}%
We now make of the following relations for fermions in terms of Berezin
Grassman integrals. Thus for any fermion operator, we have%
\begin{equation*}
\hat{A}=\iint \mathfrak{DqDp\ }A\left( \mathfrak{p,q}\right) \Delta \left( 
\mathfrak{p,q}\right)
\end{equation*}%
where%
\begin{eqnarray}
A\left( \mathfrak{p},\mathfrak{q}\right) &=&\int \mathfrak{Du\ }e^{-i2%
\mathfrak{q}\cdot \mathfrak{u}}\left\langle \mathfrak{p}+\mathfrak{u}%
\right\vert \hat{A}\left\vert \mathfrak{p}-\mathfrak{u}\right\rangle \\
\hat{\Delta}\left( \mathfrak{p},\mathfrak{q}\right) &=&\int \mathfrak{Dv\ }%
e^{i2\mathfrak{p}\cdot \mathfrak{v}}\left\vert \mathfrak{q}+\mathfrak{v}%
\right\rangle \left\langle \mathfrak{q}-\mathfrak{v}\right\vert
\end{eqnarray}%
From the above relations we can transform Eq. (\ref{incremental}) into a
Grassman path integral. We have from Eq. (\ref{incremental})%
\begin{equation*}
\hat{U}\left( t_{j}.t_{j-1}\right) =\iint \mathfrak{DqDp\ }%
U_{t_{j},t_{j-1}}\left( \mathfrak{p,q}\right) \Delta \left( \mathfrak{p,q}%
\right)
\end{equation*}%
where%
\begin{eqnarray*}
U_{t_{j},t_{j-1}}\left( \mathfrak{p,q}\right) &=&\int \mathfrak{Du\ e}^{-2i%
\mathfrak{u\cdot q}}\left\langle \mathfrak{p+u}\right\vert \hat{U}\left(
t_{j}.t_{j-1}\right) \left\vert \mathfrak{p-u}\right\rangle \\
&=&\exp \left[ -\frac{i}{\hbar }\left( t_{j}-t_{j-1}\right) H\left( 
\mathfrak{p},q\right) \right]
\end{eqnarray*}%
where $H\left( \mathfrak{p},q\right) $is the "lattice" Weyl transform of $%
\mathcal{H}$for the time interval $\left( t_{j}-t_{j-1}\right) $. \ We also
have 
\begin{eqnarray*}
\left\langle \mathfrak{q}_{j}\right\vert \Delta \left( \mathfrak{p,q}\right)
\left\vert \mathfrak{q}_{j-1}\right\rangle &=&\int \mathfrak{Dv\ e}^{2i%
\mathfrak{p\cdot v}}\ \left\langle \mathfrak{q}_{j}\right\vert \left\vert 
\mathfrak{q+v}\right\rangle \left\langle \mathfrak{q-v}\right\vert
\left\vert \mathfrak{q}_{j-1}\right\rangle \\
&=&\exp \left[ 2i\mathfrak{p\cdot }\left( \mathfrak{q}_{j}-\mathfrak{q}%
_{j-1}\right) \right] \delta \left( \frac{\mathfrak{q}_{j}+\mathfrak{q}_{j-1}%
}{2}-\mathfrak{q}\right)
\end{eqnarray*}%
Therefore%
\begin{eqnarray*}
\left\langle \mathfrak{q}_{j}\right\vert \hat{U}\left( t_{j}.t_{j-1}\right)
\left\vert \mathfrak{q}_{j-1}\right\rangle &=&\iint \mathfrak{DqDp\ }%
U_{t_{j},t_{j-1}}\left( \mathfrak{p,q}\right) \Delta \left( \mathfrak{p,q}%
\right) \\
&=&\int \mathfrak{Dp}\exp \left[ 2\frac{i}{\hbar }\mathfrak{p\cdot }\left( 
\frac{\mathfrak{q}_{j}-\mathfrak{q}_{j-1}}{2}\right) \right] \exp \left[ -%
\frac{i}{\hbar }\left( t_{j}-t_{j-1}\right) H\left( \mathfrak{p},\frac{%
\mathfrak{q}_{j}+\mathfrak{q}_{j-1}}{2}\right) \right] \\
&=&\int \mathfrak{Dp}\exp \epsilon \frac{i}{\hbar }\left[ \mathfrak{p\cdot }%
\left( \frac{\mathfrak{q}_{j}-\mathfrak{q}_{j-1}}{\epsilon }\right) -\left(
t_{j}-t_{j-1}\right) H\left( \mathfrak{p},\frac{\mathfrak{q}_{j}+\mathfrak{q}%
_{j-1}}{2}\right) \right]
\end{eqnarray*}%
where $\epsilon $is the incremental time, $t_{j}-t_{j-1}$. Therefore the
transition amplitude between a state $\Psi \left( t_{0}\right) $ and a state 
$\Phi \left( t\right) $ is given by%
\begin{eqnarray*}
&&\left\langle \Phi \left( t\right) \right\vert \left\vert \Psi \left(
t_{0}\right) \right\rangle \\
&=&\int \prod\limits_{j=1}^{n+1}\mathfrak{Dp}_{j}\prod\limits_{j=1}^{n+1}%
\mathfrak{Dq}_{j}\ \phi \left( q_{n+1},t_{n+1}\right) \exp \left\{ -\epsilon
\sum\limits_{j=1}^{n+1}\left[ \mathfrak{p}_{j}\mathfrak{\cdot }\left( \frac{%
\mathfrak{q}_{j}-\mathfrak{q}_{j-1}}{\epsilon }\right) +H\left( \mathfrak{p}%
_{j},\frac{\mathfrak{q}_{j}+\mathfrak{q}_{j-1}}{2}\right) \right] \right\}
\psi \left( q_{1},t_{0}\right)
\end{eqnarray*}%
where $\phi \left( q_{n+1},t_{n+1}\right) =\left\langle \Phi \left( t\right)
\right\vert \left\vert q_{n+1}\right\rangle $and $\psi \left(
q_{1},t_{0}\right) =\left\langle q_{1}\right\vert \left\vert \Psi \left(
t_{0}\right) \right\rangle $. Another useful result for the path integral is
the expression for the partition function, using Matsubara imaginary time 
\cite{matsubara},%
\begin{eqnarray*}
\exp \left( -\beta \Omega \right) &=&Tr\exp \left( -\beta \left( \mathcal{H}%
-\mu \hat{N}\right) \right) \\
&=&Tr\exp \left( -\beta \hat{K}\right)
\end{eqnarray*}%
which yields%
\begin{equation*}
\exp \left( -\beta \Omega \right) =\int \int \prod\limits_{j=1}^{n+1}%
\mathfrak{Dp}_{j}\mathfrak{Dq}_{j}\exp \left\{ -\epsilon
\sum\limits_{j=1}^{n+1}\left[ \mathfrak{p}_{j}\mathfrak{\cdot }\left( \frac{%
\mathfrak{q}_{j}-\mathfrak{q}_{j-1}}{\epsilon }\right) +K\left( \mathfrak{p}%
_{j},\frac{\mathfrak{q}_{j}+\mathfrak{q}_{j-1}}{2}\right) \right] \right\}
\end{equation*}%
The anti-periodic boundary condition for $\mathfrak{p}$ and $\mathfrak{q}$
is precisley what is needed to preserve the antiperiodicity in each time
variables of the Green's function in the path integral formulation of
statistical quantum field theory \cite{soper}.

\section{Unified Q-Distribution Function Theory in CMP and QFT}

We have seen that both in CMP as well as in QFT, their quantum distribution
functions of field variables are well-defined. This suggests that we can
formulate QFT in terms of Q-distribution function of field variables,
similar to those of CMP in terms of Wigner distribution functions. Some
hints in this direction has been given very early on by Guttinger \cite%
{guttinger}. More pertinent to our work in this direction has recently been
given by Drummond \cite{drummond1,drummond2}, and also by Friederich \cite%
{friederich} and by Manko et al \cite{manko}.

Clearly, following the route to the formulation of nonequilibrium superfield
theory of quantum transport \cite{buot8}, we can form a new theory of
dissipative nonequilibrium quantum field theory based on the quantum
transport equation of the Q-distribution function or the 'Wigner
distribution function' of conjugate functional field variables. This is an
interesting new formulation of quantum field theory which we hope to
investigte in details in a separate communication. A good beginning in this
direction has been given by Drummond \cite{drummond1,drummond2} and by
Friederich \cite{friederich}.

\section{Concluding Remarks}

The unification discussed in this paper is rooted in the formally identical
expression for the phase-space point projector denoted as $\hat{\Delta}%
\left( \mathfrak{p},\mathfrak{q}\right) $. Here, any quantum operator can be
expanded in terms of the $\hat{\Delta}\left( \mathfrak{p},\mathfrak{q}%
\right) $operator basis. Although $\hat{\Delta}\left( \mathfrak{p},\mathfrak{%
q}\right) $carries two equivalent expressions as basis operators, this is
made a unique and symmetrical expression, which is a again formally
identical in the unification discussed here, namely, 
\begin{equation*}
\hat{\Delta}\left( \mathfrak{p},\mathfrak{q}\right) =\left( N\right)
^{-1}\sum\limits_{\mathfrak{v},\mathfrak{u}}\exp \left\{ -i\left( \mathcal{P}%
\cdot \mathfrak{v}-\mathcal{Q}\cdot \mathfrak{u}\right) \right\} \exp
\left\{ 2i\left( \mathfrak{p}\cdot \mathfrak{v}-\mathfrak{q}\cdot \mathfrak{u%
}\right) \right\} \Omega _{\lambda \lambda ^{\prime }}
\end{equation*}%
or in terms of continuous Grassman variables,

\begin{equation*}
\hat{\Delta}\left( \mathfrak{p},\mathfrak{q}\right) =\iint \mathfrak{DvDu}%
\sum\limits_{\mathfrak{v},\mathfrak{u}}\exp \left\{ -i\left( \mathcal{P}%
\cdot \mathfrak{v}-\mathcal{Q}\cdot \mathfrak{u}\right) \right\} \exp
\left\{ 2i\left( \mathfrak{p}\cdot \mathfrak{v}-\mathfrak{q}\cdot \mathfrak{u%
}\right) \right\} \Omega _{\lambda \lambda ^{\prime }}
\end{equation*}

With the above unification all physics that follows have corresponding
formally identical expressions, such as the distribution in dual space or
phase-space and their path integral expressions. These results clearly
demonstrate the power of the mathematical language of "lattice" Weyl-Wigner
formulation of quantum physics \cite{trHn}, as shown by the authors in
previous publications. Indeed for finite fermion systems, the creation and
annihilation formalism exactly maps to that of CMP since free Bloch
electrons are fermions. Thus, CMP formalism maybe considered a bosonization
of free Bloch electrons which incorporates the Pauli exclusion principle and
Fermi-Dirac distribution.

The fact that 
\begin{equation*}
\hat{Y}\left( \mathfrak{u,v}\right) =\exp \left[ -i\left( \hat{P}\cdot 
\mathfrak{v}-\hat{Q}\cdot \mathfrak{u}\right) \right] \ \Omega
\end{equation*}%
span all operators describing fermions, bosons, and spin systems suggests
that in principle, bosonization, fermionization and Jordan-Wigner
fermionization for spin systems, as well as the Holstein--Primakoff
transformation from boson operators to the spin operator can be performed
depending on the physical situations and ease in the calculations. In
quantum physics, unitary transformation on the creation and annihilation
operators themselves is also employed, as exemplified by the Bogoliubov
transformation. Moreover, whenever $\hat{Y}\left( \mathfrak{u,v}\right)
\Longrightarrow M_{ij}$ is already expressed in matrix form, the
Jordan--Schwinger transformation is a map from matrices $M_{ij}$ to bilinear
expressions of creation and annihilation operators, e.g., of the form, $\hat{%
a}_{i}^{\dagger }M_{ij}\hat{a}_{j}$, which expedites computation of
representations.

The present unification suggests a new formulation of QFT in terms of
nonperturbative 'Wigner distribution' or Q-distribution of conjugate
functional-field variables quantum transport equations.

In the Appendix, we also present for completeness some of the well-known
bosononization and fermionization transformations. Moreover, we mention some
other very important transformations in CMP and QFT which deals with the
decoupling of the different degrees of freedom. This is exemplified, e.g.,
by the well-known Foldy-Woutheysen for relativistic Dirac electrons \cite%
{foldy}, as well as by the perturbative decoupling of energy bands to all
orders in the calculation of magnetic susceptibility of Bloch fermions \cite%
{many}, etc.

\begin{acknowledgement}
One of the authors (F.A.B.) is grateful for the hospitality of the CNU
Department of Physics, for the support of the Balik Scientist Program of the
Philippine Council for Industry, Energy and Emerging Technology Research and
Development of the Department of Science and Technology (PCIEERD-DOST).
\end{acknowledgement}

\pagebreak

\appendix\textbf{Appendices}

\section{Non-Hermitian and time-reversal breaking symmetry in quantum
transport}

In nonequiulibrium quantum transport physics, the doubling of degrees of
freedom is what endows the dual spaces, since we have an explicit time axis
to describe irreversibility in the quantum Liouville equation. We expect
that the solution to the quantum Liouville equation will have as its dual
the anti-chronological super-statevector. In what follows, we will determine
this dual super-statevector by a variational tecnique.

\subsection{Variational technique for calculating transition probability}

We present here a general technique for calculating the transition
probability of the super-Schr\"{o}dinger equation. In CMP and QFT of closed
systems, the expectation values are generally obtain as time-ordered
correlations using path integrals%
\begin{equation*}
\left\langle F\left( \phi \right) \right\rangle =\frac{\int \mathfrak{D}\phi
\ F\left( \phi \right) e^{\frac{i}{\hbar }S\left( \phi \right) }}{\int 
\mathfrak{D}\phi \ e^{\frac{i}{\hbar }S\left( \phi \right) }}
\end{equation*}%
The so-called generating functional makes use of an element of the dual
space often denoted as the source $J$. Then the geneating functional denoted
by $Z\left( J\right) $is defined by%
\begin{equation*}
Z\left( J\right) =\int \mathfrak{D}\phi \ e^{\frac{i}{\hbar }\left[ S\left(
\phi \right) +J\phi \right] }
\end{equation*}%
and therefore,%
\begin{equation*}
\frac{\delta ^{n}Z\left( J\right) }{\delta J\left( x_{1}\right) .....\delta
J\left( x_{n}\right) }=i^{n}Z\left( J\right) \left\langle \phi \left(
x_{1}\right) ....\phi \left( x_{n}\right) \right\rangle
\end{equation*}%
In what follows, we will formulate the generating functional for the
Liouville equation for the density operator for open systems also termed as
the generating super-functional by variational method.

\subsection{Time-ordered displacement operator in Liouville space}

The familiar von Neumann density-matrix operator equation in $H$-space given
by 
\begin{equation}
i\hbar \frac{\partial }{\partial t}\rho \left( t\right) =\left[ \mathcal{H}%
,\rho \right]  \label{rnceq2.1}
\end{equation}%
becomes a super-Schr\"{o}dinger equation for the super-statevector in $L$%
-space expressed as 
\begin{equation}
i\hbar \frac{\partial }{\partial t}\left. \left\vert \rho \left( t\right)
\right\rangle \right\rangle =\mathcal{L}\left. \left\vert \rho \left(
t\right) \right\rangle \right\rangle \text{.}  \label{rnceq2.2}
\end{equation}%
Note that since in general $\mathcal{L}$is non-Hermitian for open and
interacting systems, we expect to have a dual space of the solutions to Eq. (%
\ref{rnceq2.2}). This dual space super-state eigenvector is determined by
variational method in what follows.

From Eq. (\ref{rnceq2.2}), we can formally write the solution for the
super-statevector $\left. \left\vert \rho \left( t\right) \right\rangle
\right\rangle $as%
\begin{equation}
\left. \left\vert \rho \left( t\right) \right\rangle \right\rangle =T\exp 
\left[ \frac{-i}{\hbar }\int_{t_{o}}^{t}\mathcal{L}\left( t^{\prime }\right)
dt^{\prime }\right] \left. \left\vert \rho \left( t_{o}\right) \right\rangle
\right\rangle ,  \label{rnc3.1}
\end{equation}%
where $T$is the usual real-time ordering displacement operator. We take $%
t_{o}$as the time when the \textquotedblright
perturbation\textquotedblright\ Liouvillian, $\mathcal{L}^{\left( 1\right) }=%
\mathcal{L}-\mathcal{L}_{o}$, is turned on. Then we can also write%
\begin{equation}
\left. \left\vert \rho \left( t_{o}\right) \right\rangle \right\rangle
=T\exp \left[ \frac{-i}{\hbar }\int_{\xi _{o}}^{t_{o}}\mathcal{L}_{o}\left(
t^{\prime }\right) dt^{\prime }\right] \left. \left\vert \rho \left( \xi
_{o}\right) \right\rangle \right\rangle ,  \label{rnc3.2}
\end{equation}%
where the system, during the time duration from $\xi _{o}$to $t_{o}$, is
acted on only by the \textquotedblright unperturbed\textquotedblright\
Liouvillian $\mathcal{L}_{o}$.

\subsubsection{Time evolution operator $U$ and $S$-matrix}

Therefore, we can also write%
\begin{equation}
\left. \left\vert \rho \left( t\right) \right\rangle \right\rangle =T\exp 
\left[ \frac{-i}{\hbar }\int_{0}^{t}\mathcal{L}\left( t^{\prime }\right)
dt^{\prime }\right] \overline{S}\left( t,t_{o}\right) \left. \left\vert \rho
_{o}\left( 0\right) \right\rangle \right\rangle ,  \label{rnc3.3}
\end{equation}%
where%
\begin{equation}
\left. 
\begin{array}{c}
\overline{S}\left( t,t_{o}\right) =\mathcal{U}_{o}\left( 0,t\right) \mathcal{%
U}\left( t,t_{o}\right) \mathcal{U}_{o}\left( t_{o},0\right) , \\ 
\mathcal{U}_{o}\left( 0,t\right) =T\exp \left[ \frac{i}{\hbar }\int_{0}^{t}%
\mathcal{L}_{o}\left( t^{\prime }\right) dt^{\prime }\right] , \\ 
\mathcal{U}\left( t,t_{o}\right) =T\exp \left[ \frac{-i}{\hbar }%
\int_{t_{o}}^{t}\mathcal{L}\left( t^{\prime }\right) dt^{\prime }\right] ,%
\end{array}%
\right\}  \label{rnc3.4}
\end{equation}%
\begin{equation}
\left. \left\vert \rho _{o}\left( 0\right) \right\rangle \right\rangle =%
\mathcal{U}_{o}\left( 0,\xi _{o}\right) \left. \left\vert \rho \left( \xi
_{o}\right) \right\rangle \right\rangle .  \label{rnc3.5}
\end{equation}%
If we let $t_{o}\Rightarrow -\infty $, and write $\left. \left\vert \rho
_{o}\left( 0\right) \right\rangle \right\rangle =\left. \left\vert \rho
_{eq}\right\rangle \right\rangle $, then we have%
\begin{equation}
\left. \left\vert \rho \left( t\right) \right\rangle \right\rangle =T\exp 
\left[ \frac{-i}{\hbar }\int_{0}^{t}\mathcal{L}_{o}\left( t^{\prime }\right)
dt^{\prime }\right] \overline{S}\left( t,-\infty \right) \left. \left\vert
\rho _{eq}\right\rangle \right\rangle .  \label{rnc3.6}
\end{equation}%
It follows that%
\begin{equation}
\left. \left\vert \rho \left( 0\right) \right\rangle \right\rangle =%
\overline{S}\left( 0,-\infty \right) \left. \left\vert \rho
_{eq}\right\rangle \right\rangle =\left. \left\vert \rho _{\mathcal{L}%
}\right\rangle \right\rangle =\left. \left\vert \rho _{I}\left( 0\right)
\right\rangle \right\rangle ,  \label{rnc3.7}
\end{equation}%
which defines the super-Heisenberg representation $\left. \left\vert \rho _{%
\mathcal{L}}\right\rangle \right\rangle $of the super-statevector $\left.
\left\vert \rho \left( t\right) \right\rangle \right\rangle $. Note that $%
\left. \left\vert \rho _{\mathcal{L}}\right\rangle \right\rangle $is
independent of time. Equation (\ref{rnc3.4}) also leads us to define the
super-interaction representation in L-space. Thus for any superoperator $O$,
we define%
\begin{equation}
O_{I}\left( t\right) =\mathcal{U}_{o}\left( 0,t\right) \mathcal{O}\left(
t\right) \mathcal{U}_{o}\left( t,0\right)  \label{rnc3.8}
\end{equation}%
as the super-interaction representation of $O$in L-space, and%
\begin{equation}
O_{\mathcal{L}}\left( t\right) =\mathcal{U}\left( 0,t\right) \mathcal{O}%
\left( t\right) \mathcal{U}\left( t,0\right)  \label{rnc3.9}
\end{equation}%
as its super-Heisenberg representation. Therefore, we have for any
superoperator $O$,%
\begin{equation}
O_{\mathcal{L}}\left( t\right) =\overline{S}\left( 0,t\right) O_{I}\left(
t\right) \overline{S}\left( t,0\right) ,  \label{rnc3.10}
\end{equation}%
which give the relation between the super-Heisenberg representation and the
super-interaction representation. Since $\mathcal{L}$, $\widetilde{\mathcal{H%
}}_{\epsilon }$, and $\widehat{\mathcal{H}}$commute then we have for the
\textquotedblright hat\textquotedblright and \textquotedblright
tilde\textquotedblright\ quantum field superoperators the following relations%
\begin{eqnarray}
\widehat{\psi }_{\widehat{\mathcal{H}}}\left( t\right) &=&T\exp \left[ \frac{%
i}{\hbar }\int_{0}^{t}\widehat{\mathcal{H}}\left( t^{\prime }\right)
dt^{\prime }\right] \widehat{\psi }\ T\exp \left[ \frac{-i}{\hbar }%
\int_{0}^{t}\widehat{\mathcal{H}}\left( t^{\prime }\right) dt^{\prime }%
\right]  \notag \\
&=&T\exp \left[ \frac{i}{\hbar }\int_{0}^{t}\mathcal{L}\left( t^{\prime
}\right) dt^{\prime }\right] \widehat{\psi }\ T\exp \left[ \frac{-i}{\hbar }%
\int_{0}^{t}\mathcal{L}\left( t^{\prime }\right) dt^{\prime }\right]  \notag
\\
&=&\widehat{\psi }_{\mathcal{L}}\left( t\right) ,  \label{rnc3.11}
\end{eqnarray}%
\begin{eqnarray}
\widetilde{\psi }_{\widetilde{\mathcal{H}}_{\epsilon }}\left( t\right)
&=&T\exp \left[ \frac{i}{\hbar }\int_{0}^{t}\widetilde{\mathcal{H}}%
_{\epsilon }\left( t^{\prime }\right) dt^{\prime }\right] \widetilde{\psi }\
T\exp \left[ \frac{-i}{\hbar }\int_{0}^{t}\widetilde{\mathcal{H}}_{\epsilon
}\left( t^{\prime }\right) dt^{\prime }\right]  \notag \\
&=&T\exp \left[ \frac{i}{\hbar }\int_{0}^{t}\mathcal{L}\left( t^{\prime
}\right) dt^{\prime }\right] \widetilde{\psi }\ T\exp \left[ \frac{-i}{\hbar 
}\int_{0}^{t}\mathcal{L}\left( t^{\prime }\right) dt^{\prime }\right]  \notag
\\
&=&\widetilde{\psi }_{\mathcal{L}}\left( t\right) .  \label{rnc3.12}
\end{eqnarray}%
Similar relations exist for the canonically conjugate quantum field
operators: $\widehat{\psi }_{\widehat{\mathcal{H}}}^{\dagger }\left(
t\right) =\widehat{\psi }_{\mathcal{L}}^{\dagger }\left( t\right) $, and $%
\widetilde{\psi }_{\widetilde{\mathcal{H}}_{\epsilon }}^{\dagger }\left(
t\right) =\widetilde{\psi }_{\mathcal{L}}^{\dagger }\left( t\right) $.

\subsection{Super $S$-matrix theory in $L$-space}

The \textquotedblright super-Schr\"{o}dinger equation\textquotedblright\
that we have developed in L-space and its corresponding formal solution
given above are not complete. We have to determine the canonically conjugate
counterpart of the super-statevector $\left. \left\vert \rho \left( t\right)
\right\rangle \right\rangle $. This is done in what follows by a variational
technique.

\subsection{Construction of generating super-functional}

For example, it is not clear what would be the canonically conjugate
counterpart of the super-statevector $\left. \left\vert \rho \left( t\right)
\right\rangle \right\rangle $for defining generating functional, analogous
to the existing canonically conjugate pair of H-space states, $\left\vert
\psi _{+},t\right\rangle $and $\left\vert \psi _{-},t\right\rangle $, which
naturally occur in zero temperature variational quantum action principle
forming the basis for deriving the time-dependent Schr\"{o}dinger equation,
as first given by Dirac \cite{hnbkref23}.

\subsection{Time-dependent variational principle in Liouville space}

To complete the \textquotedblright super-Schr\"{o}dinger quantum
mechanics\textquotedblright , we need to construct the corresponding
variational principle in L-space. We follow the general construction of a
variational principle \cite{hnbkref24,hnbkref25} as an optimization of the
estimate for the expectation value of any operator $\mathcal{O}$at the time $%
t_{1}$from the knowledge of the initial condition for the density matrix
super-statevector $\left. \left\vert \rho \left( t_{o}\right) \right\rangle
\right\rangle $, subject to the constraint $\left. \left\vert \rho \left(
t\right) \right\rangle \right\rangle $obeys the L-space super-Schr\"{o}%
dinger equation. Therefore, we have to optimize the following
super-functional \cite{hnbkref25},%
\begin{equation}
\Phi \left( \left. \left\vert \rho \left( t\right) \right\rangle
\right\rangle ,\left. \left\vert \Lambda \left( t\right) \right\rangle
\right\rangle \right) =\left\langle \left\langle \mathcal{O}\left( t\right)
\right\vert \right. \left. \left\vert \rho \left( t_{1}\right) \right\rangle
\right\rangle -\int_{t_{o}}^{t_{1}}\left\langle \left\langle \Lambda \left(
t\right) \right\vert \right. i\hbar \frac{\partial }{\partial t}-\mathcal{L}%
\left. \left\vert \rho \left( t\right) \right\rangle \right\rangle ,
\label{rnc3.13}
\end{equation}%
where we have introduced the dual space super-statevector $\left\langle
\left\langle \Lambda \left( t\right) \right\vert \right. $as a 'Lagrange
multiplier'. Upon integration by parts, $\Phi \left( \left. \left\vert \rho
\left( t\right) \right\rangle \right\rangle ,\left. \left\vert \Lambda
\left( t\right) \right\rangle \right\rangle \right) $is also equal to the
following super-functional%
\begin{eqnarray}
\Phi \left( \left. \left\vert \rho \left( t\right) \right\rangle
\right\rangle ,\left. \left\vert \Lambda \left( t\right) \right\rangle
\right\rangle \right) &=&\left\langle \left\langle \mathcal{O}\left(
t\right) \right\vert \right. \left. \left\vert \rho \left( t_{1}\right)
\right\rangle \right\rangle -i\hbar \left\langle \left\langle \Lambda \left(
t_{1}\right) \right\vert \right. \left. \left\vert \rho \left( t_{1}\right)
\right\rangle \right\rangle  \notag \\
&&+i\hbar \left\langle \left\langle \Lambda \left( t_{o}\right) \right\vert
\right. \left. \left\vert \rho \left( t_{o}\right) \right\rangle
\right\rangle  \notag \\
&&+\int_{t_{o}}^{t_{1}}\left\langle \left\langle i\hbar \frac{\partial
\Lambda \left( t\right) }{\partial t}-\mathcal{L}\Lambda \left( t\right)
\right\vert \right. \left. \left\vert \rho \left( t\right) \right\rangle
\right\rangle .  \label{rnc3.14}
\end{eqnarray}

Thus, the optimum condition for $\Phi \left( \left. \left\vert \rho \left(
t\right) \right\rangle \right\rangle ,\left. \left\vert \Lambda \left(
t\right) \right\rangle \right\rangle \right) $occurs for $i\hbar
\left\langle \left\langle \Lambda \left( t_{1}\right) \right\vert \right.
=\left\langle \left\langle \mathcal{O}\left( t\right) \right\vert \right. $,
and when the following equations of motion are obeyed for $t_{o}<t<t_{1}$, 
\begin{equation}
\left. 
\begin{array}{c}
i\hbar \frac{\partial \left. \left\vert \rho \left( t\right) \right\rangle
\right\rangle }{\partial t}=\mathcal{L}\left. \left\vert \rho \left(
t\right) \right\rangle \right\rangle , \\ 
i\hbar \frac{\partial \left. \left\vert \Lambda \left( t\right)
\right\rangle \right\rangle }{\partial t}=\mathcal{L}\left. \left\vert
\Lambda \left( t\right) \right\rangle \right\rangle .%
\end{array}%
\right\}  \label{rnc3.15}
\end{equation}%
If we let $t_{1}\Rightarrow \infty $and $t_{o}\Rightarrow -\infty $, apply
the boundary condition: $i\hbar \left\langle \left\langle \Lambda \left(
t_{1}\right) \right\vert \right. =\left\langle \left\langle \mathcal{O}%
\left( t\right) \right\vert \right. $, and $\left. \left\vert \rho \left(
-\infty \right) \right\rangle \right\rangle =\left. \left\vert \rho
_{eq}\right\rangle \right\rangle $, then the solutions can be written as%
\begin{equation}
\left. 
\begin{array}{c}
\left. \left\vert \rho \left( t\right) \right\rangle \right\rangle =T\exp 
\left[ \frac{-i}{\hbar }\int_{0}^{t}\mathcal{L}_{o}\left( t^{\prime }\right)
dt^{\prime }\right] \overline{S}\left( t,-\infty \right) \left. \left\vert
\rho _{eq}\right\rangle \right\rangle , \\ 
\left. \left\vert \Lambda \left( t\right) \right\rangle \right\rangle
=T^{ac}\exp \left[ \frac{-i}{\hbar }\int_{t}^{0}\mathcal{L}_{o}\left(
t^{\prime }\right) dt^{\prime }\right] \overline{S}\left( t,\infty \right)
\left. \left\vert \mathcal{O}\right\rangle \right\rangle ,%
\end{array}%
\right\}  \label{rnc3.16}
\end{equation}%
where $T^{ac}$denotes anti-chronological time ordering. Therefore, we can
write a \textquotedblright transition probability\textquotedblright\ in
L-space as%
\begin{equation}
\left\langle \left\langle \Lambda \left( t\right) \right\vert \right. \left.
\left\vert \rho \left( t\right) \right\rangle \right\rangle =\frac{1}{i\hbar 
}\left\langle \left\langle \mathcal{O}\right\vert \right. \overline{S}\left(
\infty ,-\infty \right) \left. \left\vert \rho _{eq}\right\rangle
\right\rangle =\frac{1}{i\hbar }\left\langle \left\langle \mathcal{O}_{%
\mathcal{L}}\right\vert \right. \left. \left\vert \rho _{\mathcal{L}%
}\right\rangle \right\rangle ,  \label{rnc3.17}
\end{equation}%
where $\left. \left\vert \rho _{\mathcal{L}}\right\rangle \right\rangle $is
the super-Heisenberg representation of the super-statevector $\left.
\left\vert \rho \left( t\right) \right\rangle \right\rangle $defined by Eq. (%
\ref{rnc3.7}), and similarly $\left\langle \left\langle \mathcal{O}_{%
\mathcal{L}}\right\vert \right. =\left\langle \left\langle \mathcal{O}%
\right\vert \right. \overline{S}\left( \infty ,0\right) $

Note that in terms of evaluating the expectation value of the superoperator
at $t=t_{1}=\infty $, we can explicitly write this as a time-ordered
expectation value by rewriting Eq. (\ref{rnc3.17}) as%
\begin{equation}
\left\langle \left\langle \Lambda \left( t\right) \right\vert \right. \left.
\left\vert \rho \left( t\right) \right\rangle \right\rangle =\frac{1}{i\hbar 
}\left\langle \left\langle \mathcal{O}\right\vert \right. \overline{S}\left(
\infty ,-\infty \right) \left. \left\vert \rho _{eq}\right\rangle
\right\rangle =\frac{1}{i\hbar }\left\langle \left\langle 1\right\vert
\right. \mathcal{O}\overline{S}\left( \infty ,-\infty \right) \left.
\left\vert \rho _{eq}\right\rangle \right\rangle .  \label{rnc3.18}
\end{equation}%
We are particularly interested in $\mathcal{O}=i\hbar \Im $, where $\Im $is
the identity operator corresponding to an asymptotic state of \textit{%
maximum entropy} at $t=\infty $(corresponding to $\left\vert \psi
_{+},t\right\rangle $indicating \textit{asymptotic freedom} of the zero
temperature variational quantum-action principle). Then the canonically
conjugate pair $\left. \left\vert \Lambda \left( t\right) \right\rangle
\right\rangle $and $\left. \left\vert \rho \left( t\right) \right\rangle
\right\rangle $is defined by the \textquotedblright transition
probability\textquotedblright\ in L-space given by the following relation%
\begin{equation}
\left\langle \left\langle \Lambda \left( t\right) \right\vert \right. \left.
\left\vert \rho \left( t\right) \right\rangle \right\rangle =\left\langle
\left\langle 1\right\vert \right. \overline{S}\left( \infty ,-\infty \right)
\left. \left\vert \rho _{eq}\right\rangle \right\rangle =\left\langle
\left\langle \Im _{\mathcal{L}}\right\vert \right. \left. \left\vert \rho _{%
\mathcal{L}}\right\rangle \right\rangle ,  \label{rnc3.19}
\end{equation}%
where $\left\langle \left\langle \Im _{\mathcal{L}}\right\vert \right. $is
given by 
\begin{equation}
\left\langle \left\langle \Im _{\mathcal{L}}\right\vert \right.
=\left\langle \left\langle 1\right\vert \right. \overline{S}\left( \infty
,0\right) .  \label{rnc3.20}
\end{equation}

From the time-ordered expectation value deduced by Eq. (\ref{rnc3.18}), we
see that indeed $\left\langle \left\langle 1\right\vert \right. \overline{S}%
\left( \infty ,-\infty \right) \left. \left\vert \rho _{eq}\right\rangle
\right\rangle $is the analog to the transition amplitude occurring in zero
temperature time-dependent quantum mechanics \cite{hnbkref23}.

\subsection{The effective action and generating super-functional}

In order to gain further insights into the S-matrix formalism in L-space, we
formulate the constrained stationary value \cite{hnbkref26} of the
\textquotedblright super-action\textquotedblright\ given by%
\begin{equation}
\int dt\ \left\langle \left\langle \Lambda \left( t\right) \right\vert
\right. i\hbar \frac{\partial }{\partial t}-\mathcal{L}\left. \left\vert
\rho \left( t\right) \right\rangle \right\rangle ,\qquad \left\{ 
\begin{array}{c}
\left\langle \left\langle \Lambda \left( t\right) \right\vert \right. \left.
\left\vert \rho \left( t\right) \right\rangle \right\rangle =1 \\ 
\left\langle \left\langle \Lambda \left( t\right) \right\vert \right. \Psi
\left. \left\vert \rho \left( t\right) \right\rangle \right\rangle
=\left\langle \Psi \right\rangle%
\end{array}%
\right. ,  \label{rnc3.21}
\end{equation}%
where $\Psi $stands for an arbitrary quantum field superoperator (the
extension to several species of quantum field superoperators is
straightforward). Thus, we consider the stationary variation of the
super-functional 
\begin{eqnarray}
\Omega \left( \left. \left\vert \rho \left( t\right) \right\rangle
\right\rangle ,\left. \left\vert \Lambda \left( t\right) \right\rangle
\right\rangle \right) &=&\int dt\ \left\langle \left\langle \Lambda \left(
t\right) \right\vert \right. i\hbar \frac{\partial }{\partial t}-\mathcal{L}%
\left. \left\vert \rho \left( t\right) \right\rangle \right\rangle  \notag \\
&&-\int dx\ \left\langle \left\langle \Lambda \left( t\right) \right\vert
\right. \eta \left( x\right) \cdot \Psi \left. \left\vert \rho \left(
t\right) \right\rangle \right\rangle  \notag \\
&&-\int dt\ w\left( t\right) \left\langle \left\langle \Lambda \left(
t\right) \right\vert \right. \left. \left\vert \rho \left( t\right)
\right\rangle \right\rangle ,  \label{rnc3.22}
\end{eqnarray}%
where $\eta \left( x\right) $and $w\left( t\right) $are introduced as $%
\mathbb{C}
$-number Lagrange multipliers. Note that $\eta \left( x\right) $is also
playing the role of the Schwinger source field.

Carrying out the variation with respect to $\left. \left\vert \Lambda \left(
t\right) \right\rangle \right\rangle $and $\left. \left\vert \rho \left(
t\right) \right\rangle \right\rangle $, and enforcing the stationarity of $%
\Omega \left( \left. \left\vert \rho \left( t\right) \right\rangle
\right\rangle ,\left. \left\vert \Lambda \left( t\right) \right\rangle
\right\rangle \right) $, we obtain the following equations%
\begin{equation}
\left. 
\begin{array}{c}
i\hbar \frac{\partial \left. \left\vert \rho \left( t\right) \right\rangle
\right\rangle }{\partial t}-\mathcal{L}\left. \left\vert \rho \left(
t\right) \right\rangle \right\rangle -\int dx\ \eta \left( x\right) \cdot
\Psi \left. \left\vert \rho \left( t\right) \right\rangle \right\rangle -\
w\left( t\right) \left. \left\vert \rho \left( t\right) \right\rangle
\right\rangle =0, \\ 
i\hbar \frac{\partial \left. \left\vert \Lambda \left( t\right)
\right\rangle \right\rangle }{\partial t}-\mathcal{L}\left. \left\vert
\Lambda \left( t\right) \right\rangle \right\rangle -\int dx\ \eta \left(
x\right) \cdot \Psi \left. \left\vert \Lambda \left( t\right) \right\rangle
\right\rangle -\ w^{\ast }\left( t\right) \left. \left\vert \Lambda \left(
t\right) \right\rangle \right\rangle =0.%
\end{array}%
\right\}  \label{rnc3.23}
\end{equation}%
We write the solutions to these equations as%
\begin{equation}
\left. 
\begin{array}{c}
\left. \left\vert \rho \left( t\right) \right\rangle \right\rangle _{\eta
,w}=\exp \left[ \frac{-i}{\hbar }\int_{-\infty }^{t}dt^{\prime }\ w\left(
t^{\prime }\right) \right] \left. \left\vert \rho \left( t\right)
\right\rangle \right\rangle _{\eta }\text{,} \\ 
\left. \left\vert \Lambda \left( t\right) \right\rangle \right\rangle _{\eta
,w}=\exp \left[ \frac{i}{\hbar }\int_{-\infty }^{t}dt^{\prime }\ w^{\ast
}\left( t^{\prime }\right) \right] \left. \left\vert \Lambda \left( t\right)
\right\rangle \right\rangle _{\eta }\text{,}%
\end{array}%
\right\}  \label{rnc3.24}
\end{equation}%
where $\left. \left\vert \rho \left( t\right) \right\rangle \right\rangle
_{\eta }$and $\left. \left\vert \Lambda \left( t\right) \right\rangle
\right\rangle _{\eta }$obey the equations%
\begin{equation}
\left. 
\begin{array}{c}
i\hbar \frac{\partial \left. \left\vert \rho \left( t\right) \right\rangle
\right\rangle _{\eta }}{\partial t}-\mathcal{L}\left. \left\vert \rho \left(
t\right) \right\rangle \right\rangle _{\eta }-\int dx\ \eta \left( x\right)
\cdot \Psi \left. \left\vert \rho \left( t\right) \right\rangle
\right\rangle _{\eta }=0, \\ 
i\hbar \frac{\partial \left. \left\vert \Lambda \left( t\right)
\right\rangle \right\rangle _{\eta }}{\partial t}-\mathcal{L}\left.
\left\vert \Lambda \left( t\right) \right\rangle \right\rangle _{\eta }-\int
dx\ \eta \left( x\right) \cdot \Psi \left. \left\vert \Lambda \left(
t\right) \right\rangle \right\rangle _{\eta }=0.%
\end{array}%
\right\}  \label{rnc3.25}
\end{equation}

Therefore, we can calculate the \textquotedblright transition
probability\textquotedblright\ in L-space from the solutions of Eq. (\ref%
{rnc3.23}), which is given by Eq. (\ref{rnc3.24}). We obtain the following
relation 
\begin{equation}
\left\langle \left\langle \Lambda \left( t\right) \right\vert \right. \left.
\left\vert \rho \left( t\right) \right\rangle \right\rangle _{\eta
,w}=1=\exp \left[ \frac{-i}{\hbar }\int_{-\infty }^{\infty }dt^{\prime }\
w\left( t^{\prime }\right) \right] \left\langle \left\langle \Lambda \left(
t\right) \right\vert \right. \left. \left\vert \rho \left( t\right)
\right\rangle \right\rangle _{\eta }\text{,}  \label{rnc3.26}
\end{equation}%
which leads to the \textquotedblright transition
probability\textquotedblright\ in the presence of the Schwinger source term
given by%
\begin{equation}
\left\langle \left\langle \Lambda \left( t\right) \right\vert \right. \left.
\left\vert \rho \left( t\right) \right\rangle \right\rangle _{\eta }=\exp 
\left[ \frac{i}{\hbar }\int_{-\infty }^{\infty }dt^{\prime }\ w\left(
t^{\prime }\right) \right] =\exp \left[ \frac{i}{\hbar }W\right] .
\label{rnc3.27}
\end{equation}%
This is the same \textquotedblright transition
probability\textquotedblright\ as that obtained in Eq. (\ref{rnc3.19}).
Therefore, we obtain the equality%
\begin{equation}
\left\langle \left\langle 1\right\vert \right. \overline{S}\left( \infty
,-\infty \right) \left. \left\vert \rho _{eq}\right\rangle \right\rangle
=\exp \left[ \frac{i}{\hbar }W\right] .  \label{rnc3.28}
\end{equation}

Analogous to the zero temperature $S$-matrix formalism, we identify $W$as
the generating super-functional for connected $n$-point super-Green's
functions. This is supported by the time-ordered way of taking the average
value, shown by Eq. (\ref{rnc3.18}). To have a deeper appreciation of this
analogy, we take the scalar product with $\left\langle \left\langle \Lambda
\left( t\right) \right\vert \right. _{\eta ,w}$of both sides of the first
line of Eq. (\ref{rnc3.23}) and integrate over time. We obtain%
\begin{eqnarray}
W &=&\int dt^{\prime }\ w\left( t^{\prime }\right)  \notag \\
&=&\int dt\ \left\langle \left\langle \Lambda \left( t\right) \right\vert
\right. \left\{ i\hbar \frac{\partial \left. \left\vert \rho \left( t\right)
\right\rangle \right\rangle }{\partial t}-\mathcal{L}-\int dx\ \eta \left(
x\right) \cdot \Psi \right\} \left. \left\vert \rho \left( t\right)
\right\rangle \right\rangle _{\eta ,w}  \label{rnc3.29}
\end{eqnarray}%
from which we deduced the variational derivative%
\begin{equation}
\frac{\delta }{\delta \eta }W=-\left\langle \left\langle \Lambda \left(
t\right) \right\vert \right. \Psi \left. \left\vert \rho \left( t\right)
\right\rangle \right\rangle _{\eta ,w}=\left\langle \Psi \right\rangle
\qquad \left( \text{at the optimum condition}\right) .  \label{rnc3.30}
\end{equation}%
Therefore, we have the effective super-action given by%
\begin{equation}
A_{eff}=\int_{-\infty }^{\infty }dt\ \left\langle \left\langle \Lambda
\left( t\right) \right\vert \right. i\hbar \frac{\partial }{\partial t}-%
\mathcal{L}\left. \left\vert \rho \left( t\right) \right\rangle
\right\rangle _{\eta ,w}=W+\int dx\ \eta \left( x\right) \cdot \left\langle
\Psi \right\rangle  \label{rnc3.31}
\end{equation}%
from which we also deduced the variational derivative 
\begin{equation}
\frac{\delta }{\delta \left\langle \Psi \right\rangle }A_{eff}=\eta .
\label{rnc3.32}
\end{equation}

Hence, the effective super-action is stationary with respect to the
variation of $\left\langle \Psi \right\rangle $when $\eta \left( x\right) =0 
$, which is the physical situation, i.e., $W=A_{eff}$. Thus for $\eta \left(
x\right) =0$, we have%
\begin{equation}
W=-i\hbar \ \ln \left\langle \left\langle 1\right\vert \right. \overline{S}%
\left( \infty ,-\infty \right) \left. \left\vert \rho _{eq}\right\rangle
\right\rangle =\int_{-\infty }^{\infty }dt\ \left\langle \left\langle
\Lambda \left( t\right) \right\vert \right. i\hbar \frac{\partial }{\partial
t}-\mathcal{L}\left. \left\vert \rho \left( t\right) \right\rangle
\right\rangle .  \label{rnc3.33}
\end{equation}%
The relation derived here for nonequilibrium quantum-field theory between
super $\overline{S}$-matrix and the effective super-action provides a
rigorous basis supporting the functional theory of time-dependent many-body
quantum mechanics discussed by Rajagopal and Buot in a series of papers \cite%
{hnbkref27,hnbkref28,hnbkref29}.

We will refer to $\overline{S}\left( \infty ,-\infty \right) $as the
complete $S$-matrix superoperator for quantum-field theoretical methods of
nonequilibrium systems. Equation (\ref{rnc3.4}) allows us to write the
evolution equation for the $\overline{S}$-matrix operator as 
\begin{equation}
i\hbar \frac{\partial }{\partial t}\overline{S}\left( t,t_{o}\right) =%
\mathcal{L}_{I}^{\left( 1\right) }\left( t\right) \ \overline{S}\left(
t,t_{o}\right) ,  \label{rnc3.34}
\end{equation}%
where%
\begin{equation}
\mathcal{L}_{I}^{\left( 1\right) }\left( t\right) =\mathcal{U}_{o}\left(
0,t\right) \left( \mathcal{L}-\mathcal{L}_{o}\right) \mathcal{U}_{o}\left(
t,0\right) ,  \label{rnc3.35}
\end{equation}%
and%
\begin{equation}
\left( \mathcal{L}-\mathcal{L}_{o}\right) =\left[ v\left( 1\right) +u\left(
1\right) \right] \Psi \left( 1\right) +\left[ v\left( 1,2\right) +u\left(
1,2\right) \right] \Psi \left( 1\right) \Psi \left( 2\right) .
\label{rnc3.36}
\end{equation}

The Schwinger external sources \cite{hnbkref30} indicated by $u(1)$and $%
u(1,2)$contain all the time-dependent part of $v(1)$and $v(1,2)$
respectively. These Schwinger source terms are to be set equal to zero at
the end of all calculations. The solution to Eq. (\ref{rnc3.34}) immediately
yields%
\begin{equation}
\overline{S}\left( t,t_{o}\right) =1-\frac{i}{\hbar }\int_{t_{o}}^{t}%
\mathcal{L}_{I}^{\left( 1\right) }\left( t^{\prime }\right) \ \overline{S}%
\left( t^{\prime },t_{o}\right) \ dt^{\prime },  \label{rnc3.37}
\end{equation}%
which upon iteration yields%
\begin{equation}
\overline{S}\left( t,t_{o}\right) =T\exp \left[ \frac{-i}{\hbar }%
\int_{t_{o}}^{t}\mathcal{L}_{I}^{\left( 1\right) }\left( t^{\prime }\right)
\ dt^{\prime }\right] .  \label{rnc3.38}
\end{equation}%
Thus, we can explicitly exhibit $\left\langle \left\langle 1\right\vert
\right. \overline{S}\left( \infty ,-\infty \right) \left. \left\vert \rho
_{eq}\right\rangle \right\rangle $as%
\begin{eqnarray}
\left\langle \left\langle 1\right\vert \right. \overline{S}\left( \infty
,-\infty \right) \left. \left\vert \rho _{eq}\right\rangle \right\rangle
&=&\left\langle \left\langle 1\right\vert \right. T\exp \left[ \frac{-i}{%
\hbar }\int_{-\infty }^{\infty }\mathcal{L}_{I}^{\left( 1\right) }\left(
t^{\prime }\right) \ dt^{\prime }\right] \left. \left\vert \rho
_{eq}\right\rangle \right\rangle  \notag \\
&=&\exp \left[ \frac{i}{\hbar }W\right] ,  \label{rnc3.39}
\end{eqnarray}%
where $W$is the effective action. Equation (\ref{rnc3.39}) shows the
relation of Eq. (\ref{rnc3.28}) to $\mathcal{L}_{I}^{\left( 1\right) }\left(
t\right) $similar to zero-temperature many-body quantum-field theory.

\section{Some well-known fermionization and bosonization transformations}

Aside from the well-known bosonization in one-dimensional Tomanaga-Luttinger
and Hubbard model, there are other very highly-utilized fermionization and
bosonization transformations in CMP. Moreover, some decoupling
transformations are used in both QFT and CMP. For completeness, we give
these here.

\subsection{Holstein-Primakoff transformation}

The Holstein--Primakoff transformation in quantum mechanics is a mapping to
the spin operators from boson creation and annihilation operators \cite%
{holprima}. This Involves redefinition of angular momentum states $%
\left\vert j,m\right\rangle \Longrightarrow \left\vert n\right\rangle ,$ $%
n=1,2,...2j$, where the new eignvalues are the values of $n$ reminiscent of
the harmonic oscillator problem. This entails transformation of the creation
and annihilation operators and corresponding spin vector operators. This is
also reminiscent of the renormalization of position and momentum operators,
lattice-position eigenvectors and crystal-momentum eigenvectors, in
energy-band dynamics leading to discrete quantum mechanics on discrete
finite fields. In HP transformation, $S^{+}$ and $S^{-}$ are deduced from
the theory of angular momentum.%
\begin{equation*}
S^{z}\left\vert n\right\rangle =\hbar \left( j-n\right) \left\vert
n\right\rangle
\end{equation*}%
\begin{equation*}
b^{\dagger }b\left\vert n\right\rangle =n\left\vert n\right\rangle
\end{equation*}%
\begin{equation*}
S^{z}=\hbar \left( j-b^{\dagger }b\right)
\end{equation*}%
From the the theory of angular momentum%
\begin{eqnarray*}
S^{+}\left\vert j,m\right\rangle &=&\hbar \sqrt{\left( j-m\right) \left(
j+m+1\right) }\left\vert m+1\right\rangle \\
S^{+}\left\vert n\right\rangle &=&f^{+}\left( b^{\dagger },b\right)
\left\vert n\right\rangle \\
&=&\hbar \sqrt{\left( j-\left( j-n\right) \right) \left( j+\left( j-n\right)
+1\right) }\left\vert \left( j-n\right) +1\right\rangle \\
&=&\hbar \sqrt{\left( n\right) \left( 2j-\left( n-1\right) \right) }%
\left\vert \left( n-1\right) \right\rangle \\
&=&\hbar \sqrt{\left( 2j-b^{\dagger }b\right) }\sqrt{n}\left\vert \left(
n-1\right) \right\rangle \\
&=&\hbar \sqrt{\left( 2j-b^{\dagger }b\right) }b\ \left\vert \left( n\right)
\right\rangle
\end{eqnarray*}%
\begin{equation*}
S^{-}\left\vert n\right\rangle ==\hbar b^{\dagger }\sqrt{\left(
2j-b^{\dagger }b\right) }\ \left\vert \left( n\right) \right\rangle
\end{equation*}%
Holstein-Primakoff transformation is 
\begin{eqnarray*}
S_{i}^{-} &=&a_{i}^{\dagger }\left( \sqrt{2S-a_{i}^{\dagger }a_{i}}\right) \\
S_{i}^{+} &=&\left( \sqrt{2S-a_{i}^{\dagger }a_{i}}\right) a_{i} \\
S_{i}^{z} &=&S-a_{i}^{\dagger }a_{i} \\
a_{i}^{\dagger }a_{i}\, &\leq &2S
\end{eqnarray*}%
Note: instead of the pre-Bogoliubov transformation $u\Longrightarrow \sqrt{%
2S-a_{i}^{\dagger }a_{i}}$, we have,%
\begin{eqnarray*}
S_{i}^{+}S_{i}^{-} &=&\left( \sqrt{2S-a_{i}^{\dagger }a_{i}}a_{i}\right)
\left( a_{i}^{\dagger }\sqrt{2S-a_{i}^{\dagger }a_{i}}\right) \\
S_{i}^{-}S_{i}^{+} &=&\left( a_{i}^{\dagger }\sqrt{2S-a_{i}^{\dagger }a_{i}}%
\right) \left( \sqrt{2S-a_{i}^{\dagger }a_{i}}a_{i}\right) \\
&=&a_{i}^{\dagger }\left( 2S-a_{i}^{\dagger }a_{i}\right) a_{i}
\end{eqnarray*}

\subsection{Dyson--Maleev transformation}

A non-Hermitian Dyson--Maleev variant realization $J$ is related to the
above and valid for all spins,%
\begin{eqnarray*}
J_{+} &=&\hbar a^{\dagger } \\
J_{-} &=&S_{-}\sqrt{2s-a^{\dagger }a}=\hbar a^{\dagger }\left( 2s-a^{\dagger
}a\right) \\
J_{z} &=&S_{z}=\hbar \left( s-a^{\dagger }a\right)
\end{eqnarray*}%
satisfying the same commutation relations and characterized by the same
Casimir invariant with Casimir operators, $S^{2}$ and $S_{z}.$

\subsection{Jordan-Wigner transformation}

The Jordan--Wigner transformation maps spin operators onto fermionic
creation and annihilation operators \cite{nielsen, porta,
hastings,fidkowski,batista}.

\subsubsection{Derivation of Jordan-Wigner transformation}

We will show how to map a $1$-D spin chain of spin-$1/2$ particles to
fermions. Take spin-$1/2$ Pauli operators acting on a site $j$ of a $1$-D
chain, $\sigma _{j}^{+}$, $\sigma _{j}^{-}$ and $\sigma _{j}^{-}$. The
anticommutator of $\sigma _{j}^{+}$, $\sigma _{j}^{-}$, is $\left\{ \sigma
_{j}^{+},\sigma _{j}^{-}\right\} =1$, as would be expected from fermionic
creation and annihilation operators. We are tempted to set%
\begin{eqnarray}
\sigma _{j}^{+} &=&\frac{\left( \sigma _{j}^{x}+i\sigma _{j}^{y}\right) }{2}%
=f_{j}^{\dagger }  \notag \\
\sigma ^{-} &=&\frac{\left( \sigma _{j}^{x}-i\sigma _{j}^{y}\right) }{2}%
=f_{j}  \notag \\
\sigma _{j}^{z} &=&2f_{j}^{\dagger }f_{j}-1  \label{tempted}
\end{eqnarray}%
We now have the site anticommutation $\left\{ f_{j}^{\dagger },f_{j}\right\}
=1$; however commutation enters for different sites, i.e., $\left[
f_{j}^{\dagger },f_{k}\right] =0$, for $j\neq k$ instead of the desired
anticommutation for fermions. \ In 1928, Jordan and Wigner invented a
transformation which renders $\left\{ f_{j}^{\dagger },f_{k}\right\} =0$,
for $j\neq k$. The J-W transformation is given by,%
\begin{eqnarray*}
a_{j}^{+} &=&\exp \left( +i\pi \sum\limits_{k=1}^{j-1}f_{k}^{\dagger
}f_{k}\right) \cdot f_{j}^{\dagger } \\
a_{j} &=&\exp \left( -i\pi \sum\limits_{k=1}^{j-1}f_{k}^{\dagger
}f_{k}\right) \cdot f_{j} \\
a_{j}^{+}a_{j} &=&f_{j}^{\dagger }f_{j}
\end{eqnarray*}%
which can be written as,%
\begin{eqnarray*}
\exp \left( \pm i\pi \sum\limits_{k=1}^{j-1}f_{k}^{\dagger }f_{k}\right)
&=&\prod\limits_{k=1}^{j-1}e^{i\pi f_{k}^{\dagger }f_{k}} \\
&=&\prod\limits_{k=1}^{j-1}\left( 1-2f_{k}^{\dagger }f_{k}\right) \\
&=&\prod\limits_{k=1}^{j-1}\left( -\sigma ^{z}\right)
\end{eqnarray*}%
They differ from the Eq. (\ref{tempted}) only by a phase $\exp \left( \pm
i\pi \sum\limits_{k=1}^{j-1}f_{k}^{\dagger }f_{k}\right) $ with $%
k=1,2....j-1 $. The domain of $f_{k}^{\dagger }f_{k}\ \epsilon \ \left\{
0,1\right\} $. So that 
\begin{equation*}
\exp \left( \pm i\pi \sum\limits_{k=1}^{j-1}f_{k}^{\dagger }f_{k}\right)
=\prod\limits_{k=1}^{j-1}e^{i\pi f_{k}^{\dagger }f_{k}}=\left\{ 
\begin{array}{c}
1\text{ occupied }j\text{ is even} \\ 
-1\text{ occupied }j\text{ is odd}%
\end{array}%
\right.
\end{equation*}%
The number of occupied modes in $k=1,2....j-1$ determine the sign of $\exp
\left( \pm i\pi \sum\limits_{k=1}^{j-1}f_{k}^{\dagger }f_{k}\right) $. We
now have%
\begin{eqnarray*}
\left\{ a_{i}^{\dagger },a_{j}\right\} &=&a_{i}^{\dagger
}a_{j}+a_{j}a_{i}^{\dagger } \\
&=&\prod\limits_{k=1}^{i-1}\left( 1-2f_{k}^{\dagger }f_{k}\right) \cdot
f_{i}^{\dagger }\prod\limits_{k^{\prime }=1}^{j-1}\left( 1-2f_{k^{\prime
}}^{\dagger }f_{k^{\prime }}\right) \cdot f_{j} \\
&&+\prod\limits_{k^{\prime }=1}^{j-1}\left( 1-2f_{k^{\prime }}^{\dagger
}f_{k^{\prime }}\right) \cdot f_{j}\prod\limits_{k=1}^{i-1}\left(
1-2f_{k}^{\dagger }f_{k}\right) \cdot f_{i}^{\dagger }
\end{eqnarray*}%
which yields,%
\begin{eqnarray*}
&&\left( 1-2f_{k}^{\dagger }f_{k}\cdot f_{i}^{\dagger }\right) \left(
1-2f_{k^{\prime }}^{\dagger }f_{k^{\prime }}\cdot f_{j}\right) \\
&=&f_{i}^{\dagger }\cdot \left( 1-2f_{k}^{\dagger }f_{k}\right) \left(
1-2f_{k^{\prime }}^{\dagger }f_{k^{\prime }}\cdot f_{j}\right) \\
&=&f_{i}^{\dagger }\cdot \left( 1-2f_{k}^{\dagger }f_{k}-2f_{k^{\prime
}}^{\dagger }f_{k^{\prime }}+4f_{k}^{\dagger }f_{k}f_{k^{\prime }}^{\dagger
}f_{k^{\prime }}\right) \cdot f_{j}
\end{eqnarray*}%
The method of representing one type of operator (e.g. spin, boson, fermions)
in terms of another is one of the fundamental aspects of theoretical quantum
physics.

Morover, in one dimension, there is a close connection between the physics
of fermions, bosons, and spins which is lacking in higher dimensions. The
most important aspect of one dimensional physics that distinguishes it from
higher dimensional ones lies in particle statistics, where particle
exchanges are only possible if particle pass through each other (i.e.,
collide), which is not true in higher dimension. One dimensional systems are
now commonplace with nanofabricated materials, e.g. quantum nanowires,
nanotubes, and some organic compounds. The physics of one-dimensional
systems is important because some higher dimensional problems can be reduced
to one-dimensional, e.g. in Kondo effect the low-energy physics is basically
dealt with spherically symmetric $s$ channel so that the problem is
effectively radial and provides a solvable system of interacting quantum
problem.

The Jordan-Wigner (J-W) transformation is the simplest statistics-changing
transformation in one dimension. The $S_{i}^{z}$ essentially initiate the
J-W transformation\footnote{%
Just as $S_{i}^{z}$ initiate the Holstein-Primakoff transformation.} and is
local in the site index. We have%
\begin{equation*}
S_{i}^{z}=\frac{2c_{i}^{\dagger }c_{i}}{2}
\end{equation*}%
To effect \textit{self-induced} propagation, we need a nonlocal \textit{%
string} of lowering and raising operators, whereby spin at different sites
commute while fermions anticommute.

We use the following the following representation,%
\begin{eqnarray*}
S_{i}^{+} &=&c_{i}^{\dagger }\prod\limits_{j<i}\left( 1-2c_{j}^{\dagger
}c_{j}\right) \\
S_{i}^{-} &=&\prod\limits_{j<i}\left( 1-2c_{j}^{\dagger }c_{j}\right) c_{i}
\end{eqnarray*}%
By identifying%
\begin{equation*}
\vec{\sigma}=\vec{S}
\end{equation*}%
it is easy to see that%
\begin{eqnarray*}
\left[ \sigma _{i}^{+},\sigma _{j}^{-}\right] &=&\delta _{ij}\sigma _{i}^{z}
\\
\left[ \sigma _{i}^{z},\sigma _{j}^{\pm }\right] &=&\pm 2\delta _{ij}\sigma
_{i}^{\pm }
\end{eqnarray*}%
and ordinary fermionic commutation relations. The J-W transform works such
that the string is cooked up so that it changes sign from $+1$ to $-1$
depending on whether the number of fermions to the left of site $i$ is even
or odd.

J-W transformation is vital to numerical Monte Carlo simulation of $1$-D
systems because of the absence of 'fermion sign problem'. Quantum Monte
Carlo methods calculates an integral for a quantity like partition function
or correlation function by random sampling. The problem with fermions is
that because of the change of sign under exchange of any two wavefunctions,
series expansion must generate terms of both signs. This causes difficulty
since partial results fluctuate wildly when the actual quantity to be
calculated is much smaller in magnitude (as often the case). On the other
hand if the resulting values calculated by this sampling process are of the
same sign (so that the overall magnitude of the answer is necesarilly much
larger than of each individual term) then the truncation errors are much
less severe.

Other numerical methods like the density-matrix renomalization group (DMRG),
which is an extension of Wilson iterative numerical RG approach to $1$-D
chains, are also quite successful for low energy states. This DMRG method
has been able to calculate the Haldane gap in spin-$1$ chain to many decimal
places.

\subsubsection{J-W transformation to solve the $XX$ chain}

We now use the above J-W transformation to solve the so-called $XX$ chain.
This is like the Heisenberg model with no $z$ coupling:%
\begin{equation*}
H_{XX}=\frac{J}{4}\sum\limits_{i}\left( \sigma _{i}^{+}\sigma
_{i+1}^{-}+\sigma _{i}^{-}\sigma _{i+1}^{+}\right)
\end{equation*}%
Performing the J-W transformation, we have%
\begin{equation*}
H_{XX}=\frac{J}{4}\sum\limits_{i}\left( 
\begin{array}{c}
\left( c_{i}^{\dagger }\prod\limits_{j<i}\left( 1-2c_{j}^{\dagger
}c_{j}\right) \right) \left( \prod\limits_{j<i+1}\left( 1-2c_{j}^{\dagger
}c_{j}\right) c_{i+1}\right) \\ 
+\left( \prod\limits_{j<i}\left( 1-2c_{j}^{\dagger }c_{j}\right)
c_{i}\right) \left( c_{i+1}^{\dagger }\prod\limits_{j<i+1}\left(
1-2c_{j}^{\dagger }c_{j}\right) \right)%
\end{array}%
\right)
\end{equation*}

\subsection{The Bogoliubov-Valatin transformation for superfluidity}

Consider the Bogoliubov transformation. We have,%
\begin{equation*}
\Gamma =UA_{k}
\end{equation*}%
\begin{equation*}
U=\left( 
\begin{array}{cc}
u & -\nu \\ 
\nu & u%
\end{array}%
\right)
\end{equation*}%
where 
\begin{equation*}
\left\vert u\right\vert ^{2}+\left\vert \nu \right\vert ^{2}=1
\end{equation*}%
where%
\begin{equation*}
u^{\ast }\nu -\nu ^{\ast }u=0
\end{equation*}%
by orthogonality condition.

The unitary operator of transformation $S\left( r_{k}\right) $ can be
recognized to be the squeezing operator,%
\begin{equation*}
S\left( r\right) =\exp \left[ \sum\limits_{k\neq 0}\frac{r_{k}}{2}\left(
a_{k}^{\dagger }a_{-k}^{\dagger }+a_{-k}a_{k}\right) \right]
\end{equation*}%
where $r=\left\{ r_{k}\right\} $. Note: $H=\sum\limits_{k}\left[
E_{k}a_{k}^{\dagger }a_{k}+\frac{g_{k}}{2}\left( a_{k}a_{-k}+h.c.\right) %
\right] $ and 
\begin{eqnarray*}
\gamma _{k,\uparrow } &=&ua_{k,\uparrow }-va_{-k,\downarrow }^{\dagger } \\
\gamma _{-k,\downarrow } &=&ua_{-k,\downarrow }^{\dagger }+va_{k,\uparrow }
\end{eqnarray*}%
Consider the nonunitary tranformation, 
\begin{equation*}
e^{\alpha }a_{k}e^{-\beta }=e^{\alpha -\beta }a_{k}=\tilde{a}_{k}
\end{equation*}%
We take $e^{\alpha -\beta }=u$. Then we can have%
\begin{eqnarray*}
\alpha -\beta &=&\ln u\left( \cosh ^{2}x-\sinh ^{2}x\right) \\
&=&\ln u
\end{eqnarray*}%
\begin{eqnarray*}
\alpha &=&\ln u\cosh ^{2}x \\
\beta &=&\ln u\sinh ^{2}x
\end{eqnarray*}%
Let the first non-unitary transformation be 
\begin{eqnarray*}
\tilde{a}_{k,\uparrow } &=&ua_{k,\uparrow } \\
\tilde{a}_{-k,\downarrow }^{\dagger } &=&ua_{-k,\downarrow }^{\dagger }
\end{eqnarray*}%
after which we apply the second transformation: $S\tilde{a}_{k}S^{-1}=b_{k}$%
. This can either be made unitary by affixing $i$ or making $r_{k}$ a pure
imaginary in the exponent or non-unitary by just using real exponent but
canonical tranformation. Even with this two options, the overall
transformation is non-unitary because of the first transformation to 'tilde'
operators.

\begin{proof}
\begin{eqnarray*}
S\tilde{a}_{k}S^{-1} &=&b_{k} \\
&=&\exp \left[ \sum\limits_{k\neq 0}\frac{r_{k}}{2}a_{k}^{\dagger
}a_{-k}^{\dagger }+a_{-k}a_{k}\right] \tilde{a}_{k}\exp \left[
-\sum\limits_{k\neq 0}\frac{r_{k}}{2}a_{k}^{\dagger }a_{-k}^{\dagger
}+a_{-k}a_{k}\right]
\end{eqnarray*}%
To prove, we make use of the following well-known relations,%
\begin{eqnarray*}
\exp \left( \lambda A\right) B\exp \left( -\lambda A\right)
&=&\sum\limits_{n=0}^{\infty }\frac{\lambda ^{n}}{n!}\left\{ A^{n},B\right\}
\\
&=&\left[ A,.....\left[ A,\left[ A,B\right] \right] ....\right] ,\text{ \ }n%
\text{ commutations}
\end{eqnarray*}%
with special relation for $n=0$%
\begin{equation*}
\left\{ A^{0},B\right\} =B\text{.}
\end{equation*}%
If%
\begin{equation*}
\left[ A,B\right] =cI
\end{equation*}%
then 
\begin{equation*}
\exp \left( \lambda A\right) B\exp \left( -\lambda A\right) =B+\lambda \left[
A,B\right]
\end{equation*}%
and higher-order commutators are zero.%
\begin{equation*}
\left[ a,a^{\dagger }\right] =1\text{, }\left[ a,a\right] =\left[ a^{\dagger
},a^{\dagger }\right] =0\text{,}
\end{equation*}

Now consider the commutator%
\begin{eqnarray*}
&&\left[ \left( \sum\limits_{k\neq 0}\frac{r_{k}}{2}a_{k}^{\dagger
}a_{-k}^{\dagger }+a_{-k}a_{k}\right) ,\tilde{a}_{k^{\prime }}\right] \\
&=&\left( \sum\limits_{k\neq 0}\frac{r_{k}}{2}a_{k}^{\dagger
}a_{-k}^{\dagger }+a_{-k}a_{k}\right) \tilde{a}_{k^{\prime }}-\tilde{a}%
_{k^{\prime }}\left( \sum\limits_{k\neq 0}\frac{r_{k}}{2}a_{k}^{\dagger
}a_{-k}^{\dagger }+a_{-k}a_{k}\right)
\end{eqnarray*}%
We have%
\begin{eqnarray*}
\left( \sum\limits_{k\neq 0}\frac{r_{k}}{2}a_{k}^{\dagger }a_{-k}^{\dagger
}+a_{-k}a_{k}\right) \tilde{a}_{k^{\prime }} &=&\left( \sum\limits_{k\neq 0}%
\frac{r_{k}}{2}a_{k}^{\dagger }a_{-k}^{\dagger }\tilde{a}_{k^{\prime
}}+a_{-k}a_{k}\tilde{a}_{k^{\prime }}\right) \\
&=&\left( u\sum\limits_{k\neq 0}\frac{r_{k}}{2}a_{k}^{\dagger
}a_{-k}^{\dagger }a_{k^{\prime }}+a_{k^{\prime }}a_{-k}a_{k}\right) \\
&=&\left( u\sum\limits_{k\neq 0}\frac{r_{k}}{2}a_{k}^{\dagger }\left(
a_{k^{\prime }}a_{-k}^{\dagger }-\delta _{k+k^{\prime }}\right)
+a_{k^{\prime }}a_{-k}a_{k}\right) \\
&=&\left( \sum\limits_{k\neq 0}\frac{r_{k}}{2}\left( \left[ a_{k^{\prime
}}a_{k}^{\dagger }a_{-k}^{\dagger }-a_{-k}^{\dagger }\delta _{k-k^{\prime }}%
\right] -a_{k}^{\dagger }\delta _{k+k^{\prime }}\right) +a_{k^{\prime
}}a_{-k}a_{k}\right) \\
&=&\left( \sum\limits_{k\neq 0}\frac{r_{k}}{2}\left( a_{k^{\prime
}}a_{k}^{\dagger }a_{-k}^{\dagger }+a_{k^{\prime
}}a_{-k}a_{k}-a_{-k}^{\dagger }\delta _{k-k^{\prime }}-a_{k}^{\dagger
}\delta _{k+k^{\prime }}\right) \right)
\end{eqnarray*}%
Thus, with $r_{k}=r_{-k}$ 
\begin{eqnarray*}
\left( \sum\limits_{k\neq 0}\frac{r_{k}}{2}a_{k}^{\dagger }a_{-k}^{\dagger
}+a_{-k}a_{k}\right) a_{k^{\prime }} &=&a_{k^{\prime }}\left(
\sum\limits_{k\neq 0}\frac{r_{k}}{2}a_{k}^{\dagger }a_{-k}^{\dagger
}+a_{-k}a_{k}\right) -\left( \sum\limits_{k\neq 0}\frac{r_{k}}{2}\left(
a_{-k}^{\dagger }\delta _{k-k^{\prime }}+a_{k}^{\dagger }\delta
_{k+k^{\prime }}\right) \right) \\
&=&a_{k^{\prime }}\left( \sum\limits_{k\neq 0}\frac{r_{k}}{2}a_{k}^{\dagger
}a_{-k}^{\dagger }-a_{-k}a_{k}\right) -\left( \frac{r_{k^{\prime }}}{2}%
\left( a_{-k^{\prime }}^{\dagger }+a_{k^{\prime }}^{\dagger }\right) \right)
\end{eqnarray*}%
So%
\begin{eqnarray*}
\left[ \left( \sum\limits_{k\neq 0}\frac{r_{k}}{2}a_{k}^{\dagger
}a_{-k}^{\dagger }-a_{-k}a_{k}\right) ,a_{k^{\prime }}\right] &=&\left( -%
\frac{r_{k^{\prime }}}{2}\left( a_{-k^{\prime }}^{\dagger }+a_{-k^{\prime
}}^{\dagger }\right) \right) \\
&=&-ur_{k^{\prime }}a_{-k^{\prime }}^{\dagger }
\end{eqnarray*}%
and second commutator is 
\begin{equation*}
\left[ \left( \frac{r_{k^{\prime }}}{2}\left( a_{-k^{\prime }}^{\dagger
}+a_{-k^{\prime }}^{\dagger }\right) \right) ,a_{k^{\prime }}\right] =\left( 
\frac{r_{k^{\prime }}}{2}\left( a_{-k^{\prime }}^{\dagger }a_{k^{\prime
}}+a_{-k^{\prime }}^{\dagger }a_{k^{\prime }}\right) \right) -\left( \frac{%
r_{k^{\prime }}}{2}\left( a_{k^{\prime }}a_{-k^{\prime }}^{\dagger
}+a_{k^{\prime }}a_{-k^{\prime }}^{\dagger }\right) \right)
\end{equation*}%
\begin{equation*}
\left( \frac{r_{k^{\prime }}}{2}\left( a_{-k^{\prime }}^{\dagger
}a_{k^{\prime }}+a_{-k^{\prime }}^{\dagger }a_{k^{\prime }}\right) \right) =%
\frac{r_{k^{\prime }}}{2}\left( \left\{ a_{k^{\prime }}a_{-k^{\prime
}}^{\dagger }\right\} +a_{k^{\prime }}a_{-k^{\prime }}^{\dagger }\right)
\end{equation*}%
which yields%
\begin{equation*}
\left[ \left( \frac{r_{k^{\prime }}}{2}\left( a_{-k^{\prime }}^{\dagger
}+a_{-k^{\prime }}^{\dagger }\right) \right) ,a_{k^{\prime }}\right] =0
\end{equation*}%
Therefore%
\begin{eqnarray*}
S\tilde{a}_{k}S^{-1} &=&b_{k} \\
&=&\exp \left[ \sum\limits_{k\neq 0}\frac{r_{k}}{2}a_{k}^{\dagger
}a_{-k}^{\dagger }-a_{-k}a_{k}\right] a_{k}\exp \left[ -\sum\limits_{k\neq 0}%
\frac{r_{k}}{2}a_{k}^{\dagger }a_{-k}^{\dagger }-a_{-k}a_{k}\right] \\
b_{k} &=&a_{k}-r_{k^{\prime }}a_{-k}^{\dagger }
\end{eqnarray*}%
Similary, we have%
\begin{eqnarray*}
&&\left[ \left( \sum\limits_{k\neq 0}\frac{r_{k}}{2}a_{k}^{\dagger
}a_{-k}^{\dagger }+a_{-k}a_{k}\right) ,\tilde{a}_{-k^{\prime }}^{\dagger }%
\right] \\
&=&\left( u\sum\limits_{k\neq 0}\frac{r_{k}}{2}a_{k}^{\dagger
}a_{-k}^{\dagger }+a_{-k}a_{k}\right) a_{-k^{\prime }}^{\dagger
}-a_{-k^{\prime }}^{\dagger }\left( \sum\limits_{k\neq 0}\frac{r_{k}}{2}%
a_{k}^{\dagger }a_{-k}^{\dagger }+a_{-k}a_{k}\right)
\end{eqnarray*}%
Consider 
\begin{eqnarray*}
&&\left( \sum\limits_{k\neq 0}\frac{r_{k}}{2}\left( a_{k}^{\dagger
}a_{-k}^{\dagger }a_{-k^{\prime }}^{\dagger }+a_{-k}a_{k}a_{-k^{\prime
}}^{\dagger }\right) \right) \\
&=&\left( \sum\limits_{k\neq 0}\frac{r_{k}}{2}\left( a_{k}^{\dagger
}a_{-k}^{\dagger }a_{-k^{\prime }}^{\dagger }+a_{k}a_{-k}a_{-k^{\prime
}}^{\dagger }\right) \right) \\
&=&\left( \sum\limits_{k\neq 0}a_{-k^{\prime }}^{\dagger }\frac{r_{k}}{2}%
\left( a_{k}^{\dagger }a_{-k}^{\dagger }+a_{k}\left[ a_{-k^{\prime
}}^{\dagger }a_{-k}+\delta _{k-k^{\prime }}\right] \right) \right) \text{ }
\\
&=&\left( \sum\limits_{k\neq 0}a_{-k^{\prime }}^{\dagger }\frac{r_{k}}{2}%
\left( a_{k}^{\dagger }a_{-k}^{\dagger }+\left[ \left\{ a_{-k^{\prime
}}^{\dagger }a_{k}a_{-k}+\delta _{k+k^{\prime }}a_{-k^{\prime }}\right\}
+a_{k}\delta _{k-k^{\prime }}\right] \right) \right) \\
&=&a_{-k^{\prime }}^{\dagger }\sum\limits_{k\neq 0}\frac{r_{k}}{2}\left(
a_{k}^{\dagger }a_{-k}^{\dagger }+a_{-k}a_{k}\right) +\frac{r_{k}}{2}\left(
a_{k^{\prime }}+a_{k^{\prime }}\right)
\end{eqnarray*}%
Therefore%
\begin{eqnarray*}
\left[ \left( \sum\limits_{k\neq 0}\frac{r_{k}}{2}a_{k}^{\dagger
}a_{-k}^{\dagger }-a_{-k}a_{k}\right) ,a_{-k^{\prime }}^{\dagger }\right]
&=&ua_{-k^{\prime }}^{\dagger }+ur_{k^{\prime }}a_{k^{\prime }} \\
&=&b_{-k^{\prime }}^{\dagger }
\end{eqnarray*}%
Summarizing, we have%
\begin{eqnarray*}
b_{k} &=&\left( ua_{k}-ur_{k}a_{-k}^{\dagger }\right) \\
b_{-k}^{\dagger } &=&\left( ua_{-k}^{\dagger }+ur_{k}a_{k}\right)
\end{eqnarray*}%
or in matrix form%
\begin{equation*}
\left( 
\begin{array}{c}
b_{k} \\ 
b_{=k}^{\dagger }%
\end{array}%
\right) =\left( 
\begin{array}{cc}
u & -ur_{k} \\ 
ur_{k} & u%
\end{array}%
\right) \left( 
\begin{array}{c}
a_{k} \\ 
a_{-k}^{\dagger }%
\end{array}%
\right)
\end{equation*}%
Puting $r_{k}$ a pure imaginary and $u$ a $c$-number, i.e., $\left\{
r_{k},u\right\} \ \epsilon \ 
\mathbb{C}
$, we put%
\begin{equation*}
ur_{k}=u\frac{v}{u}=v\text{ is a pure imaginary}
\end{equation*}%
We can also put $\left\{ r_{k},u\right\} \ \epsilon \ 
\mathbb{R}
$ and the transformation can still be canonical. Therefore the whole
transformation idea is not unique and can at most only be partly unitary. We
take $v$ as pure imaginary 
\begin{eqnarray*}
\left( 
\begin{array}{c}
b_{k} \\ 
b_{-k}^{\dagger }%
\end{array}%
\right) &=&\left( 
\begin{array}{cc}
u & -v \\ 
-v^{\ast } & u^{\ast }%
\end{array}%
\right) \left( 
\begin{array}{c}
a_{k} \\ 
a_{-k}^{\dagger }%
\end{array}%
\right) \\
&=&\left( 
\begin{array}{cc}
u & -v \\ 
v & u^{\ast }%
\end{array}%
\right) \left( 
\begin{array}{c}
a_{k} \\ 
a_{-k}^{\dagger }%
\end{array}%
\right)
\end{eqnarray*}%
Taking the inverse, we have [diagonal of product $=$ determinant of original
matrix] 
\begin{eqnarray*}
&&\left( 
\begin{array}{cc}
u & -v \\ 
v & u%
\end{array}%
\right) \left( 
\begin{array}{cc}
u & v \\ 
-v & u%
\end{array}%
\right) \\
&=&\left( 
\begin{array}{cc}
\left\vert u\right\vert ^{2}+\left\vert v\right\vert ^{2} & \left(
uv-vu\right) \\ 
\left( vu-uv\right) & \left\vert u\right\vert ^{2}+\left\vert v\right\vert
^{2}%
\end{array}%
\right) \\
&=&\left( 
\begin{array}{cc}
1 & 0 \\ 
0 & 1%
\end{array}%
\right)
\end{eqnarray*}%
The make the whole transformation non-unitary. The condition 
\begin{equation*}
u^{2}+v^{2}=1
\end{equation*}%
makes the transformation canonical. Whereas, the condition $uv-vu=0$ occurs
by virtue of othogonality of the eigenfunctions. Thus, 
\begin{equation*}
\left( 
\begin{array}{c}
a_{k} \\ 
a_{-k}^{\dagger }%
\end{array}%
\right) =\left( 
\begin{array}{cc}
u & v \\ 
-v & u%
\end{array}%
\right) \left( 
\begin{array}{c}
b_{k} \\ 
b_{-k}^{\dagger }%
\end{array}%
\right)
\end{equation*}
\end{proof}

\subsubsection{Check for microcausality}

\begin{equation*}
\left( 
\begin{array}{c}
b_{k} \\ 
b_{-k}^{\dagger }%
\end{array}%
\right) =\left( 
\begin{array}{cc}
u & -v \\ 
v & u^{\ast }%
\end{array}%
\right) \left( 
\begin{array}{c}
a_{k} \\ 
a_{-k}^{\dagger }%
\end{array}%
\right)
\end{equation*}%
\begin{eqnarray*}
\left[ b_{k},b_{k}^{\dagger }\right] &=&1 \\
&&u^{2}\left[ \left( a_{k}-r_{k}a_{-k}^{\dagger }\right) ,\left(
a_{k}^{\dagger }+r_{k}a_{-k}\right) \right] \\
&=&\left( a_{k}-r_{k}a_{-k}^{\dagger }\right) \left( a_{k}^{\dagger
}+r_{k}a_{-k}\right) -\left( a_{k}^{\dagger }+r_{k}a_{-k}\right) \left(
a_{k}-r_{k}a_{-k}^{\dagger }\right) \\
&=&\left( a_{k}\right) \left( a_{k}^{\dagger }\right) -\left(
r_{k}^{2}a_{-k}^{\dagger }\right) \left( a_{-k}\right) -\left(
r_{k}a_{-k}^{\dagger }\right) \left( a_{k}^{\dagger }\right) +r_{k}\left(
a_{k}\right) \left( a_{-k}\right) \\
&&-a_{k}^{\dagger }a_{k}+\left( r_{k}^{2}a_{-k}\right) \left(
a_{-k}^{\dagger }\right) +r_{k}a_{k}^{\dagger }\left( a_{-k}^{\dagger
}\right) -r_{k}\left( a_{-k}\right) a_{k} \\
&=&u^{2}\left[ a_{k},a_{k}^{\dagger }\right] +u^{2}r_{k}^{2}\left[
a_{-k},a_{-k}^{\dagger }\right] =u^{2}+u^{2}r_{k}^{2}\text{, } \\
&=&\text{ }u_{k}^{2}+u_{k}^{2}r_{k}^{2}\text{, let }r_{k}^{2}=\frac{v_{k}^{2}%
}{u_{k}^{2}}
\end{eqnarray*}%
Hyperbolic Functions

\begin{eqnarray*}
&\Longrightarrow &u^{2}+v_{k}^{2}=1=\cosh ^{2}x_{k}-\sinh ^{2}x_{k}\text{
for }v_{k}\ \epsilon \ 
\mathbb{C}
_{imaginary} \\
&=&\cos ^{2}x+\sin ^{2}x\text{ for }v_{k}\ \epsilon \ 
\mathbb{R}
\\
\left[ b_{k},b_{k}\right] &=&0 \\
\left[ b_{k}^{\dagger },b_{k}^{\dagger }\right] &=&0
\end{eqnarray*}%
\begin{eqnarray*}
\tanh \left( 2r_{k}\right) &=&\frac{g_{k}}{E_{k}} \\
\frac{\cosh \left( 2r_{k}\right) }{\sinh \left( 2r_{k}\right) } &=&\frac{%
g_{k}}{E_{k}}
\end{eqnarray*}%
\begin{equation*}
E_{k}\cosh \left( 2r_{k}\right) -g_{k}\sinh \left( 2r_{k}\right) =0
\end{equation*}%
Note that the low-energy excitation modes in the superfluid are found to be
proportional to the $k$ exhibiting the property of sound waves. In this case
Bogoliubov transformation is not defined for $k=0.$ This infrared divergence
is identified by $r_{o}\sim O\left( \ln N\right) $ so that it is divergent
actually only in the thermodynamic limit.

\section{Decoupling degrees of freedom in CMP and QFT}

Another important transformations which has to do with separating degrees of
freedom are the so-called decoupling transformations. We will only present
here the basic principle which goes back to Foldy and Woutheysen \cite%
{foldy,dirac,silinco}, decoupling positive and negative energy states of the
relativistic Dirac electrons. The technique is an essentially iterative
procedure to all orders in decoupling by adding to the transformed
Hamiltonian at each order of the calculation a \textit{suitable} decoupling
perturbation. Similar technique has been employed in the calculation of the
magnetic susceptibility of interacting Bloch fermions in solids \cite{many}.
More recently, this technique appears in somewhat similar form in the
literature which address suppressing decoherence in quantum information
systems \cite{seth}.

\subsection{Decoupling transformation for Dirac energy bands}

The decoupling of energy bands of $4\times 4$ Hamiltonian of the
relativistic Dirac equation was initiated by Foldy-Woutheysen
transformation. This work perhaps marks the birth of dressing transformation
in quantum physics, i.e., diagonalization or removal of off-diagonal terms
to arbitrary order.

The Dirac equation is given by,%
\begin{equation}
i\frac{\partial \Psi }{\partial t}=\left( \beta m+\vec{\alpha}\cdot \vec{p}%
\right) \Psi  \label{eq1}
\end{equation}%
where $p$ is the momentum operator, $\alpha $ and $\beta $ are the
well-known Dirac matrices (in the usual representation with $\beta $%
diagonal) and units in which $\hbar =c=1$. Restoring dimensional units of $c$
and $\hbar $ we have%
\begin{equation*}
i\hbar \frac{\partial \Psi }{\partial t}=\left( \beta mc^{2}+c\vec{\alpha}%
\cdot \vec{p}\right) \Psi
\end{equation*}%
where%
\begin{eqnarray*}
\beta &=&\left( 
\begin{array}{cc}
I_{2} & 0 \\ 
0 & -I_{2}%
\end{array}%
\right) =\gamma ^{o}\text{, 'even' since diagonal only} \\
\alpha _{\mu } &=&\left( 
\begin{array}{cc}
0 & \sigma _{\mu } \\ 
\sigma _{\mu } & 0%
\end{array}%
\right) \text{, 'odd' since diagonal is zero}
\end{eqnarray*}%
Notice:%
\begin{eqnarray*}
\frac{1+\beta }{2} &=&\frac{1}{2}\left( 
\begin{array}{cc}
I_{2} & 0 \\ 
0 & I_{2}%
\end{array}%
\right) +\frac{1}{2}\left( 
\begin{array}{cc}
I_{2} & 0 \\ 
0 & -I_{2}%
\end{array}%
\right) \\
&=&\left( 
\begin{array}{cc}
I_{2} & 0 \\ 
0 & 0%
\end{array}%
\right) \left( 
\begin{array}{c}
\Phi ^{\prime } \\ 
X^{\prime }%
\end{array}%
\right) =\left( 
\begin{array}{c}
\Phi ^{\prime } \\ 
0%
\end{array}%
\right) \\
\frac{1-\beta }{2} &=&\frac{1}{2}\left( 
\begin{array}{cc}
I_{2} & 0 \\ 
0 & I_{2}%
\end{array}%
\right) -\frac{1}{2}\left( 
\begin{array}{cc}
I_{2} & 0 \\ 
0 & -I_{2}%
\end{array}%
\right) \\
&=&\left( 
\begin{array}{cc}
0 & 0 \\ 
0 & I_{2}%
\end{array}%
\right) \left( 
\begin{array}{c}
\Phi ^{\prime } \\ 
X^{\prime }%
\end{array}%
\right) =\left( 
\begin{array}{c}
0 \\ 
X^{\prime }%
\end{array}%
\right)
\end{eqnarray*}%
Therefore%
\begin{equation*}
\frac{1+\beta }{2}\Psi +\frac{1+\beta }{2}\Psi =\Psi
\end{equation*}%
Therefore%
\begin{eqnarray*}
i\frac{\partial }{\partial t}\Psi &=&H^{\prime }\Psi \\
i\frac{\partial }{\partial t}\left( 
\begin{array}{c}
\Phi ^{\prime } \\ 
X^{\prime }%
\end{array}%
\right) &=&\sqrt{p^{2}+m^{2}}%
\begin{array}{cc}
I_{2} & 0 \\ 
0 & -I_{2}%
\end{array}%
\left( 
\begin{array}{c}
\Phi ^{\prime } \\ 
X^{\prime }%
\end{array}%
\right) \\
&=&\left( 
\begin{array}{c}
\sqrt{p^{2}+m^{2}}\Phi ^{\prime } \\ 
-\sqrt{p^{2}+m^{2}}X^{\prime }%
\end{array}%
\right) \text{, this is a decoupled bands!}
\end{eqnarray*}%
or%
\begin{eqnarray*}
i\frac{\partial }{\partial t}\Phi ^{\prime } &=&\sqrt{p^{2}+m^{2}}\Phi
^{\prime } \\
i\frac{\partial }{\partial t}X^{\prime } &=&-\sqrt{p^{2}+m^{2}}X^{\prime }
\end{eqnarray*}%
\begin{eqnarray*}
\beta \alpha _{\mu } &=&\left( 
\begin{array}{cc}
I_{2} & 0 \\ 
0 & -I_{2}%
\end{array}%
\right) \left( 
\begin{array}{cc}
0 & \sigma _{\mu } \\ 
\sigma _{\mu } & 0%
\end{array}%
\right) =\left( 
\begin{array}{cc}
0 & \sigma _{\mu } \\ 
-\sigma _{\mu } & 0%
\end{array}%
\right) \\
\alpha _{\mu }\beta &=&\left( 
\begin{array}{cc}
0 & \sigma _{\mu } \\ 
\sigma _{\mu } & 0%
\end{array}%
\right) \left( 
\begin{array}{cc}
I_{2} & 0 \\ 
0 & -I_{2}%
\end{array}%
\right) =\left( 
\begin{array}{cc}
0 & -\sigma _{\mu } \\ 
\sigma _{\mu } & 0%
\end{array}%
\right) \\
\beta \alpha _{\mu }\beta &=&\left( 
\begin{array}{cc}
0 & \sigma _{\mu } \\ 
-\sigma _{\mu } & 0%
\end{array}%
\right) \left( 
\begin{array}{cc}
I_{2} & 0 \\ 
0 & -I_{2}%
\end{array}%
\right) =\left( 
\begin{array}{cc}
0 & -\sigma _{\mu } \\ 
-\sigma _{\mu } & 0%
\end{array}%
\right) =-\alpha _{\mu } \\
\left\{ \alpha _{\mu },\beta \right\} &=&0
\end{eqnarray*}%
\begin{eqnarray*}
\beta ^{2} &=&\left( 
\begin{array}{cc}
I_{2} & 0 \\ 
0 & -I_{2}%
\end{array}%
\right) \left( 
\begin{array}{cc}
I_{2} & 0 \\ 
0 & -I_{2}%
\end{array}%
\right) =I \\
\vec{\alpha}\cdot \vec{\alpha} &=&\sum\limits_{\mu }\left( 
\begin{array}{cc}
0 & \sigma _{\mu } \\ 
\sigma _{\mu } & 0%
\end{array}%
\right) \left( 
\begin{array}{cc}
0 & \sigma _{\mu } \\ 
\sigma _{\mu } & 0%
\end{array}%
\right) =\sum\limits_{\mu }\left( 
\begin{array}{cc}
\sigma _{\mu }^{2} & 0 \\ 
0 & \sigma _{\mu }^{2}%
\end{array}%
\right) =\sum\limits_{\mu =1}^{3}\left( 
\begin{array}{cc}
I_{2} & 0 \\ 
0 & I_{2}%
\end{array}%
\right)
\end{eqnarray*}%
\begin{eqnarray*}
\sigma _{x}^{2} &=&\left( 
\begin{array}{cc}
0 & 1 \\ 
1 & 0%
\end{array}%
\right) \left( 
\begin{array}{cc}
0 & 1 \\ 
1 & 0%
\end{array}%
\right) =\left( 
\begin{array}{cc}
1 & 0 \\ 
0 & 1%
\end{array}%
\right) \\
\sigma _{y}^{2} &=&\left( 
\begin{array}{cc}
0 & -i \\ 
i & 0%
\end{array}%
\right) \left( 
\begin{array}{cc}
0 & -i \\ 
i & 0%
\end{array}%
\right) =\left( 
\begin{array}{cc}
1 & 0 \\ 
0 & 1%
\end{array}%
\right) \\
\sigma _{z}^{2} &=&\left( 
\begin{array}{cc}
1 & 0 \\ 
0 & -1%
\end{array}%
\right) \left( 
\begin{array}{cc}
1 & 0 \\ 
0 & -1%
\end{array}%
\right) =\left( 
\begin{array}{cc}
1 & 0 \\ 
0 & 1%
\end{array}%
\right)
\end{eqnarray*}%
\begin{eqnarray*}
\sigma _{x}\sigma _{y} &=&i\sigma _{z} \\
\sigma _{y}\sigma _{z} &=&i\sigma _{x} \\
\sigma _{x}\sigma _{z} &=&-i\sigma _{y}
\end{eqnarray*}%
Thus%
\begin{equation*}
H_{Dirac}=\left( 
\begin{array}{cc}
mc^{2} & c\vec{p}\cdot \vec{\sigma} \\ 
c\vec{p}\cdot \vec{\sigma} & -mc^{2}%
\end{array}%
\right)
\end{equation*}

The eigenfunction obeys%
\begin{eqnarray*}
i\frac{\partial \psi }{\partial t} &=&\left( \beta m+\vec{\alpha}\cdot \vec{p%
}\right) \psi \\
&=&E\psi
\end{eqnarray*}%
The eigenfunctions are of the the form of Bloch function$\sim u\left( \vec{p}%
\right) e^{-\vec{p}.\vec{x}}$. For each value of $\vec{p}$ there are $4$
linearly independent spinors $u\left( \vec{p}\right) $%
\begin{equation*}
u\left( p\right) =\left( 
\begin{array}{c}
u_{+\uparrow } \\ 
u_{+\downarrow } \\ 
u_{-\uparrow } \\ 
u_{-\downarrow }%
\end{array}%
\right)
\end{equation*}%
corresponding to the two energy eigenvalues $\pm \sqrt{m^{2}+p^{2}}$ and two
eigenvalues $\pm 1$ for the $z$-component related to spin. Equation (\ref%
{eq1}) contains odd operators, specifically the components of $\alpha .$ It
is possible to perform canonical transformation which remove all odd
operators. This is the Foldy-Woutheysen transformation.

If $S$ is an Hermitian operator, then%
\begin{eqnarray*}
\Psi ^{\prime } &=&e^{iS}\Psi \\
H^{\prime } &=&e^{iS}He^{-iS}-e^{iS}\left( i\frac{\partial }{\partial t}%
\right) e^{-iS}
\end{eqnarray*}%
leaves%
\begin{eqnarray*}
\left( i\frac{\partial }{\partial t}\right) e^{-iS}e^{iS}\Psi
&=&He^{-iS}e^{iS}\Psi \\
\left( i\frac{\partial }{\partial t}\right) e^{-iS}e^{iS}\Psi
&=&He^{-iS}e^{iS}\Psi \\
\left[ e^{iS}\left( i\frac{\partial }{\partial t}\right) e^{-iS}\right]
e^{iS}\Psi +\left( i\frac{\partial }{\partial t}\right) e^{iS}\Psi &=&\left(
e^{iS}He^{-iS}\right) e^{iS}\Psi \\
\left( i\frac{\partial }{\partial t}\right) e^{iS}\Psi &=&\left[ \left(
e^{iS}He^{-iS}\right) -e^{iS}\left( i\frac{\partial }{\partial t}\right)
e^{-iS}\right] e^{iS}\Psi \\
\left( i\frac{\partial }{\partial t}\right) \Psi ^{\prime } &=&H^{\prime
}\Psi ^{\prime }
\end{eqnarray*}%
Let $S$ be a non-explicity time-dependent operator $\left[ \left( \beta m+%
\vec{\alpha}\cdot \vec{p}\right) \right] $: $i\frac{\partial \Psi }{\partial
t}=\left( \beta m+\vec{\alpha}\cdot \vec{p}\right) \Psi $ 
\begin{equation*}
S=-i\left( \frac{1}{2m}\right) \beta \vec{\alpha}\cdot \vec{p}\ \omega
\left( \frac{\vec{p}}{m}\right)
\end{equation*}%
where the function\footnote{$Interpretation:$%
\par
$T\left( \vec{p}\right) \Psi \left( x\right) =\frac{1}{\left( 2\pi \hbar
\right) ^{3}}\iint d\vec{p}d\vec{x}T\left( p^{\prime }\right) \exp \left[ 
\frac{i}{\hbar }p^{\prime }\cdot \left( x-x^{\prime }\right) \Psi \left(
x^{\prime }\right) \right] $} $\omega \left( \frac{\vec{p}}{m}\right) $ is
to be determined such that $H^{\prime }$ is free of odd operators.

If $\left\{ H,S\right\} =0$, then%
\begin{eqnarray*}
H^{\prime } &=&e^{iS}He^{-iS}=e^{2iS}H \\
&=&e^{2iS}H=\cos \left( 2S+i\sin 2S\right)
\end{eqnarray*}%
Since 
\begin{eqnarray*}
2iS &=&\left( \frac{1}{m}\right) \beta \vec{\alpha}\cdot \vec{p}\ \omega
\left( \frac{\vec{p}}{m}\right) =\Omega \ \beta \vec{\alpha}\cdot \vec{p} \\
\Omega &=&\left[ \left( \frac{1}{m}\right) \omega \left( \frac{\vec{p}}{m}%
\right) \right]
\end{eqnarray*}%
\begin{eqnarray*}
H^{\prime } &=&\exp \left( \left( \frac{1}{m}\right) \beta \vec{\alpha}\cdot 
\vec{p}\ \omega \left( \frac{\vec{p}}{m}\right) \right) H \\
&=&\left\{ 1+\Omega \beta \vec{\alpha}\cdot \vec{p}-\Omega ^{2}\frac{\left( 
\vec{\alpha}\cdot \vec{p}\right) ^{2}}{2}-\frac{\Omega ^{3}}{3!}\left( \vec{%
\alpha}\cdot \vec{p}\right) ^{2}\left( \beta \vec{\alpha}\cdot \vec{p}%
\right) +\frac{\Omega ^{4}\left( \vec{\alpha}\cdot \vec{p}\right) ^{4}}{4!}+%
\frac{\Omega ^{5}\left( \vec{\alpha}\cdot \vec{p}\right) ^{4}}{5!}\left(
\beta \vec{\alpha}\cdot \vec{p}\right) -..\right\} H \\
&=&\left[ 1-\Omega ^{2}\frac{\left( \vec{\alpha}\cdot \vec{p}\right) ^{2}}{2}%
+\frac{\Omega ^{4}\left( \vec{\alpha}\cdot \vec{p}\right) ^{4}}{4!}-\frac{%
\Omega ^{6}\left( \vec{\alpha}\cdot \vec{p}\right) ^{6}}{6!}+..\right] H \\
&&+\left[ \Omega \beta \vec{\alpha}\cdot \vec{p}-\frac{\Omega ^{3}}{3!}%
\left( \vec{\alpha}\cdot \vec{p}\right) ^{2}\left( \beta \vec{\alpha}\cdot 
\vec{p}\right) +\frac{\Omega ^{5}\left( \vec{\alpha}\cdot \vec{p}\right) ^{4}%
}{5!}\left( \beta \vec{\alpha}\cdot \vec{p}\right) -\frac{\Omega ^{7}\left( 
\vec{\alpha}\cdot \vec{p}\right) ^{6}}{6!}\left( \beta \vec{\alpha}\cdot 
\vec{p}\right) +..\right] H \\
&=&\left\{ Cos\left( p\Omega \right) +\sin \left( p\Omega \right) \frac{%
\beta \vec{\alpha}\cdot \vec{p}}{p}\right\} H
\end{eqnarray*}%
\begin{eqnarray*}
\left\{ Cos\left( p\Omega \right) \beta m+\sin \left( p\Omega \right) \frac{%
\beta \vec{\alpha}\cdot \vec{p}}{p}\beta m\right\} &=&\left\{ Cos\left(
p\Omega \right) \beta m-\sin \left( p\Omega \right) \frac{\vec{\alpha}\cdot 
\vec{p}}{p}m\right\} \\
\left\{ Cos\left( p\Omega \right) \vec{\alpha}\cdot \vec{p}+\sin \left(
p\Omega \right) \frac{\beta \vec{\alpha}\cdot \vec{p}}{p}\vec{\alpha}\cdot 
\vec{p}\right\} &=&\left\{ pCos\left( p\Omega \right) \vec{\alpha}\cdot \vec{%
p}+\sin \left( p\Omega \right) \beta p\right\}
\end{eqnarray*}%
Grouping, we obtained by separating the odd terms, $\vec{\alpha}\cdot \vec{p}%
,$ 
\begin{equation*}
H^{\prime }=\beta \left\{ m\cos \left( p\Omega \right) +p\sin \left( p\Omega
\right) \right\} +\frac{\vec{\alpha}\cdot \vec{p}}{p}\left[ pCos\left(
p\Omega \right) -m\sin \left( p\Omega \right) \right]
\end{equation*}%
Therefore free of odd terms if 
\begin{equation*}
\left[ p\cos \left( p\Omega \right) -m\sin \left( p\Omega \right) \right] =0
\end{equation*}%
\begin{eqnarray*}
\frac{p}{m} &=&\frac{\sin \left( p\Omega \right) }{\cos \left( p\Omega
\right) }=\tan ^{-1}\left( p\Omega \right) \\
&=&\tan ^{-1}\left( \frac{p}{m}\omega \left( \frac{\vec{p}}{m}\right) \right)
\end{eqnarray*}%
\begin{eqnarray*}
H^{\prime } &=&\beta \left\{ m\cos \left( p\Omega \right) +p\sin \left(
p\Omega \right) \right\} \\
&=&\beta \left\{ m\frac{m}{\sqrt{p^{2}+m^{2}}}+p\frac{p}{\sqrt{p^{2}+m^{2}}}%
\right\} \\
&=&\beta \sqrt{p^{2}+m^{2}}=\beta E_{p}
\end{eqnarray*}%
\begin{equation*}
\cos \left( p\Omega \right) =\frac{m}{\sqrt{p^{2}+m^{2}}}
\end{equation*}

Appendix

Now the powers of 
\begin{eqnarray*}
\left( \beta \vec{\alpha}\cdot \vec{p}\right) ^{2} &=&\left( \beta \vec{%
\alpha}\cdot \vec{p}\right) \left( \beta \vec{\alpha}\cdot \vec{p}\right)
=\left( \beta \vec{\alpha}\beta \cdot \vec{p}\right) \left( \vec{\alpha}%
\cdot \vec{p}\right) =-\left( \vec{\alpha}\cdot \vec{p}\right) ^{2} \\
\left( \beta \vec{\alpha}\cdot \vec{p}\right) ^{3} &=&\left( \beta \vec{%
\alpha}\cdot \vec{p}\right) \left( \beta \vec{\alpha}\cdot \vec{p}\right)
\left( \beta \vec{\alpha}\cdot \vec{p}\right) =-\left( \vec{\alpha}\cdot 
\vec{p}\right) ^{2}\left( \beta \vec{\alpha}\cdot \vec{p}\right) \\
\left( \beta \vec{\alpha}\cdot \vec{p}\right) ^{4} &=&\left( \beta \vec{%
\alpha}\cdot \vec{p}\right) \left( \beta \vec{\alpha}\cdot \vec{p}\right)
\left( \beta \vec{\alpha}\cdot \vec{p}\right) \left( \beta \vec{\alpha}\cdot 
\vec{p}\right) =\left( \vec{\alpha}\cdot \vec{p}\right) ^{4} \\
\left( \beta \vec{\alpha}\cdot \vec{p}\right) ^{5} &=&\left( \vec{\alpha}%
\cdot \vec{p}\right) ^{4}\left( \beta \vec{\alpha}\cdot \vec{p}\right) \\
\left( \beta \vec{\alpha}\cdot \vec{p}\right) ^{6} &=&\left( \vec{\alpha}%
\cdot \vec{p}\right) ^{4}\left( \beta \vec{\alpha}\cdot \vec{p}\right)
^{2}=-\left( \vec{\alpha}\cdot \vec{p}\right) ^{6} \\
\left( \beta \vec{\alpha}\cdot \vec{p}\right) ^{2n} &=&\left( -1\right) ^{%
\frac{2n}{2}}\left( \vec{\alpha}\cdot \vec{p}\right) ^{2n} \\
\left( \beta \vec{\alpha}\cdot \vec{p}\right) ^{2n+1} &=&\left( -1\right) ^{%
\frac{2n}{2}}\left( \vec{\alpha}\cdot \vec{p}\right) ^{2n}\left( \beta \vec{%
\alpha}\cdot \vec{p}\right)
\end{eqnarray*}

We have to show that indeed, $\left\{ H,S\right\} =0.$%
\begin{eqnarray*}
\left\{ H,S\right\} &=&HS+SH \\
&=&\left( \beta m+\vec{\alpha}\cdot \vec{p}\right) \left[ -i\left( \frac{1}{%
2m}\right) \beta \vec{\alpha}\cdot \vec{p}\ \omega \left( \frac{\vec{p}}{m}%
\right) \right] \\
&&+\left[ -i\left( \frac{1}{2m}\right) \beta \vec{\alpha}\cdot \vec{p}\
\omega \left( \frac{\vec{p}}{m}\right) \right] \left( \beta m+\vec{\alpha}%
\cdot \vec{p}\right) \\
&=&\left\{ 
\begin{array}{c}
\left[ -i\left( \frac{\beta }{2}\right) \beta \vec{\alpha}\cdot \vec{p}\
\omega \left( \frac{\vec{p}}{m}\right) \right] \\ 
+\left[ -i\left( \frac{\vec{\alpha}\cdot \vec{p}\beta }{2m}\right) \vec{%
\alpha}\cdot \vec{p}\ \omega \left( \frac{\vec{p}}{m}\right) \right]%
\end{array}%
\right\} \\
&&+\left\{ 
\begin{array}{c}
\left[ -i\left( \frac{\beta }{2}\right) \vec{\alpha}\cdot \vec{p}\ \beta
\omega \left( \frac{\vec{p}}{m}\right) \right] \\ 
+\left[ -i\left( \frac{\beta \vec{\alpha}\cdot \vec{p}}{2m}\right) \ \vec{%
\alpha}\cdot \vec{p}\omega \left( \frac{\vec{p}}{m}\right) \right]%
\end{array}%
\right\} \\
&=&0\text{, since }\left\{ \beta \vec{\alpha}\cdot \vec{p}+\vec{\alpha}\cdot 
\vec{p}\ \beta \right\} =0
\end{eqnarray*}

\subsection{\protect\normalsize Effective band theory of magnetic
susceptibility of relativistic Dirac fermions}

{\normalsize In this section, we formulate the magnetic susceptibility of
relativistic Dirac fermions analogous to energy-band dynamics of crystalline
solids. The Hamiltonian of free relativistic Dirac fermions is of the form 
\begin{equation}
\mathcal{H}=\beta \Delta +c\vec{\alpha}\cdot \vec{P}.  \label{eq4}
\end{equation}%
We designate quantum operators in capital letters and their corresponding
eigenvalues in small letters. The equation for the eigenfunctions and
eigenvalues is 
\begin{equation}
\mathcal{H}b_{\lambda }(x,p)=E_{\lambda }(p)b_{\lambda }(x,p),  \label{eq7}
\end{equation}%
where $E_{\lambda }(p)=\pm E(p)$, and $E_{\lambda }(\vec{q}^{\ \prime }-q)=%
\frac{1}{(2\pi \hbar )^{3}}\int d\vec{p}\ e^{(\frac{i}{\hbar })\vec{p}\cdot (%
\vec{q}^{\ \prime }-q)}E_{\lambda }(p)$, $\lambda $ labels the band index: $%
\pm $ spin band for positive energy states and $\pm $ spin band for negative
energy states. 
\begin{equation*}
E_{\lambda }(p)=\pm \sqrt{(cp)^{2}+(mc^{2})^{2}}.
\end{equation*}%
The doubly degenerate bands is reminiscent of the Kramer conjugates in
bismuth and Bi-Sb alloys. The localized function $a_{\lambda }(\vec{x}-\vec{q%
}^{\prime })$ is the `Wannier function' for relativistic Dirac fermions,
defined below. }

{\normalsize In the absence of magnetic field we may define the Wannier
function and Bloch function of a relativistic Dirac fermions as 
\begin{align*}
b_{\lambda}(x,p) & = \frac{1}{(2 \pi \hbar)^{\frac{3}{2}}} e^{(\frac{i}{\hbar%
}) \vec{p} \cdot \vec{x}} \ u_{\lambda}(\vec{p}), \\
a_{\lambda}(\vec{x} - \vec{q}) & = \frac{1}{(2\pi \hbar)^{\frac{3}{2}}} \int
d \vec{p} \ e^{(\frac{i}{\hbar}) \vec{p} \cdot \vec{q}} \ b_{\lambda}(x,p),
\end{align*}
where $b_{\lambda}(x,p)$ is the Bloch function, and $a_{\lambda (\vec{x} - 
\vec{q})}$ the corresponding Wannier function. $u_{\lambda}(\vec{p})$ is a
four-component function. The $u_{\lambda }(\vec{p})$'s are related to the $%
u_{\lambda}(0)$'s by a unitary transformation, $S$, which also transforms
the Dirac Hamiltonian into an \textit{even} form, i.e., no longer have
interband terms or the negative and positive energy states are decoupled.
This is equivalent to the transformation from Kohn-Luttinger basis to Bloch
functions in $\vec{k} \cdot \vec{p}$ theory. We have 
\begin{equation*}
S = \frac{E + \beta \mathcal{H}}{\sqrt{2E(E + \Delta)}},
\end{equation*}
which can be written in matrix form as 
\begin{equation*}
S = 
\begin{pmatrix}
\sqrt{\frac{(E + \Delta)}{2E}} & \frac{c\{\vec{\sigma} \cdot \vec{p}
\}^{\ast}}{\sqrt{2E(E + \Delta)}} \\ 
-\frac{c\{\vec{\sigma} \cdot \vec{p} \}^{\ast}}{\sqrt{2E(E + \Delta)}} & 
\sqrt{\frac{(E + \Delta)}{2E}}%
\end{pmatrix}
,
\end{equation*}
where the entries are $2 \times 2$ matrices, $\Delta = mc^{2}$, and all
matrix elements may be viewed as matrix elements of $S$ between the $%
u_{\lambda}(0)$'s, which are the spin functions in the Pauli representation.
The transformed Hamiltonian is 
\begin{equation}
\mathcal{H} = S \mathcal{H} S^{\dagger} = \beta E \Big( \vec{P} \Big).
\label{eq5}
\end{equation}
The $a_{\lambda}(\vec{x} - \vec{q})$ is not a $\delta$-function because of
the dependence of $u_{\lambda}(\vec{p})$ on $\vec{p}$; it is spread out over
a region of the order of the Compton wavelength, $\frac{\hbar}{mc}$, of the
electron and no smaller, as pointed out first by Newton and Wigner \cite%
{newtonwigner}, Foldy and Wouthuysen\cite{foldy} and by Blount. \cite{blount}
}

{\normalsize The Weyl correspondence for the momentum and coordinate
operator giving the correct dynamics of quasiparticles is given by the
prescription that the momentum operator $\vec{P}$ and coordinate operator $%
\vec{Q}$ be defined with the aid of the Wannier function and the Bloch
function as 
\begin{align*}
\vec{P}b_{\lambda }(x,p)& =\vec{p}b_{\lambda }(x,p), \\
\vec{Q}a_{\lambda }(\vec{x}-\vec{q})& =\vec{q}a_{\lambda }(\vec{x}-\vec{q}),
\end{align*}%
and the uncertainty relation follows in the formalism, 
\begin{equation*}
\lbrack Q_{i},P_{j}]=i\hbar \delta _{ij}.
\end{equation*}%
From Eq. (\ref{eq7}), we have 
\begin{align*}
\frac{1}{(2\pi \hbar )^{\frac{3}{2}}}\int dq\ e^{(-\frac{i}{\hbar })\vec{p}%
\cdot \vec{q}}\mathcal{H}a_{\lambda }(\vec{x}-\vec{q})& =E_{\lambda }(p)%
\frac{1}{(2\pi \hbar )^{\frac{3}{2}}}\int dq\ e^{(-\frac{i}{\hbar })\vec{p}%
\cdot \vec{q}}a_{\lambda }(\vec{x}-\vec{q}), \\
\mathcal{H}a_{\lambda }(\vec{x}-\vec{q}^{\ \prime })& =\int dq\ E_{\lambda }(%
\vec{q}^{\ \prime }-q)a_{\lambda }(\vec{x}-\vec{q}).
\end{align*}%
These relations allows us to transform the `bare' Hamiltonian operator to an
`effective Hamiltonian' expressed in terms of the $\vec{P}$ operator and the 
$\vec{Q}$ operator. This is conveniently done by the use of the `lattice'
Weyl transform (`lattice' Weyl transform and Weyl transform will be used
interchangeably for infinite translationally invariant system including
crystaline solids). Thus, any operator $A(\vec{P},\vec{Q})$ which is a
function of $\vec{P}$ and $\vec{Q}$ can be obtained from the matrix elements
of the `bare' operator, $A_{op}^{b}$, between the Wannier functions or
between the Bloch functions as, 
\begin{align*}
A(\vec{P},\vec{Q})& =\sum_{\lambda \lambda ^{\prime }}\int d\vec{v}\ d\vec{u}%
\ a_{\lambda \lambda ^{\prime }}(\vec{u},\vec{v})\exp \Bigg[\bigg(-\frac{i}{%
\hbar }\bigg)\bigg(\vec{Q}\cdot \vec{u}+\vec{P}\cdot \vec{v}\bigg)\Bigg]%
\Omega _{\lambda \lambda ^{\prime }}, \\
a_{\lambda \lambda ^{\prime }}(\vec{u},\vec{v})& =h^{-8}\int d\vec{p}\ d\vec{%
q}\ a_{\lambda \lambda ^{\prime }}(\vec{p},\vec{q})\exp \Bigg[\bigg(-\frac{i%
}{\hbar }\bigg)\bigg(\vec{q}\cdot \vec{u}+\vec{p}\cdot \vec{v}\bigg)\Bigg],
\\
a_{\lambda \lambda ^{\prime }}(\vec{p},\vec{q})& =\int d\vec{v}e^{\frac{i}{%
\hbar }\vec{p}\cdot \vec{v}}\bigg\langle\vec{q}-\frac{1}{2}\vec{v},\lambda %
\bigg|A_{op}^{b}\bigg|\vec{q}+\frac{1}{2}\vec{v},\lambda ^{\prime }%
\bigg\rangle \\
& =\int d\vec{u}e^{\frac{i}{\hbar }\vec{q}\cdot \vec{u}}\bigg\langle\vec{p}+%
\frac{1}{2}\vec{u},\lambda \bigg|A_{op}^{b}\bigg|\vec{p}-\frac{1}{2}\vec{u}%
,\lambda ^{\prime }\bigg\rangle,
\end{align*}%
where $|\vec{p},\lambda \rangle $ and $|\vec{q},\lambda \rangle $ are the
state vectors representing the Bloch functions and Wannier functions,
respectively, and 
\begin{align*}
\Omega _{\lambda \lambda ^{\prime }}& =\int d\vec{p}\ |\vec{p},\lambda
\rangle \langle \vec{p},\lambda ^{\prime }| \\
& =\int \ d\vec{q}|\vec{q},\lambda \rangle \langle \vec{q},\lambda ^{\prime
}|.
\end{align*}%
}

{\normalsize 
}

\subsection{\protect\normalsize Dressing of conjugate variables in
energy-band quantum dynamics}

{\normalsize A few more words about $\vec{Q}$ and $\vec{P}$. The use of $%
\vec{Q}$, conjugate to the operator $\vec{P}$ of the Hamiltonian in even
form, is preferred in the band-dynamical formalism.\cite{wannier} The reason
we now associate $\vec{Q}$ with the operator $\vec{P}$ of the Hamiltonian in
even form is that this momentum operator now belongs to the respective bands
(each of infinite width) of the \textit{decoupled} Dirac Hamiltonian. This
operator is now analogous to the crystal momentum operator in crystalline
solids. For the original Dirac Hamiltonian $\dot{x}=c$ [from Eq. (\ref{eq4}%
)] leading to a complex \textit{zitterbewegung} motion in $x$-space, whereas
for the Hamiltonian in even form $\dot{Q}=v$ [from Eq. (\ref{eq5})], $c$ is
the speed of light and $v$ the velocity of a wave packet in the classical
limit, and thus $Q$ is more closely related to the band dynamics of fermions
than $x$. Moreover, on the cognizance that the continuum is the limit when
the lattice constant of an array of lattice points goes to zero, there is a
more compelling fundamental basis for using the lattice-position operator $Q$%
. Since quantum mechanics is the mathematics of measurement processes,\cite%
{schwinger} the most probable measured values of the positions are the
lattice-point coordinates. Indeed, these lattice points, or atomic sites,
are where the electrons spend some time in crystalline solids. Therefore the 
\textit{lattice points} and \textit{crystal momentum} are clearly the 
\textit{observables} of the theory and $q$ and $p$ constitute the
eigenvalues of the lattice-point position operator $Q$ and crystal momentum
operator $P$, respectively. Thus, $Q$ is considered here as the generalized
position operator in quantum theory for describing energy-band quantum
dynamics, canonical conjugate to `crystal' momentum operator $\vec{P}$ of
the Hamiltonian in even form. Although the `bare' operator $x$ can still be
used as position operator it only unnecessarily renders very complicated and
almost intractable resulting expressions, \cite{blount, suttorpdegroot}
since this does not directly reflect the appropriate obsevables in band
dynamics as first enunciated by Newton and Wigner\cite{newtonwigner} and by
Wannier several decades ago.\cite{wannier} Thus, in understanding the
dynamics of Dirac relativistic quantum mechanics succinctly, position space
should be defined at discrete points $q$ which are eigenvalues of the
operator $Q$. }

{\normalsize 
}

\subsection{\protect\normalsize The Even Form of Dirac Hamiltonian in a
Uniform Magnetic Field}

{\normalsize The Dirac Hamiltonian for an electron with anomalous magnetic
moment in a magnetic field is 
\begin{equation*}
\mathcal{H}_{op} = \vec{\alpha} \cdot \vec{\Pi}_{op} + \beta mc^{2} - \frac{1%
}{2} (g - 2) \mu_{B} \beta \vec{\sigma} \cdot \vec{B},
\end{equation*}
where 
\begin{align*}
\vec{\Pi}_{op} & = c \vec{P}_{op} - e \vec{A} \bigg( \vec{Q}_{op} \bigg), \\
\mu_{B} & = \frac{e \hbar}{2mc}.
\end{align*}
}

{\normalsize The transformed Hamiltonian in even form $\mathcal{H}%
_{B}^{\prime }$ is given by Ericksen and Kolsrud \cite{erikkols} 
\begin{equation}
\mathcal{H}_{B}^{\prime }=\beta \bigg[m^{2}c^{4}+\Pi ^{2}-e\hbar c(1+\lambda
^{\prime })\vec{\sigma}\cdot \vec{B}+\beta \bigg(\frac{\lambda ^{\prime
}e\hbar }{2mc}\bigg)\sigma \cdot (B\times \Pi -\Pi \times B)\bigg]^{\frac{1}{%
2}},  \label{eq46-2}
\end{equation}%
where $\lambda ^{\prime }=\frac{1}{2}\ (g-2)$, and 
\begin{align*}
\tilde{\Pi}& =cP-eA(Q)-eA(r) \\
& =cP-eA(Q+r), \\
A(Q+r)& =\frac{1}{2}B\times (Q+r), \\
r& =\beta \bigg(\frac{\lambda ^{\prime }\hbar }{mc}\bigg)\sigma .
\end{align*}%
The above Hamiltonian can be written as 
\begin{align}
\mathcal{H}_{B}^{\prime }& =\beta \bigg[m^{2}c^{4}+\Pi ^{2}-e\hbar
c(1+\lambda ^{\prime })\vec{\sigma}\cdot \vec{B}-2\bigg(\frac{1}{2}B\times
r\cdot \Pi \bigg)\bigg]^{\frac{1}{2}}  \notag \\
& =\beta \bigg[m^{2}c^{4}+\Pi ^{2}-e\hbar c(1+\lambda ^{\prime })\vec{\sigma}%
\cdot \vec{B}-2A(r)\cdot \Pi \bigg]^{\frac{1}{2}}  \notag \\
& =\beta \bigg[m^{2}c^{4}+\tilde{\Pi}^{2}-e\hbar c(1+\lambda ^{\prime })\vec{%
\sigma}\cdot \vec{B}-A^{2}(r)\bigg]^{\frac{1}{2}}  \notag \\
& =\beta \bigg[m^{2}c^{4}+\tilde{\Pi}^{2}-e\hbar c(1+\lambda ^{\prime })\vec{%
\sigma}\cdot \vec{B}-\bigg(\frac{\lambda ^{\prime }e\hbar }{2mc}\bigg)%
^{2}B^{2}\bigg]^{\frac{1}{2}}.
\end{align}%
%
%
%
%
%
%
%
%
%
%
%
%
%
%
%
}

\subsection{{\protect\normalsize \label{Tmag}Translation operator, $T_{M}(q)$%
, under uniform magnetic fields}}

{\normalsize In the presence of a uniform magnetic field, magnetic Wannier
Functions, $A_{\lambda}(x-q)$, and magnetic Bloch functions, $%
B_{\lambda}(x,p)$, exist. This is proved by using symmetry arguments. In
general, these two basis functions are complete and span all the
eigensolutions of the magnetic Hamiltonian belonging to a band index $%
\lambda $. The magnetic Wannier Functions $A_{\lambda}(x-q)$ and magnetic
Bloch functions $B_{\lambda}(x,p)$ are related by similar unitary
transformation in the absence of magnetic field, namely, 
\begin{align*}
B_{\lambda}(x,p) & = \frac{1}{(2 \pi \hbar)^{\frac{3}{2}}} e^{(\frac{i}{\hbar%
}) \vec{p} \cdot \vec{x}} \ u_{\lambda}(\vec{p}), \\
A_{\lambda (\vec{x} - \vec{q})} & = \frac{1}{(2 \pi \hbar)^{\frac{3}{2}}}
\int d \vec{p} \ e^{(\frac{i}{\hbar}) \vec{p} \cdot \vec{q}} \
B_{\lambda}(x,p),
\end{align*}
where $\vec{p}$ and $\vec{q}$ are quantum labels. }

{\normalsize Under a uniform magnetic fields, we have for a translation
operator, $T_{M}(q)$, obeying the relation, 
\begin{align}
\nabla_{r} T_{M}(q) & = [P,T_{M}(q)]  \notag \\
& = \frac{ie}{\hbar c} A(q) T_{M}(q).  \label{eq6}
\end{align}
Therefore, 
\begin{equation*}
T_{M}(q) = \exp \bigg( \frac{-ie}{\hbar c} A(r) \cdot q \bigg) C(q),
\end{equation*}
where $C_{0}(q)$ is an operator which do not depend explicitly on $r$. Since 
$T_{M}(q)$ is a translation operator by amount $q$ leads us to write 
\begin{equation*}
C_{0}(q) = \exp (-q \cdot \nabla r), \qquad \text{a pure displacement
operator by amount} - q.
\end{equation*}%
Equation (\ref{eq6}) means that $[P,T_{M}(q)]$ is diagonal if $T_{M}(q)$ is
diagonal, and therefore they have the same eigenfunctions and the same
quantum label. Therefore displacement operator in a translationally
symmetric system under a uniform magnetic field acquire the so-called `%
\textit{Peierls phase factor}'. }

{\normalsize Clearly, bringing the wavepacket or Wannier function around a
closed loop, or around plaquette in the tight-binding limit, would acquire a
phase equal to the magnetic flux through the area defined by the loop. This
is the so-called Bohm-Aharonov effect or Berry phase. Thus, the concept of
Berry phase has actually been floating around in the theory of band dynamics
since the time of Peierls. Berry \cite{berry} has brilliantly generalized
the concept to parameter-dependent Hamiltonians even in the absence of
magnetic field through the so-called \textit{Berry connection}, \textit{%
Berry curvature}, and \textit{Berry flux}. }

{\normalsize The magnetic translation operator generates all magnetic
Wannier functions belonging to band index $\lambda$ from a given magnetic
Wannier function centered at the origin, $A_{\lambda}^{0}(r-0)$, as 
\begin{align*}
A_{\lambda}(r-q) & = T_{M}(q) A_{\lambda}^{0}(r - 0) \\
& = \exp \bigg( \frac{-ie}{\hbar c} A(r) \cdot q \bigg) A_{\lambda}^{0}(r-q).
\end{align*}
We also have the following relation, 
\begin{equation*}
T_{M}(q) T_{M}(\rho) = \exp \bigg( \frac{ie}{\hbar c} A(q) \cdot \rho \bigg) %
T_{M}(q + \rho),
\end{equation*}
\begin{align*}
[T_{M}(q),T_{M}(\rho)] & = \exp \bigg( \frac{ie}{\hbar c} A(q) \cdot \rho %
\bigg) T_{M}(q + \rho) - \exp \bigg( \frac{ie}{\hbar c} A(\rho) \cdot q %
\bigg) T_{M}(\rho + q) \\
& = 2i \sin \bigg( \frac{e}{\hbar c} A(q) \cdot \rho \bigg) T_{M}(q + \rho).
\end{align*}
Moreover, we have, 
\begin{align}
\mathcal{H} B_{\lambda}(x,p) & = E_{\lambda} \Big( p - \frac{e}{c} A(q) %
\Big) B_{\lambda}(x,p),  \notag \\
\mathcal{H} A_{\lambda}(\vec{x} - \vec{q}^{\ \prime}) & = \int dq \ e^{i 
\frac{e}{c} A(q^{\ \prime}) \cdot q} E_{\lambda}(\vec{q}^{^{\prime}} - q)
A_{\lambda}(\vec{x} - \vec{q}),  \label{eq25-2}
\end{align}
and the lattice Weyl transform of any operator, $A_{op}$, is 
\begin{equation}
a_{\lambda \lambda^{\prime}}(p,q) = \int d \vec{v} \ e^{\frac{i}{\hbar} \vec{%
p} \cdot \vec{v}} \bigg\langle A_{\lambda} \bigg( \vec{q} - \frac{1}{2} \vec{%
v} \bigg) \bigg| A_{op} \bigg| A_{\lambda^{\prime}} \bigg( \vec{q} + \frac{1%
}{2} \vec{v} \bigg) \bigg\rangle.  \label{eq15}
\end{equation}
The Weyl transform of the Hamiltonian operator is easily calculated using
Eq. (\ref{eq25-2}) and Eq. (\ref{eq15}). The reader is referred to Ref. (%
\cite{buot3, buot4}) for details of the derivation. Applying Eq. (\ref{eq15}%
) to the even form of the Dirac Hamiltonian, we have 
\begin{align*}
h_{B}^{\prime}(\vec{p},\vec{q})_{\lambda \lambda^{\prime}} & = \int d \vec{v}
\ e^{\frac{i}{\hbar} \vec{p} \cdot \vec{v}} \bigg\langle A_{\lambda} \bigg( 
\vec{q} - \frac{1}{2} \vec{v} \bigg) \bigg| \mathcal{H}_{B}^{\prime} \bigg| %
A_{\lambda^{\prime}} \bigg( \vec{q} + \frac{1}{2} \vec{v} \bigg) \bigg\rangle
\\
& = \int d \vec{v} \ \exp \bigg[ \frac{i}{\hbar} \bigg( p - \frac{e}{c} A(q) %
\bigg) \cdot v \bigg] \tilde{E}_{\lambda}(v;B) \delta_{\lambda
\lambda^{\prime}} \\
& = E_{\lambda} \bigg( \vec{p} - \frac{e}{c} A(q); B \bigg) \delta_{\lambda
\lambda^{\prime}}.
\end{align*}
}

{\normalsize 
}

\subsubsection{{\protect\normalsize The function $E_{\protect\lambda }(\vec{p%
}-\frac{e}{c}A(q);B)\protect\delta _{\protect\lambda \protect\lambda %
^{\prime }}$}}

{\normalsize The function $E_{\lambda}(\vec{p} - \frac{e}{c} A(q);B)$ is the
Weyl transform of $\beta [\mathcal{H}^{2}]^{\frac{1}{2}}$, where the matrix $%
\beta$ served to designate the four bands. In order to calculate $\chi$ we
only need the knowledge of $E_{\lambda}(\vec{p} - \frac{e}{c} A(q);B)$ as an
expansion up to second order in the coupling constant $e$ and after a change
of variable [this is effected by setting $A(q) = 0, p = \hbar k$ in the
expansion], we obtain the expression of $E_{\lambda}(\vec{p} - \frac{e}{c}
A(q); B) |_{A(q) = 0}$, where the dependence in the field $B$ is beyond the
vector potential, 
\begin{equation*}
E_{\lambda} \Big( \vec{k}; B \Big) = E_{\lambda} \Big( \vec{k}; B \Big) +
BE_{\lambda}^{(1)} \Big( \vec{k} \Big) + B^{2} E_{\lambda}^{(2)} \Big( \vec{k%
} \Big) + \cdots
\end{equation*}
The function $E_{\lambda}(\vec{p} - \frac{e}{c} A(q); B) |_{A(q) = 0}$ which
includes the anomalous magnetic moment of the electron is obtained as 
\begin{align*}
E_{\lambda}(k;B) = & \beta \Bigg\{ E - \frac{ec}{2E} \vec{L}_{c.m.} \cdot 
\vec{B} - \frac{(1 + \lambda^{\prime})}{2E} \ e \hbar c \vec{\sigma} \cdot 
\vec{B} - \frac{(1 + \lambda^{\prime})^{2}}{8E^{3}} \bigg( e \hbar c \vec{%
\sigma} \cdot \vec{B} \bigg)^{2} \\
& + \frac{(e\hbar c)^{2} \epsilon^{2}}{8E^{5}} B^{2} \bigg[ 1 + \bigg( \frac{%
\lambda^{\prime} E}{mc^{2}} \bigg)^{2} \bigg] + O(e^{3}) \Bigg\},
\end{align*}
where 
\begin{align*}
\vec{L}_{c.m.} & = \beta \bigg( \frac{\lambda^{\prime} \hbar}{mc} \bigg) 
\vec{\sigma} \times \vec{p}, \\
\epsilon^{2} & = m^{2} c^{4} + c^{2} \hbar^{2} k_{z}^{2}, \\
E \Big( \vec{k} \Big) & = \Big[ m^{2} c^{4} + c^{2} \hbar^{2} k^{2} \Big]^{%
\frac{1}{2}}.
\end{align*}
}

{\normalsize The term, $\vec{L}_{c.m.}$, is a magnetodynamic effect, i.e.,
due to hidden average angular momentum $\vec{L}_{c.m.}$ of a moving
electron. Thus, the introduction of the Pauli anomalous term in $\mathcal{H}$
at the outset endows a rigid-body behavior to the electron, and its angular
momentum about the origin $\vec{L}_{0}$ is 
\begin{equation*}
\vec{L}_{0} = \vec{L}_{MO} + \vec{L}_{c.m.},
\end{equation*}
where $\vec{L}_{MO}$ is the angular momentum about the origin of the system
of charge concentrated as a point at the center of mass and $\vec{L}_{c.m.}$
is the average angular momentumof the system, as a spread-out distribution
of charge about the center of mass. Thus, 
\begin{align*}
\vec{L}_{0} = & \vec{q} \times \vec{p} + \Bigg\langle \sum_{i} \vec{r}_{i}
\times \vec{p}_{i} \Bigg\rangle, \\
\Bigg\langle \sum_{i} \vec{r}_{i} \times \vec{p}_{i} \Bigg\rangle & = \beta %
\bigg( \frac{\lambda^{\prime} \hbar}{mc} \bigg) \vec{\sigma} \times \vec{p},
\end{align*}
\begin{align}
M & = -\Big[ 2E_{\lambda}^{(2)} \Big( \vec{k} \Big) B \Big]_{sp}  \notag \\
& = -\frac{(e \hbar c)^{2} \epsilon^{2}}{4 \Big[ E_{\lambda} \Big( \vec{k} %
\Big) \Big]^{5}} \Bigg[ 1 + \bigg( \frac{\lambda^{\prime} E}{mc^{2}} \bigg)%
^{2} \Bigg]B.  \label{eq50}
\end{align}
The induced magnetic moment due to a distribution of electric charge is 
\begin{equation}
M = -\frac{Be^{2} \langle r^{2} \rangle}{4mc^{2}},  \label{eq51}
\end{equation}
where $\langle r^{2} \rangle$ is the average of the square of the spatial
spread of the distribution normal to the magnetic field. Equating Eqs. (\ref%
{eq50}) with (\ref{eq51}) we obtain 
\begin{equation}
\langle r^{2} \rangle = \frac{mc^{2}(\hbar c)^{2} \epsilon ^{2}}{%
[E_{\lambda} (k)]^{5}} \Bigg[ 1 + \bigg( \frac{\lambda^{\prime} E}{mc^{2}} %
\bigg)^{2} \Bigg].  \label{eq52}
\end{equation}
For positive energy states $E_{\lambda}(k) = (c^{2} \hbar^{2} k^{2} + m^{2}
c^{4})^{\frac{1}{2}}$ and in the nonrelativistic limit, Eq. \ref{eq52})
reduces to 
\begin{equation*}
\langle r^{2} \rangle = (1 + \lambda^{\prime 2}) \bigg( \frac{\hbar}{mc} %
\bigg)^{2},
\end{equation*}
and thus the effective spread of the electron at rest, and for $%
\lambda^{\prime} = 0$, is precisely equal to the Compton wavelength. }

{\normalsize 
}

\subsection{\protect\normalsize Magnetic Susceptibility of Dirac Fermions}

{\normalsize The magnetic susceptibility is given by 
\begin{align*}
\chi = & -\frac{1}{48 \pi ^{3}} \Big( \frac{e}{\hbar c} \Big)^{2}
\sum_{\lambda} \int d \vec{k} \left\{ \frac{\partial^{2} E_{\lambda} \Big( 
\vec{k}; 0 \Big)}{\partial k_{x}^{2}} \frac{\partial^{2} E_{\lambda} \Big( 
\vec{k}; 0 \Big)}{\partial k_{y}^{2}} - \Bigg( \frac{ \partial^{2}
E_{\lambda} \Big( \vec{k}; 0 \Big)}{\partial k_{x} \partial k_{y}} \Bigg)%
^{2} \right\} \frac{\partial f(E_{\lambda})}{\partial E_{\lambda}} \\
& -\bigg( \frac{1}{2 \pi} \bigg)^{3} \sum_{\lambda} \int d \vec{k} \Big[ %
E_{\lambda}^{(1)}(k) \Big]^{2} \frac{\partial f(E_{\lambda})}{\partial
E_{\lambda}} - \bigg( \frac{1}{2 \pi} \bigg)^{3} \sum_{\lambda} \int d \vec{k%
} \ 2E_{\lambda}^{(2)}(k) f(E_{\lambda}).
\end{align*}
Using the following change of variable of integration, 
\begin{equation*}
(\hbar c)^{3} \int d \vec{k} = \int_{-\infty}^{\infty} d \eta \int_{0}^{2
\pi} d \phi \ E \Big( \vec{k} \Big) \ dE \Big( \vec{k} \Big),
\end{equation*}
where 
\begin{equation*}
\eta = \hbar ck_{z},
\end{equation*}
we obtain for the positive energy states the expression for $\chi$ which can
be divided into more physically meaningful terms as 
\begin{equation*}
\chi = \chi_{_{LP}} + \chi_{_{P}} + \chi_{_{sp}} + \chi_{_{g}} +
\chi_{_{MD}},
\end{equation*}
where 
\begin{align}
\chi _{_{LP}} = & \frac{1}{24 \pi^{3}} \bigg( \frac{e}{\hbar c} \bigg)^{2}
\int_{-\infty}^{\infty} d \eta \int_{\epsilon}^{\infty} \frac{\epsilon^{2}}{%
E^{3}} \frac{\partial f(E)}{\partial E} dE,  \label{chi1} \\
\chi_{_{P}} = & -\frac{(1 + \lambda^{\prime})^{2}}{8 \pi^{2}} \bigg( \frac{e%
}{\hbar c} \bigg)^{2} \int_{-\infty}^{\infty} d \eta
\int_{\epsilon}^{\infty} \frac{1}{E} \frac{\partial f(E)}{\partial E} dE,
\label{chi2} \\
\chi_{_{sp}} = & -\frac{1}{8 \pi^{2}} \bigg( \frac{e^{2}}{\hbar c} \bigg) %
\int_{-\infty}^{\infty} d \eta \int_{\epsilon}^{\infty} \frac{\epsilon^{2}}{%
E^{4}} \Bigg[ 1 + \bigg( \frac{\lambda^{\prime} E}{mc^{2}} \bigg)^{2} \Bigg] %
f(E) dE,  \label{chi3} \\
\chi_{_{g}} = & \frac{(1 + \lambda^{\prime})^{2}}{8 \pi^{2}} \bigg( \frac{%
e^{2}}{\hbar c} \bigg) \int_{-\infty}^{\infty} d \eta
\int_{\epsilon}^{\infty} \frac{f(E)}{E^{2}} dE,  \label{chi4} \\
\chi_{_{MD}} = & -\frac{\lambda^{\prime 2}}{8 \pi^{2}} \bigg( \frac{e^{2}}{%
\hbar c} \bigg) \int_{-\infty}^{\infty} d \eta \int_{\epsilon}^{\infty} 
\frac{(E^{2} - \epsilon^{2})}{(mc^{2})^{2} E} \frac{\partial f(E)}{\partial E%
} dE,  \label{chi5}
\end{align}
where 
\begin{align*}
\bigg( \frac{ec}{2E} \vec{L}_{c.m.} \bigg)_{z}^{2} & = \bigg( \frac{%
\lambda^{\prime} e \hbar c}{2mc^{2}} \bigg) \frac{(E^{2} - \epsilon^{2})}{%
E^{2}}, \\
\vec{B} & = B \frac{\vec{z}}{|\vec{z}|}.
\end{align*}
The total susceptibility for the positive energy states is 
\begin{equation}
\chi = \frac{1}{(2 \pi)^{2}} \bigg( \frac{e^{2}}{\hbar c} \bigg) \bigg[ (1 +
\lambda^{\prime})^{2} - \frac{1}{3} \bigg] \int_{0}^{\infty } d \eta \ \frac{%
f(\epsilon)}{\epsilon}\frac{1}{(2 \pi)^{2}} \bigg( \frac{e^{2}}{\hbar c} %
\bigg) \bigg( \frac{\lambda^{\prime}}{mc^{2}} \bigg)^{2} \int_{0}^{\infty} d
\eta \ G(\epsilon - \mu),  \label{eq62}
\end{equation}
where 
\begin{align*}
G(\epsilon - \mu) = & k_{B} T \ln \bigg\{ 1 + \exp \bigg[ - \frac{(\epsilon
- \mu)}{k_{B} T} \bigg] \bigg\} \\
& = \int_{\epsilon}^{\infty} f(E) dE.
\end{align*}
The contributions of the holes is obtained by replacement of $f(\epsilon)$
and $G(\epsilon - \mu)$ in Eq. (\ref{eq62}) by $(1 - f(-\epsilon))$ and $%
G(\epsilon + \mu)$, respectively. }

{\normalsize The relative importance of terms that made up $\chi$ at $T = 0$
of Dirac fermions, where $n$ is the electron density, $k_{F} = (3 \pi^{2}
n)^{\frac{1}{3}}$, $\eta_{F} = \hbar ck_{F}$ and $E_{F} = (\Delta^{2} +
\eta_{F}^{2})^{\frac{1}{2}}$, is summarized below. }

\begin{center}
{\normalsize \ {\footnotesize \ 
\begin{tabular}{|l|l|l|}
\hline
Various Contributions to $\chi_{Dirac}$ at ${T = 0}$ & Nonrelativistic, $%
\frac{\eta_{F}}{\Delta} \ll 1$ & Ultrarelativistic, $\frac{\eta_{F}}{\Delta}
\gg 1$ \\ \hline
$\chi_{_{LP}} = -\frac{1}{12 \pi^{2}} \Big( \frac{e^{2}}{\hbar c} 
\Big) 
    \frac{1}{E_{F}^{3}} \Big( \frac{\eta_{F}^{3}}{3} + \Delta^{2} \eta_{F} %
\Big)$ & $-\frac{1}{12 \pi^{2}} \Big( \frac{e}{mc^{2}} \Big) k_{F}$ & $-%
\frac{1}{12 \pi^{2}} \Big( \frac{e^{2}}{\hbar c} \Big) \frac{1}{3}$ \\ \hline
$\chi_{_{P}} = \frac{1}{4 \pi^{2}}(1 + \lambda^{\prime})^{2} \Big( 
    \frac{e^{2}}{\hbar c} \Big) \frac{\eta_{F}}{E_{F}}$ & $\frac{1}{4 \pi^{2}%
}(1 + \lambda^{\prime})^{2} \Big( \frac{e}{mc^{2}} \Big) k_{F}$ & $\frac{1}{%
4 \pi^{2}}(1 + \lambda^{\prime})^{2} \Big( \frac{e^{2}}{\hbar c} \Big)$ \\ 
\hline
$\chi_{_{MD}} = -\frac{\lambda^{\prime 2}}{4 \pi^{2}} \Big( 
    \frac{e^{2}}{\hbar c} \Big) \frac{1}{\Delta^{2}} \bigg[ \frac{1}{E_{F}} %
\Big( \frac{\eta_{F}^{3}}{3} + \Delta^{2} \eta_{F} \Big) - \eta_{F} E_{F} %
\bigg]$ & $\Longrightarrow 0$ & $\frac{\lambda^{\prime 2}}{4 \pi^{2}} 
\Big( 
    \frac{e^{2}}{\hbar c} \Big) \frac{2}{3} \Big( \frac{\eta_{F}}{\Delta} %
\Big)^{2}$ \\ \hline
$\chi_{_{spread}} = -\frac{1}{12 \pi^{2}} \Big( \frac{e^{2}}{\hbar c} 
\Big) 
    \sinh^{-1} \Big( \frac{\eta_{F}}{\Delta} \Big) - \chi_{_{LP}}$ &  & $-%
\frac{1}{12 \pi^{2}} \Big( \frac{e^{2}}{\hbar c} \Big) \Big[ \ln \frac{2
\eta_{F}}{\Delta} - \frac{1}{3} \Big]$ \\ \hline
\qquad $-\frac{\lambda^{\prime 2}}{4 \pi^{2}} \Big( \frac{e^{2}}{\hbar c} %
\Big) \frac{1}{\Delta^{2}} \bigg[ \frac{\eta_{F}(\eta_{F}^{2} + \Delta^{2})^{%
\frac{1}{2}}}{2} + \frac{\Delta^{2}}{2} \sinh^{-1} \Big( 
    \frac{\eta_{F}}{\Delta} \Big) \bigg]$ &  &  \\ \hline
\qquad $+ \frac{\lambda^{\prime 2}}{4 \pi^{2}} \Big( \frac{e^{2}}{\hbar c} %
\Big) \frac{1}{E_{F}} \Big( \frac{\eta_{F}^{3}}{3} + \Delta^{2} \eta_{F} %
\Big) \frac{1}{\Delta^{2}}$ & $\Longrightarrow 0$ & $-\frac{\lambda^{\prime
2}}{4 \pi^{2}} \Big( \frac{e^{2}}{\hbar c} \Big) \Big[ \frac{1}{6} 
\big( 
    \frac{\eta_{F}}{\Delta} \big)^{2} + \frac{1}{2} \ln \frac{2 \eta_{F}}{%
\Delta} \Big]$ \\ \hline
$\chi_{_{g}} = \frac{1}{4 \pi^{2}}(1 + \lambda^{\prime})^{2} \Big( 
    \frac{e^{2}}{\hbar c} \Big) \Big[ \sinh^{-1} \Big( \frac{\eta_{F}}{\Delta%
} \Big) - \frac{\eta_{F}}{E_{F}} \Big]$ & $\Longrightarrow 0$ & $\frac{1}{4
\pi^{2}}(1 + \lambda^{\prime})^{2} \Big( \frac{e^{2}}{\hbar c} \Big) 
\Big[ 
    \ln \frac{2 \eta_{F}}{\Delta} - 1 \Big]$ \\ \hline
\end{tabular}
}}
\end{center}

{\normalsize 
}

\subsection{\protect\normalsize Displacement Operator under Uniform High
External Electric Fields}

{\normalsize To complement Sec. \ref{Tmag}, we give the translation operator
for uniform electric field case, $\mathcal{H} = \mathcal{H} - e \vec{F}
\cdot \vec{x}$. We have for the displacement operator, $T_{E}(q)$, obeying
the relation, 
\begin{align*}
i \hbar \dot{T}_{E}(q) & = [T_{E}(q), \mathcal{H}], \\
\dot{T}_{E}(q) & = \frac{ie}{\hbar} F \cdot q \ T_{E}(q).
\end{align*}
Therefore 
\begin{equation*}
T_{E}(q) = C_{0}(q,\tau) \exp \bigg( \frac{ie}{\hbar}\bigg) Ft \cdot q,
\end{equation*}
where $C_{0}(q,\tau)$ is an operator which do not depend explicitly on time, 
$t$. $T_{E}(q)$, being a displacement operator in space and time lead us to
write the operator 
\begin{equation*}
C_{0}(q,\tau) = \exp \Bigg( q \cdot \frac{\partial}{\partial r} + \tau \frac{%
\partial}{\partial t} \Bigg).
\end{equation*}
}

{\normalsize $T_{E}(q)$ plays critical role similar to $T_{M}(q)$ for
establishing the phase space quantum transport dynamics at very high
electric fields, where we consider realistic transport problems as
time-dependent many-body problems. For zero field case we are dealing with
biorthogonal Wannier functions and Bloch functions because the Hamiltonian
is no longer Hermetian due to the presence of energy variable, $z$, in the
self-energy. This means that $[T_{E}(q), \mathcal{H}]$ is diagonal in the
bilinear expansion if $T_{E}(q)$ is diagonal. The eigenfunction of the
`lattice' translation operator $T_{E}(q)$ must then be labeled by a
wavenumber $\vec{k}$ which is varying in time as 
\begin{equation*}
\vec{k} = \vec{k}_{0} + \frac{e \vec{F}}{\hbar} t,
\end{equation*}
and $\mathcal{H}$ is also diagonal in $\vec{k}$. Similarly, the energy
variable, $z$, in the Hamiltonian must also vary as 
\begin{equation*}
z = z_{0} + e \vec{F} \cdot \vec{q}.
\end{equation*}%
}

{\normalsize Similar developments for translationally invariant many-body
system subjected to a uniform electric field allows us to define the
corresponding electric Bloch functions and electric Wannier functions, in a
unifying manner for both magnetic and electric fields. This electric-field
version allows us to derive the quantum transport equation of the particle
density at very high electric fields. This will be discussed in another
communication dealing with quantum transport in many-body systems. \ A more
general displacement operator is recently given by Buot \cite{covar}.}

{\normalsize 
}

\end{document}